\documentclass[titlesmallcaps, copyrightpage]{uqthesis}
\pdfoutput=1

\usepackage{amsmath}
\usepackage{braket}
\usepackage{graphics}
\usepackage[usenames,dvipsnames]{color}
\usepackage{natbib}
\usepackage{pdfpages}
\usepackage{graphicx}
\usepackage{eurosym}
\usepackage{appendix}
\usepackage{play}
\usepackage[grey,times]{quotchap}
\usepackage{makeidx}
\makeindex
\usepackage{hyperref}
\usepackage{listings}
\usepackage[nottoc,numbib]{tocbibind}
\usepackage{verbatim}
\usepackage{wrapfig}
\usepackage{booktabs}
\usepackage{amsfonts}
\usepackage{array}
\usepackage[acronym,nomain,nonumberlist]{glossaries}
\usepackage{listings}
\usepackage{setspace}
\usepackage[linewidth=1pt]{mdframed}
\usepackage{sverb}
\usepackage{aas_macros}
\usepackage{ctable}
\usepackage{bookmark}
\bookmarksetup{
  numbered, 
  open,
}
\usepackage{epstopdf}

\makeglossaries
\renewcommand*{\glossaryentrynumbers}[1]{}

\newcommand\abs[1]{\left|#1\right|}

\newcommand{\camb}{\textsc{camb}}

\newcommand{\kmsmpc}{km\,s$^{-1}$\,Mpc$^{-1}$}

\newcommand{\halofit}{\textsc{halofit}}
\newcommand{\specialcell}[2][c]{\begin{tabular}[#1]{@{}c@{}}#2\end{tabular}}

\begin{document}

\hypersetup{pageanchor=true}
\pdfbookmark{Title}{toc}

\title{Undergraduate Thesis\\ \vspace{0.5 cm} Extraction of cosmological information from WiggleZ}
\author{Samuel Hinton}
\department{Science}

\renewcommand{\degreetext}{in partial fulfilment of the Bachelor of Science (Honours) degree\\ in the
discipline of Physics}

\frontmatter

\titlepage


\begin{flushright}
Samuel Hinton\\ 41966855\\ 30 Matingara St, Chapel Hill, QLD 4069\\
\end{flushright}

\noindent \today \\
\noindent Associate Professor Tim McIntyre\\
Head of Physics\\
School of Mathematics and Physics\\
The University of  Queensland\\
St Lucia QLD 4072\\

\noindent Dear Associate Professor Tim McIntyre,\\ \\
In accordance with the requirement of the Degree of Bachelor of Science (Honours) in the School of Mathematics and Physics, I submit the following thesis entitled:

\begin{center}
  \emph{``Extraction of cosmological information from WiggleZ''}
\end{center}

\noindent The thesis was performed under the supervisor Prof. Tamara Davis and Chris Lidman. I declare that the work submitted in thesis is my own, except as acknowledge in the text and footnotes, and has not been previously submitted for a degree at the University of Queensland or any other institution. \\

\noindent Yours sincerely \\ \\ 

\noindent \line(1,0){250} \\

\noindent Samuel Hinton


\chapter{Acknowledgements}

I would like to thank my supervisors Tamara Davis and Chris Lidman for their significant help during the construction of this thesis. In particular, I am grateful for Tamara's patience with my stream of questions - some of them well thought out - more of them perhaps asked too quickly before I had thought them through properly. I am also grateful for the suggestions and feedback on fitting methodology provided by Chris Blake throughout the year.\\

I am grateful to St John's College and the AAO for the assistance their scholarships have provided in the creation of this thesis.\\

I would also like to thank Joshua Calcino, Carolyn Wood and Sarah Thompson for their emotional support during the year!\\

\chapter{Abstract}

In this thesis, I analyse the 2D anisotropic Baryon Acoustic Oscillation (BAO) signal present in the final WiggleZ dataset. I utilise newly released covariance matrices from the WizCOLA simulations and follow well tested methodologies used in prior analyses of the BAO signal in alternative datasets. The WiggleZ data is presented in two forms - in multipole expansion and in data wedges - both of which are analysed in this thesis. I test my model against previous one dimensional analyses of the BAO signal in the WiggleZ data and against WizCOLA simulations, and find it to be free of bias or systematic offsets. However, the analysis on the wedge data format was unable to properly constrain cosmological parameters, and thus discarded in favour of the multipole analysis. The multipole analysis determined $\Omega_c h^2$, $H(z)$ and $D_A(z)$ for three redshift bins, of effective redshifts $z = 0.44, 0.60$ and $0.73$. The respective constraints on $\Omega_c h^2$ are $0.119^{+0.029}_{-0.026}$, $0.151^{+0.038}_{-0.025}$ and $0.140^{+0.036}_{-0.022}$. The fits for $H(z)$ are respectively  $87 \pm 16$, $90 \pm 15$ and $82 \pm 13$ \kmsmpc, and for $D_A(z)$ I find values of $1300 \pm  160$, $1300 \pm  180$ and $1350 \pm  160$ Mpc. These cosmological constraints are consistent with Flat $\Lambda$CDM cosmology.

\hypersetup{pageanchor=true}
\pdfbookmark[chapter]{Table of Contents}{toc}

\tableofcontents


\mainmatter

\chapter{Introduction}

Modern cosmological observations have given strict constraints on cosmological parameters and model viability, and indicate a late time accelerated expansion of the universe \citep{RiessFilippenko1998, PerlmutterAldering1999, SpergelVerde2003, RiessStrolger2004, TegmarkBlanton2004, SanchezBaugh2006, SpergelBean2007, Komatsu2009, RiessMacri2009, PercivalReid2010, ReidPercival2010,BlakeKazin2011}. This accelerating expansion is one of the foremost problems in cosmology, and efforts to determine the expansion history of the universe will allow differentiation between many proposed models \citep{AlbrechtBernstein2006, SanchezScoccola2012}. One area of promising development is detecting Baryon Acoustic Oscillations (BAO) in the large scale structure of the universe, as the BAO signal provides a robust and precise measurement of the history of the universe's expansion rate and size \citep{BlakeGlazebrook2003,HuHaiman2003,Linder2003,SeoEisenstein2003}. Analysis of the BAO signal has been performed on modern cosmology surveys, providing tight constraints on cosmological parameters \citep{Gaztanaga2009, SanchezKazinBeutler2013, AndersonAubourg2014}. The constraints BAO measurements provide are highly complimentary to, and can be used in conjunction with, constraints derived from measurements on the Cosmic Microwave Background (CMB) \citep{BennettHalpern2003, Planck201416}, weak lensing \citep{VanWaerbeke2000,WittmanTyson2000,KaiserWilson2000} and supernova data \citep{KowalskiRubin2008, KesslerBeckerCinabro2009, BetouleKessler2014}.\\

From this motivation, I attempt to extract useful cosmological information from the BAO signal present in the WiggleZ dataset \citep[WiggleZ;][]{Drinkwater2010} beyond the analyses already completed by the WiggleZ team in \citet{BlakeBroughColless2011, BlakeDavis2011, BlakeGlazebrook2011, BlakeKazin2011, Parkinson2012}. In this document, I layout the sections as follows: In Chapter 2 I introduce relevant modern cosmology for any non-technical audience. Chapter 3 contains a summary of prior literature in which the BAO signal has been used to constrain cosmological parameters in this dataset and others previously. In Chapter 4 I construct my BAO model and test it against prior studies and WizCOLA mocks, and in Chapter 5 this model is then applied to the WiggleZ dataset. Chapter 6 presents my conclusions.

\chapter{Background}
\label{ch:back}

\section{Modern Cosmology}

Due to advances in modern technology, modern cosmology is an area of rapid scientific growth. Underpinning modern cosmology is one fundamental assumption, called the Cosmological Principle, which states that on sufficiently large scales ($\sim 150$ Mpc), the universe is both isotropic and homogeneous. These assumptions have been tested and found to be in good agreement with observations of the universe \citep{Lahav2001,HansenBanday2004, HoggEisenstein2005, ScrimgeourDavis2012,SchwarzBacon2015}. From the cosmological principle and the Theory of General Relativity, Friedmann derived the dynamics of the universe in terms of energy content \citep{RydenPeterson2010}. Before detailing the Friedmann equations, one must understand the metrics and basic cosmology involved.

\subsection{Friedmann-Lema\^itre-Robertson-Walker Cosmology} \label{sec:frw}

The common metric used in to describe an expanding universe in modern cosmology is the Friedmann-Lema\^itre-Robertson-Walker metric, commonly abbreviated to the FLRW metric or the FRW metric. In spherical form, the metric can be written as
\begin{align}
ds^2 = -c^2 dt^2 + R(t)^2 \left[ d\chi^2 + S_\kappa(\chi)^2 d\Omega^2 \right], 
\end{align}
where $s$ denotes the proper distance, $c$ the speed of light, $R(t)$ the time dependent radius of the universe, $\chi$ the dimensionless radial distance, $d\Omega$ the spherical coordinates such that $d\Omega^2 \equiv d\theta^2 + \sin^2\theta d\phi^2$ and $S_\kappa(\chi)$ is dependent on the geometry of the universe, such that
\begin{align}
S_\kappa(\chi) = \begin{cases}
    \sin(\chi), & (\kappa = +1)\\
    \chi, & (\kappa = 0) \\
    \sinh (\chi) & (\kappa = -1),
  \end{cases}
\end{align}
where $\kappa$ (representing the curvature of the universe) only needs to be given for three distinct values due to the ability to scale the metric without changing the underlying physics. As such, $\kappa = 1$ corresponds to a closed (spherical) universe, $\kappa = 0$ is a flat universe, and $\kappa = -1$ represents an open (hyperbolic) universe. Curvature can be thought of in multiple ways, where perhaps the two most conceptually simple methods of understanding relate to parallel lines and the angles in a triangle. Consider two rays of light emitted at some point parallel to one another. In a closed universe, these lines would eventually converge, in a flat universe they would stay parallel, and in an open universe they would diverge. For a real world example, consider the surface of the Earth, which is closed (spherical); if one were to draw to parallel lines towards the North Pole, they would converge at said pole. 
\begin{wrapfigure}{r}{0.4\textwidth}
  \begin{center}
    \includegraphics[width=0.4\textwidth]{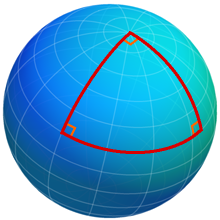}
  \end{center}
  \caption{An illustration of how angles in triangles on a surface with positive curvature exceed 180 degrees, courtesy of \citet{LegnerMathigones}.}
  \label{fig:triangle}
\end{wrapfigure}
Another useful way to conceptualise the curvature of space is to sum the angles on a triangle. In flat space, they will add to 180 degrees, as expected. In open space, they would sum to less, and would sum to more in closed space. Again we can use the Earth as a good starting point from this - it is possible to draw a triangle with three ninety degree angles by having one point at the north pole and two other points on the equator, as illustrated in Figure \ref{fig:triangle}. \\

In order to simplify explanations, we will be working with flat geometry in this document, as cosmological observations highly support a flat universe \citep{DavisMortsell2007,Mortonson2009,Planck201416}. With this simplification, we see that the metric reduces down to
\begin{align} \label{eq:flatmetric}
ds^2 = -c^2 dt^2 + R(t)^2 \left[ d\chi^2 + \chi^2 d\Omega^2 \right].
\end{align}
If we wish to find the distance between two objects at time $t_0$, one can simply transform the coordinate system such that $d\Omega$ vanishes and then integrate, giving that $D = R(t_0) \chi \equiv R_0 \chi$. Thus $\chi$ represents a distance independent of the the scale of the universe, which is denoted the comoving distance \citep{CarrollOstlie2006}.  As such, $R_0 \chi$ represents the distance $D$ between two objects if measured in the present day.\\

If we do not restrict ourselves to the present day, we have $D = R(t) \chi$, which we can also write $D = a(t) R_0 \chi$, where $a(t)$ is now a scaling factor normalised to $1$ for the present day. We can also see that, as $a(t)$ is explicitly time dependent, its time derivative is non-zero. From this fact we can recover the famous Hubble's law \citep{Hubble1929}, such that we consider the rate of change of proper distance between two objects with no peculiar velocity due to relative motion through space (i.e., their recession velocity). Note that, as discussed previously, $\chi$ for comoving objects is independent of time and scalefactor, and is thus treated as a constant.
\begin{align}
\dot{D} &= \dot{a} R_0 \chi  = \frac{\dot{a}}{a} (a R_0 \chi) = \frac{\dot{a}}{a} D\\
v_{\text{rec}} &= H D,
\end{align}
where $v_{\text{rec}} \equiv \dot{D}$ and $H \equiv \dot{a}/a$ is Hubble's constant - the ratio of the rate at which the universe is currently expanding relative to its size. We should note that it is possible for $v_{\text{rec}}$ to exceed the speed of light. This has been the cause of some confusion in the past, however as special relativity says nothing can travel \textit{through} space greater than the speed of light, and recession velocity is not due to travelling through space, but instead space expanding, this result is allowed. For more details on this, please see \citet{DavisLineweaver2004}. Hubble's constant is traditionally given in units of \kmsmpc, but can easily be written simply in terms of inverse time, so $H^{-1}$ has units of time and is known as Hubble time. Similarly, the Hubble distance is defined as $D_H = c/H$, and this length corresponds to the distance at which recession velocity due to the expansion of space is the speed of light.\\

The expansion of space has an important effect on light travelling through it, in that the wavelength of the light expands along with space, causing light to be progressively redshifted as it travels through the cosmos. Redshift, denoted $z$, is defined as 
\begin{align}
z \equiv \frac{\lambda_{\text{ob}} - \lambda_{\text{em}}}{\lambda_{\text{em}}},
\end{align}
where $\lambda_{\text{ob}}$ is the wavelength of light that is observed and $\lambda_{\text{em}}$ is the wavelength of light emitted from the source. As the scalefactor is linked with wavelength, we also find
\begin{align}
1 + z = \frac{a(t_{\text{ob}})}{a(t_{\text{em}})} = \frac{1}{a(t_{\text{em}})} \quad\rightarrow \quad z = \frac{1}{a(t_{\text{em}})} - 1,
\end{align}
if we set $t_{\text{ob}} = t_0$. For a more rigorous derivation of this relationship, see \citet[Ch 3.4]{RydenPartridge2004}. Redshift is what we observe in cosmological surveys, and the ability to link the expansion of the universe to redshift is thus fundamental to our ability to do precision cosmology, and by measuring the redshifts of various targets we are able to map the expansion dynamics of the universe. It should also be noted that there are two primary methods of determining redshift - spectroscopically or photometrically. Without going into detail, spectroscopic measurements use a spectrum and are far more accurate but far slower to gather than photometric redshifts, which use images taken in several broad colour filters.\\

These dynamics were formalised by Friedmann in the two eponymous equations given below \citep{RydenPartridge2004}:
\begin{align}
\left(\frac{\dot{a}}{a}\right)^2 &= \frac{8\pi G}{3} \rho(t) - \frac{\kappa}{R_0^2} \frac{1}{a(t)^2} + \frac{\Lambda}{3} \\
\frac{\ddot{a}}{a} &= - \frac{4\pi G}{3} (\rho(t) + 3p) + \frac{\Lambda}{3},
\end{align}
where $\Lambda$ is Einstein's cosmological constant (a contender for dark energy), $\rho(t)$ is the density of the fluid, $G$ is Newton's gravitational constant, $R_0$ is the radius of curvature of the universe and $\kappa$ is the curvature parameter encountered previously. One can write the cosmological constant in terms of density with a change of variables, such that we find
\begin{align}
\left(\frac{\dot{a}}{a}\right)^2 &= \frac{8\pi G}{3} (\rho_m + \rho_\Lambda) - \frac{\kappa}{R_0^2} \frac{1}{a(t)^2},
\end{align}
where $\rho_\Lambda = \Lambda / 8\pi G$. Setting a critical density $\rho_c$ such that $\kappa = 0$, and substituting in $H = \dot{a}/a$, we have 
\begin{align}
\rho_c = \frac{3H^2}{8\pi G}.
\end{align}
This is done so that we can move to dimensionless fractions of critical density, such that $\Omega_x = \rho_x / \rho_c$. We should distinguish that this $\Omega$ is distinct from the prior usage of $\Omega$ as spherical coordinates. From this, we can easily separate out contributions to total energy density from different sources (such as matter, cosmological constant and radiation), which allows us to find the difference from a critical density $\Omega_k$, formally
\begin{align} \label{eq:omk}
\Omega_k = 1 - \sum_x \Omega_x.
\end{align}
Building upon this we can combine the fluid equation and acceleration equation with the Friedman equation \citep[see][Ch 4.2, 4.3, for full derivation]{RydenPartridge2004} to model the equation of state for each contributing fluid (matter, radiation, cosmological constant are all modelled as perfect fluids), with the equation of state $w$ defined as $w \equiv p/\rho$ (pressure over density). The evolution dynamics are given such that the density fraction evolves as
\begin{align} \label{eq:hz}
H(t)^2 = H_0^2 \sum_x \Omega_x a^{-3(1+w_x)}.
\end{align}
Non-relativistic matter (also known as cold matter) has an equation of state of $w = 0$, meaning its density evolves as $a^{-3}$, or inversely proportional to volume, as one would expect when treating matter as pressureless dust. In other words, the energy per particle is simply the rest mass energy from $E=mc^2$, so the energy should drop in proportion to density, which is inversely proportional to volume. As discussed previously, light becomes redshifted during expansion. If we imagine a sea of photons, we can see that photon density would drop proportionally to volume, and in addition, as the energy of a photon is given by $E=hc/\lambda$, the increase in $\lambda$ proportional to the increase in $a$ causes each individual photon to lose energy as well. Combining these two factors, the equation of state of radiation is $w = 1/3$ and thus evolves as $a^{-4}$. In $\Lambda$CDM cosmology, dark energy represents the energy density of the vacuum, and is thus constant, giving $w = -1$. Finally the curvature of the universe $\Omega_k$ has $w=-1/3$ and thus evolves as $a^{-2}$, although we should note that this does not represent a physical energy as the other terms, it simply comes from the mathematical formalism found in equation \eqref{eq:omk}. Together, this gives
\begin{align} \label{eq:dynamics}
H(t)^2 = H_0^2 \left( \Omega_m a^{-3} + \Omega_r a^{-4} + \Omega_k a^{-2} + \Omega_\Lambda \right).
\end{align}
For the present universe, radiation pressure has dropped sufficiently for it to often be discarded as negligible ($\Omega_r < 10^{-4}$), however this was not the case in the early universe \citep{RydenPartridge2004, Planck201416}. As we shall be primarily dealing with the Flat $\Lambda$CDM model, and due to the measured flatness of the universe ($\Omega_k = 0$ within error), we can further simplify the above equation to
\begin{align}
H(t)^2 = H_0^2 \left( \Omega_m a^{-3} + \Omega_\Lambda \right).
\end{align}
We can also take the ratio of $H(z)$ to $H_0$, denoted $E(z)$, giving
\begin{align}
E(z) = H(z) / H_0 = \sqrt{\sum_x \Omega_x a^{-3(1+w_x)}}.
\end{align}
Using the metric from equation \eqref{eq:flatmetric} along the radial component such that $ds = d\Omega = 0$, we can show (via $da = -a^2 dz$) that 
\begin{align}
D(t,z) = R_0 a(t) \chi = c \int_0^z \frac{d z^\prime}{H(z^\prime)} = \frac{c}{H_0} \int_0^z \frac{dz^\prime}{E(z^\prime)}.
\end{align}
For small $\Delta z$ in which we can approximate $H(z)$ as constant and $\Delta v \approx c \Delta z$, we can thus write the radial distance as $\Delta D_\parallel = c\Delta z/H(z)$. We can also express transverse distance $\tilde{D}(t,z) = R(t) S_\kappa(\chi)$, and further define the angular diameter distance $D_A \equiv \tilde{D} / (1 + z)$. For flat universes, this reduces to $D_A = R(t) \chi(z) / (1 + z)$.\\

For a treatment and review of other cosmological models, from dynamical dark energy \citep{PeeblesRatra1988} to exotic models such as Chaplygin gas \citep{BentoBertolami2003, Benaoum2012} please see \citet{PeeblesRatra2003,DavisMortsell2007, FriemanTurnerHuterer2008, GottSlepian2011}. My analysis in thesis is mainly concerned with testing and parametrizing the Flat $\Lambda$CDM model, as it is the primarily favoured cosmological model \citep{SanchezKazinBeutler2013, Planck201416}, and I will be explicit about any extensions I make to the model that deviate from Flat $\Lambda$CDM. Equation \eqref{eq:dynamics} gives the dynamical evolution of the universe, and in order to understand the origin and significance of Baryon Acoustic Oscillations I will elaborate on the early history of the universe in which they were formed.\\

\subsection{A brief history of the universe}

The origins of the BAO stretch back to the beginning of the universe, and so we delve into Big Bang cosmology to provide a sufficient background. The Big Bang refers to a point in space-time at which our models break down due to a predicted singularity, and immediately after the Big Bang the universe was an extraordinarily dense, hot, soup of energy. Importantly, this soup was not completely homogeneous, for it contained tiny perturbations in energy density thought to be the result of quantum fluctuations expanded to macroscopic size via the process of inflation and then expansion. Three minutes in, the universe had expanded (and thus cooled) enough for the first nucleons to form - Hydrogen, Helium and Lithium. Radiation density is still sufficiently high that baryonic matter and light is strongly coupled via the process of Thomson scattering \citep{PeeblesYu1970, SunyaevZeldovich1970, Doroshkevich1978}.

\begin{figure}[h!]
  \begin{center}
    \includegraphics[width=\textwidth]{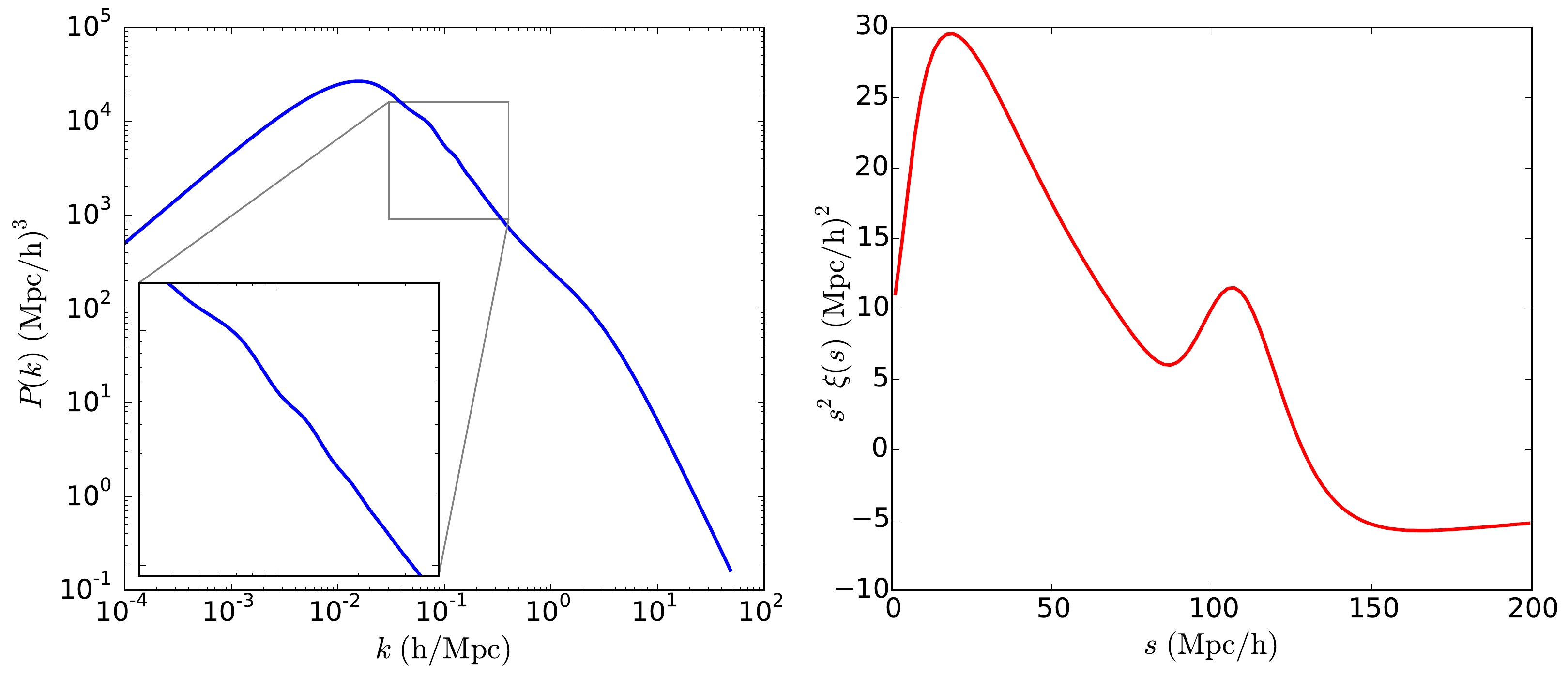}
  \end{center}
  \caption{The left hand panel represents the power spectrum after recombination of the early universe, created using parameters from \citet{Planck201416}. The angle average correlation function is displayed in the right hand panel, and the rise at approximately 110 Mpc/$h$ is the BAO peak. Notice that due to the small amplitude of the BAO peak, correlation functions are traditionally displayed not with power $\xi(s)$, but with $s^2 \xi(s)$. As such, the amplitude of the BAO peak is visually presented approximately ten thousand times stronger than it actually is.}
  \label{fig:Backgroundpk2xi}
\end{figure}

As the universe continues to expand, the density fluctuations move through the ultra-relativistic matter-photon fluid as acoustic waves. As the universe continues to expand and cool, at approximately 3000K free electrons bind to atomic nuclei, and we define the point at which the mean free path length of light is the Hubble distance as the point of recombination. We observe the light from this period as the Cosmic Microwave Background (CMB), and many cosmological studies have utilised measurements on the CMB to constrain cosmological parameters and models \citep{BoggessMather1992,BennettLarson2013,Planck20151}. As the universe expands further still, the radiation density continues to decrease faster than the matter density, and as the pressure on matter from photons drops, eventually the mean free path of atomic nuclei exceeds the Hubble length, indicating that the influence of radiation on particle dynamics is now at an end. This point is known as the drag epoch, and represents the point where acoustic waves freeze out, unable to propagate without sufficient coupling to light. The drag epoch occurs after the point of recombination, when the universe was approximately 2\% larger than than the point of recombination, and corresponds to a redshift of $z_d \sim 1060$ compared to the redshift of the CMB at $z_* \sim 1090$ \cite{Planck2015Parameters}. The length scale at which these acoustic oscillations end up is the characteristic size of large scale structure, and the final pattern of acoustic oscillations are known as Baryon Acoustic Oscillations (BAO). As such, BAO refer to a preferred length scale in large-scale structure formation that corresponds to the density fluctuations imprinted in the universe at the end of the drag epoch \citep{BondEfstathiou1984, Holtzman1989, HuSugiyama1996, EisensteinHu1998, MeiksinWhitePeacock1999}. The comoving size of this characteristic length remains constant throughout the evolution of the universe, and by examining the galaxy distribution in the universe with a two point correlation function, this increased density of structure at the characteristic length is revealed statistically as a single well-defined peak in the matter correlation function \citep{Matsubara2004}. An example power spectrum and its associated correlation function are given in Figure \ref{fig:Backgroundpk2xi}.

Furthermore, due to the finite speed of light, looking further out in the universe represents a look into the past, and thus by measuring the BAO signal at different redshifts in the universe, we have a method of determining the expansion history of the universe. For more detail on early universe physics, please see \citet{BashinskyBertschinger2001,BashinskyBertschinger2002}.

\subsection{Baryon Acoustic Oscillations - 1D and 2D}

As discussed above, one dimensional baryon acoustic oscillations can be measured via the creation of a two point correlation function, where the distribution of real-space comoving distance $\chi$ between pairs of objects reveals the BAO peak. Alternatively, the comoving distance $\chi$ can be broken into component vectors $\sigma$ and $\pi$, which respectively give the distance between the object perpendicular to the line of sight and parallel to the line of sight. Decomposing the BAO signal into two dimensions offers greater ability to constrain cosmology at the cost of requiring larger data sets. Whilst it is expected that the physical BAO signal is isotropic, anisotropic observational effects introduce warping into the observable BAO signal, and the information contained in these anisotropies can be used for constraining cosmology. \\

\begin{figure}[h!]
  \begin{center}
    \includegraphics[width=\textwidth]{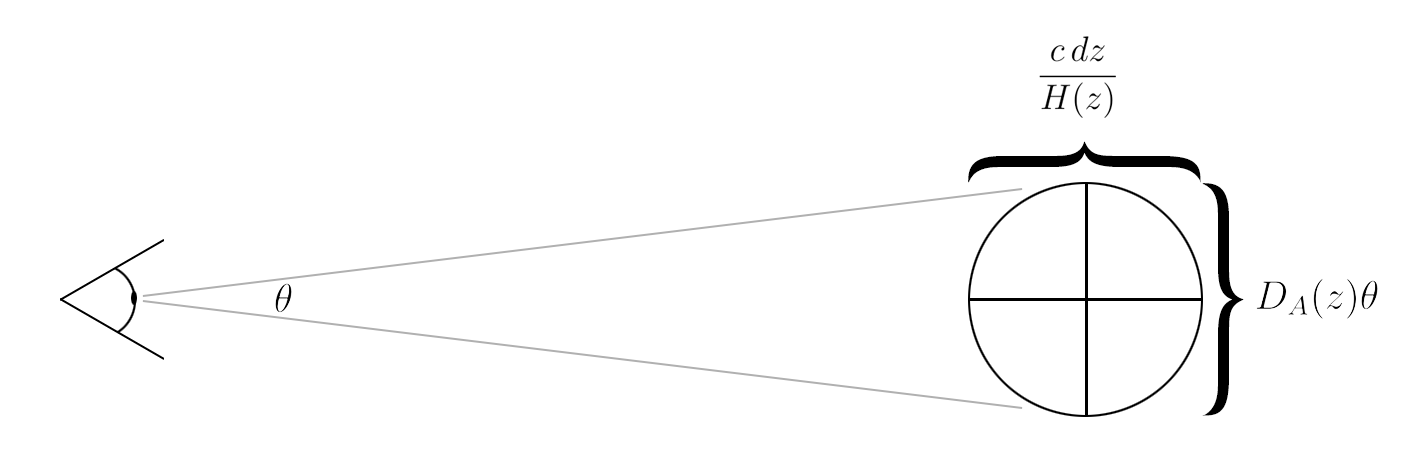}
  \end{center}
  \caption{A spherical `standard volume' with known diameter can be used to constrain distance measures both along the line of sight and transverse to the line of sight.}
  \label{fig:ap}
\end{figure}

These anisotropic effects in our observation are caused by both the our choice of fiducial cosmology and by physical, observational effects. The real (physical) comoving distance between objects is not directly observable, and instead the measured value in cosmological surveys is galaxy redshift (and angular position in the sky). We go from redshift and angle to a distance via use of a fiducial cosmology, however the difference between actual cosmology and fiducial cosmology introduces anisotropic warping, which we aim to quantify in the Alcock-Paczynski test. As the standard ruler provided by the BAO signal is valid in all directions, it can be used to provide a standard measurement in the directions parallel to the line of sight, and perpendicular to the line of sight. As shown in Figure \ref{fig:ap}, we can use this standard rule to provide constraints on $c dz /H(z)$ and $D_A(z)\theta$. \\

\begin{figure}[h!]
  \begin{center}
    \includegraphics[width=\textwidth]{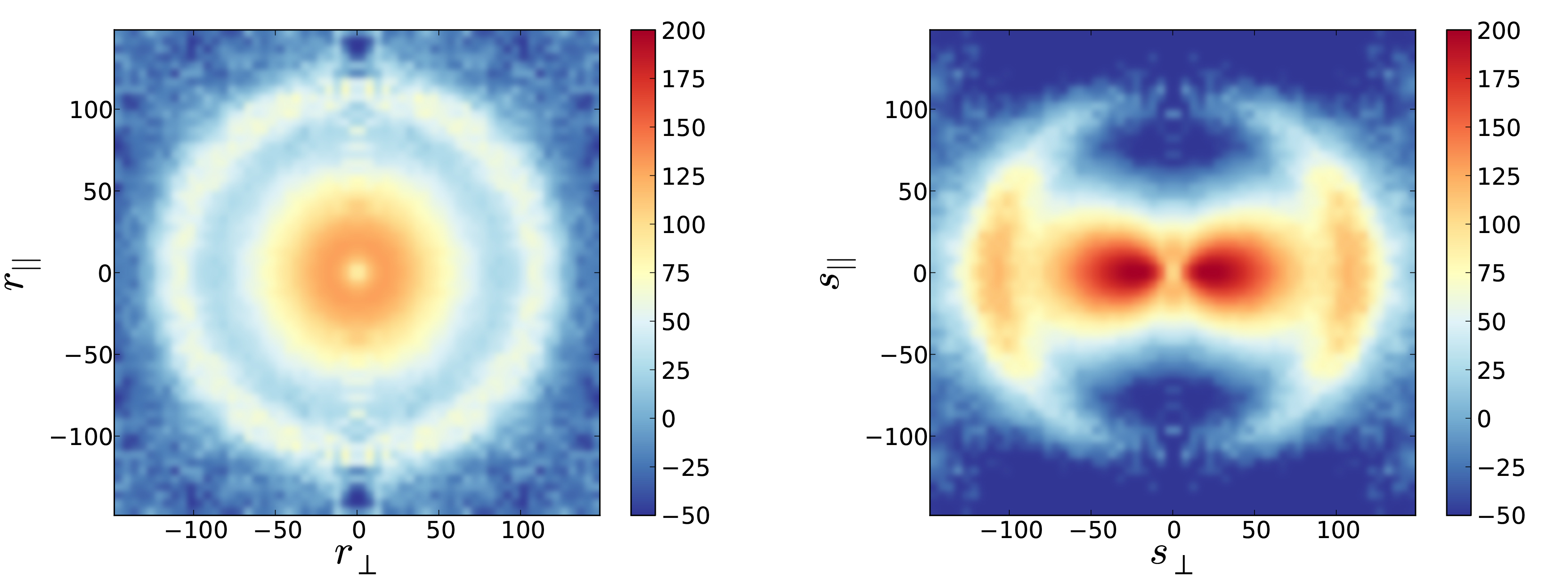}
  \end{center}
  \caption{The LasDamas galaxy correlation functions separated into perpendicular to line of sight distance $\perp$ and parallel to the line of sight distance $\parallel$. The left hand panel shows the physical (real space) correlation function, and is highly isotropic. The right hand panel shows the redshift space correlation function (the calculated distances when redshifts are converted using a fiducial cosmology), and it can be seen that the redshift space correlation function undergoes anisotropic warping. Figure panels from \citet{PadmanabhanXuEisenstein2012}.}
  \label{fig:ani}
\end{figure}

Other anisotropic effects are introduced in the form of redshift-space distortions. For example, a galaxy with peculiar velocity towards us would have two contributions to its redshift: the expansion of space stretching light, and a Doppler shift due to its peculiar velocity. These are not easily separable, and as such when the observed redshift is turned into a distance measurement, we would conclude the galaxy was closer to us than it actually was due to the contribution by the Doppler shift. The reverse is true if the galaxy has a peculiar velocity away from us. The coherent movement of matter out of voids and into overdensities, known as the Kaizer effect, `squishes' the BAO signal along the line of sight. In addition, velocity dispersion - such as that caused by virial motion of galaxies in a galaxy cluster - produce phenomenon known as Fingers of God, and produces a distension of the BAO signal narrowly along the line of sight. There are several sources of anisotropy in computed correlation functions, and an effective way to illustrate these effects is compare universe simulations (as we possess information on the real space distance) and what one would observe in said simulated universe. This is illustrated  in Figure \ref{fig:ani}.\\

\begin{figure}[h!]
  \begin{center}
    \includegraphics[width=\textwidth]{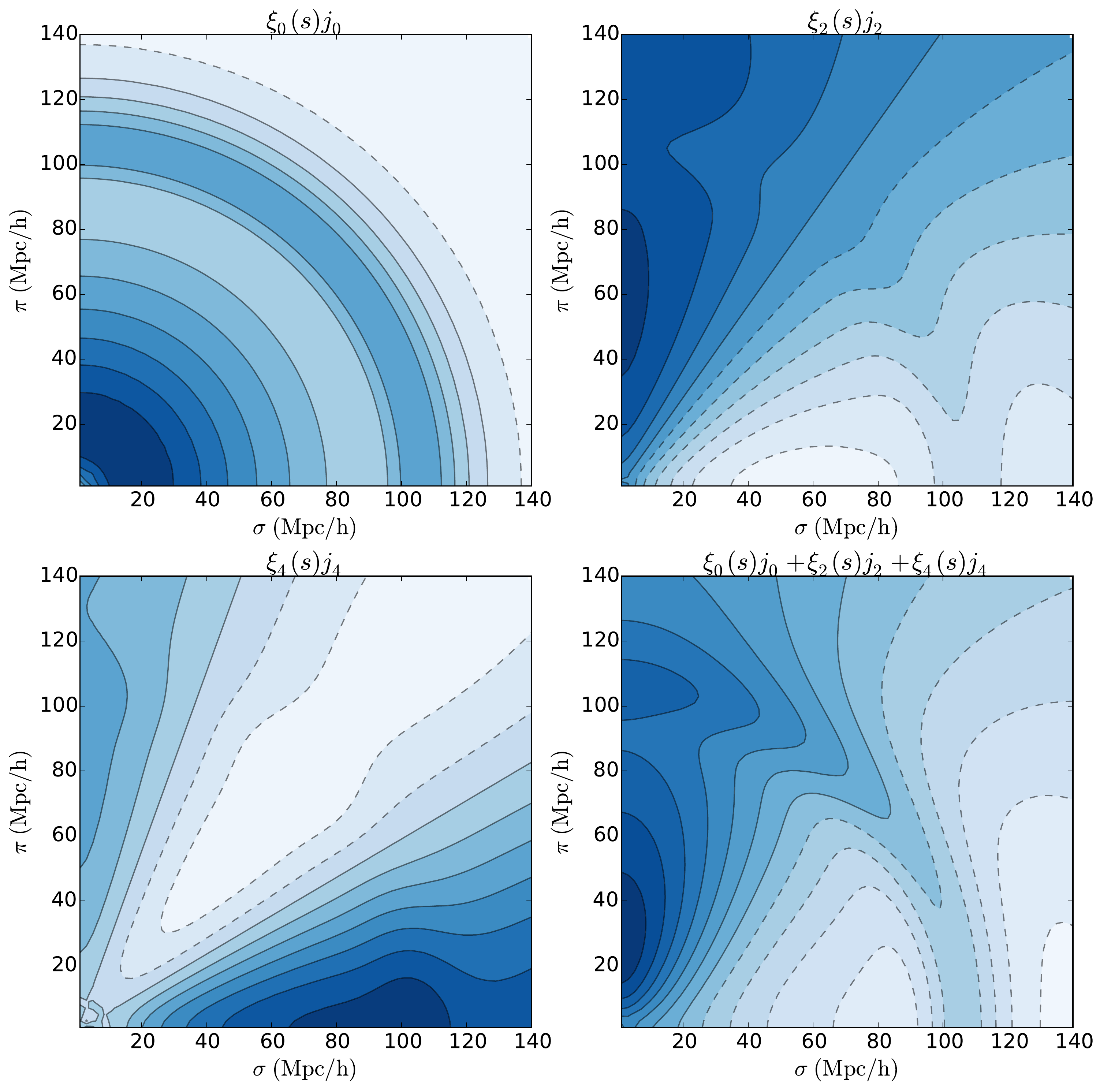}
  \end{center}
  \caption{A power spectrum generated using best fitting \citet{Planck201416} parameters has been decomposed into monopole, quadrupole and hexadecapole moments and converted to a power spectrum via Fourier transformation. The top left panel shows the monopole moment $\xi_0(S)$ multiplied by the spherical Bessel function $j_0$. The top right panel shows the quadrupole moment $\xi_2(s)$ with the second order spherical Bessel function $j_2$. The hexadecapole and forth order spherical Bessel function are shown in the bottom left, and the sum of all moments is shown in the bottom right panel.}
  \label{fig:xi2d}
\end{figure}

The anisotropies present in the correlation function can be decomposed using multipole expansion, as illustrated in Figure \ref{fig:xi2d}. In prior studies that examine the one dimensional angle-averaged BAO peak, this refers to the monopole component of the correlation function. Given advances in renormalised perturbation theory (RPT), and progress in accurately modelling non-linear growth \citep{CrocceScoccimarro2006, Matsubara2008Resumming, Matsubara2008, TaruyaNishimichi2009}, it is now possible to correct for some sources of anisotropy, allowing a one dimensional BAO signal to be reconstructed from two dimensional galaxy distribution data that provides better signal than the unreconstructed data \citep{EisensteinSeoSirko2007, SeoEckelEisenstein2010, PadmanabhanXuEisenstein2012, KazinKoda2014}. \\

Constructing a correlation function using an underlying cosmology is computationally expensive, and thus not desired when running numerical fits to the correlation function. Fortunately, one does not have to recompute the correlation function when doing model testing for each new parametrization of the model, one can instead create a correlation function using an underlying fiducial cosmology, and test similar cosmologies by introducing scaling factors (such as scaling distance or amplitude), where the the best fit values to the scaling parameters can be used to determine the correct perturbation to the fiducial cosmology that recovers the actual cosmology in the universe \citep{SanchezScoccola2012}. The process of constraining cosmology therefore involves utilising a set fiducial model, extracting the correlation function from the galaxy distribution, fitting a cosmological model to this correlation function, and then combining the fit results with the fiducial model to get the final cosmological constraints.\\

Decomposing the BAO signal into its tangential and parallel to line-of-sight components can be used to simultaneously extract $D_A$ and $H(z)$ \citep{BlakeGlazebrook2003, SeoEisenstein2003, Wang2006}. As discussed in \S\ref{sec:frw}, transverse and radial distances can be given in terms of $(1+z) D_A(z)$ and $c z/H(z)$ respectively, and the imprinted length scale of the BAO can be used to constrain both of these distilled parameters. When one analyses the BAO without separating out parallel and perpendicular to line of sight distances, one can constrain the parameter $D_V$, 
\begin{align} \label{eq:dv}
D_V \equiv \left[ (1+z)^2 D_A(z)^2 \frac{cz}{H(z)}\right]^{1/3},
\end{align}
which is simply a product of the transverse constraint on $(1+z) D_A$ and radial constraint on $cz/H(z)$, weighted by the two transverse directions and the one line of sight direction. This parameter has degeneracy with the matter density of the universe $\Omega_m$, and so therefore constraints are often given on the acoustic parameter $A(z)$ introduced by \citet{EisensteinZehavi2005}, which is given by
\begin{align}
A(z) \equiv D_V(z) \frac{\sqrt{\Omega_m H_0^2}}{cz}.
\end{align}

Decomposing the BAO signal into the line of sight and tangential components has only recently become possible as doing so requires a greater amount of data than many prior surveys have gathered (for example, \citet{OkumuraMatsubara2008} concluded DR3 of Sloan Digital Sky Survey was not sufficient for robust detection of the BAO peak). For robust detection survey criteria should have the number of targets on the order of $10^5$ or more, and span a volume of at least a cubic Gpc \citep{Tegmark1997,BlakeGlazebrook2003,BlakeParkinson2006}. Modern technological advances are helping increase galaxy survey counts rapidly, which will make BAO analysis even more important in the next generation of surveys. To illustrate the growth in survey counts over time, consider a short chronology of completed, undergoing and proposed cosmological surveys:
\begin{itemize}
\item The Point Source Catalog Redshift Survey surveyed $\sim 15\,000$ galaxies using the Infrared Astronomical Satellite \citep{SaundersSutherland2000}.
\item The 2dF Galaxy Redshift Survey got determined redshifts of $221\,414$ galaxies \citep{CollessPeterson2003} over an area of 1500 square degrees.
\item The WiggleZ final data release has redshifts for $225\,415$ galaxies with a total volume of 1 Gpc$^3$ in redshift range $z < 1$ \citep{Drinkwater2010, Parkinson2012}.
\item The tenth data release of the Sloan Digital Sky Survey III (SDSS-III) Baryon Oscillation Spectroscopic Survey (BOSS) contains $1\,507\,954$ redshifted spectra over an area of 6000 square degrees \citep{AhnAlexandroff2014}.
\item The proposed space based telescope Euclid plans to photometrically and spectroscopically redshift scores of millions of galaxy targets \citep{CimattiRobberto2009,WangPercival2010}.
\item The proposed DESI survey plans to gather approximately 22 million galaxy redshifts and 2 million quasar redshifts over a volume of (Gpc/$h$)$^3$ \citep{LeviBebek2013}.
\end{itemize}

\newpage
\section{WiggleZ} \label{sec:wigglez}

The WiggleZ Dark Energy Survey was carried out between 2006 to 2011 at the Australian Astronomical Observatory over the course of 276 nights \citep{Drinkwater2010}. The survey measured redshifts of $225\,415$ galaxy spectra, targeting blue emission-line galaxies in a redshift range of $0.2 < z < 1.0$. The target selection function is summarised in \citet{BlakeDavis2011}, and explained in detail in \citet{BlakeBrough2010}. \\

A variety of prior analyses have been conducted on the WiggleZ dataset. As the survey meets the criteria for being able to detect the BAO signal - volumes of order 1 Gpc$^3$ with order of $10^5$ redshifted galaxies \citep{Tegmark1997,BlakeGlazebrook2003,BlakeParkinson2006} - this includes analyses of the BAO signal.\\

The one dimensional BAO signal was analysed for all data in \citet{BlakeDavis2011}, and this analysis was refined by subdividing the data into redshift bins in \citet{BlakeKazin2011}. A final analysis of the 1D BAO signal involving reconstruction of the BAO peak was performed by \citet{KazinKoda2014}. Analyses that use properties of the 2D data (but not the BAO peak) include the utilisation of redshift space distortions to measure growth rate of structure \citep{BlakeBroughColless2011,ContrerasBlake2013} and using the Alcock-Paczynski test on galaxy clustering to measure expansion history \citep{BlakeGlazebrook2011}. Cosmological results from the WiggleZ papers were combined with other surveys and datasets in \citet{Parkinson2012}. For further publications using the WiggleZ dataset, see the publication list linked to from the WiggleZ home site.\footnote{\url{http://wigglez.swin.edu.au/site/}}\\

One investigation that has not been undertaken with the WiggleZ data is a two dimensional analysis of full BAO signal across all available redshift bins. Whilst this may not give tight cosmological constraints due to the number of galaxies being only above the bare minimum needed to detect the 2D BAO peak, the methodology used in such an analysis is directly applicable to future surveys.\\

A recent improvement to the WiggleZ survey is the creation of accurate mock catalogues from the WizCOLA simulations \citep{KodaBlake2015}. The simulations provide covariance estimates at greater accuracy than the log-normal realisations used in previous analyses, and serve the added purpose of allowing me to check that my model correlation function is sufficiently accurate that I can recover the simulation cosmology.\\

\begin{figure}[h!]
  \begin{center}
    \includegraphics[width=\textwidth]{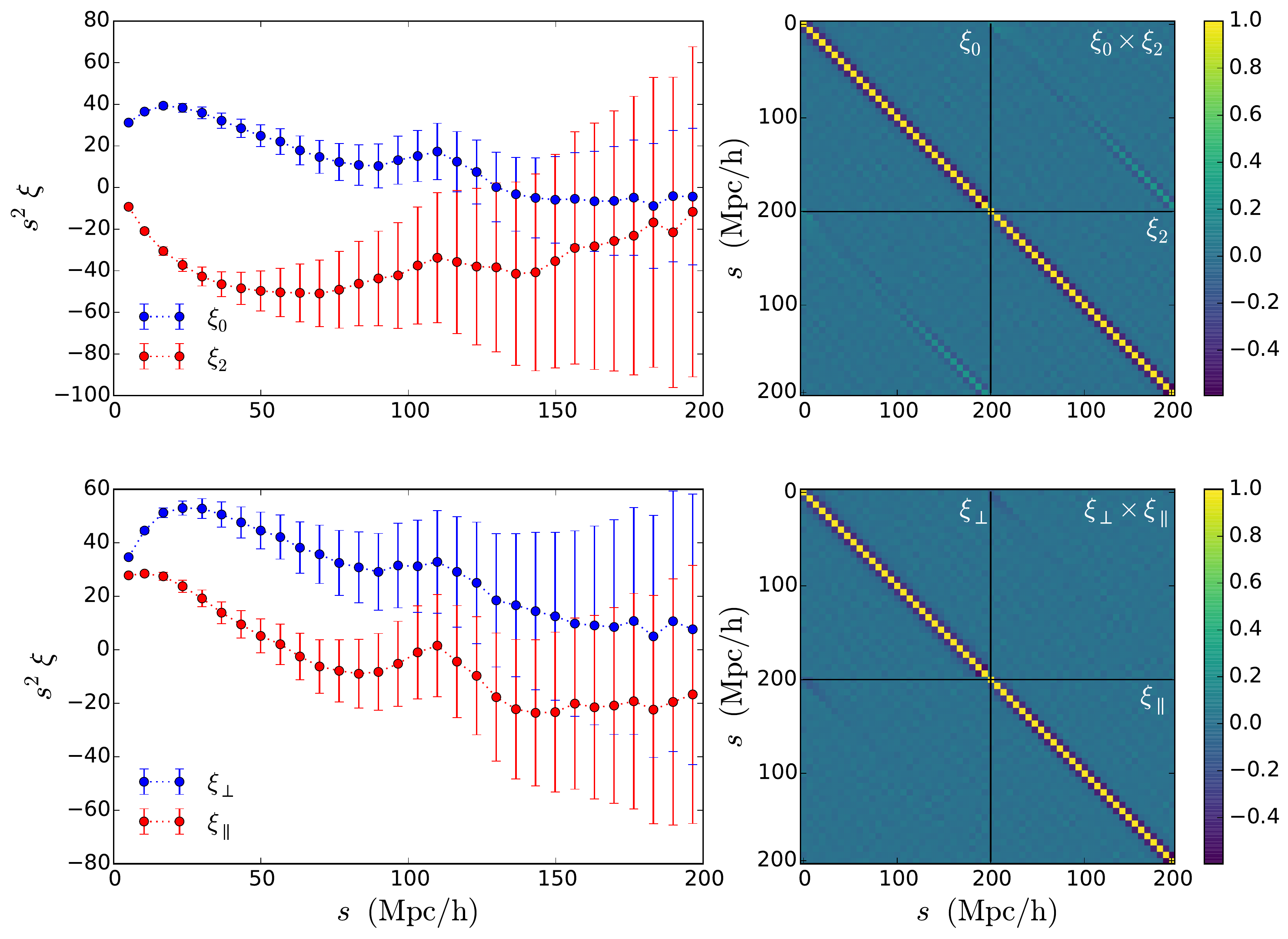}
  \end{center}
  \caption{The mean data points and covariance matrices for both the wedge and multipole expression of the WizCOLA data \citep{KazinKoda2014,KodaBlake2015}.}
  \label{fig:wizcola}
\end{figure}

\section{Markov Chain Monte Carlo}

When fitting a model, the simplest approach is to generate a grid of parameters in which to search and fit each point in the grid. Unfortunately, this approach becomes infeasible as the number of parameters in your model increases, due to grid search scaling as a function of $\mathcal{O}(r^n)$, where $r$ represents the grid resolution and $n$ the number of parameters in the model.  As cosmological models often have a high number of parameters (also referred to as dimensions or degrees of freedom), they require a different fitting approach. A popular solution do this problem is to use Markov Chain Monte Carlo (MCMC) methods, which allow fitting to high dimension models without the rapidly expanding search size found in grid searches.\\

In general, a Markov chain is a stochastic process that satisfies the Markov property - that the probability  of the next state is dependent only on the current state and no prior states \citep{Markov1988theory}. This property, also known as the chain being memoryless, forms the core of the MCMC algorithm. To complete the definition, a Monte Carlo method is a class of algorithms that utilise distributions of random sampling results to produce results. There are many different classes of MCMC algorithms, but we shall only consider the popular Metropolis-Hastings (MH) algorithm used in the model fitting in this document. The probability distributions obtained from performing an MH MCMC analysis are useful specifically because the method of selecting or rejecting potential points is chosen such that the resultant probability distribution is proportional to the (unknown) underlying probability distribution for the model \citep{Hobson2010bayesian, Ivezic2013}. Given a point in chain $\theta_i$, a random variable $u$ (which generates a random number between zero and one), and the likelihood function for an arbitrary potential point $y$ as $Q(y|\theta_i)$, the Metropolis-Hastings algorithm gives the next sample in the chain as 
\begin{align}
\theta_{i+1} = \begin{cases}
y & \rm{if } \ Q(y|\theta_i) \geq u \\
\theta_i & \rm{otherwise.}
\end{cases}
\end{align}
It is through the function $Q(y|\theta_i)$ that we can constrain the distribution of samples to reflect the underlying likelihood distribution. Taking the likelihood of a model as $\exp(-\chi^2 / 2)$ \citep{Press1992}, where $\chi^2$ is given as a sum over all data points $x$,
\begin{align}
\chi^2 = \sum_x \left(\frac{\mathrm{model}_x - \mathrm{observed}_x}{\mathrm{variance}_x}\right)^2,
\end{align}
we select $Q(y|\theta_i) = \exp(-\Delta \chi^2 / 2)$. Proposing points close to the current point generally results in a small $\Delta \chi^2$ and thus a high acceptance ratio of new points, whilst selecting points `far away' often gives rise to a large $\Delta \chi^2$, which, due to the exponential nature of the selection function, are often rejected. The method for selecting proposal points is to use a Gaussian random variable for each parameter, centred at the current point. The width of this Gaussian can then be tuned to ensure that an optimal rejection rate is achieved. Rejecting too many points results in a distribution with less points, whilst having too high an acceptance ratio often makes the walk too slow to converge. Due to the selection function, it can be seen that the walks (the chain of points) tend to walk towards lower $\chi^2$ values. The initial process of starting at a random point and walking down the $\chi^2$ slope until the sample becomes stationary (the chain is irreducible, aperiodic, and positive recurrent) is known as the burn-in period, and must be removed from the final distribution. Similarly, consecutive points give rise to traces of auto-correlation in the final distribution, and a thinning of the walk samples is also normally undertaken to make samples independent \citep{Gilks1995markov}. The final distribution should then be proportional to the underlying probability surface, and as such the distribution of parameters in the chain can be used to determine parameter constraints.\\

To ensure the final distribution is accurate, many tests can be applied to the distribution. For these tests, most require that more than one walk was run, such that it becomes possible to confirm that all walks converged to the same distribution. These convergence diagnostics are varied, from the Gelman-Rubin statistic, Geweke diagnostic, Raftery and Lewis's diagnostic and the Heidelberg and Welch diagnostic \citep{Gilks1995markov,CowlesCarlin1996}. To ensure chain convergence I test all fits generated in the development of this document using the Gelman-Rubin statistic, which calculates the ratio between the variance in separate chains and the variance of the total distribution, where a value divergent from unity indicates unsatisfactory mixing and convergence between different walks.\\

Due to the ability for MCMC algorithms to handle models with a high number of dimensions, it is a very popular choice in cosmological model fitting and testing. A program called \textsc{cosmomc} was written to generate MCMC walks using cosmological data sets \citep{LewisBridle2002}, in which a FORTRAN program generates walks, and a Python module is used to extract results from these distributions. Many prior studies utilise this software, however for this analysis I wrote my own MCMC code that does not utilise \textsc{cosmomc}.

\chapter{Prior Literature}

Initial detection of the BAO signal is not limited to the latest generation of surveys; \citet{PercivalBaugh2001}
 and \citet{ColePercival2005} detected hints of the BAO signal in 2-degree Field Galaxy Redshift Survey, and \citet{MillerNichol2001} combined smaller datasets and also detected the BAO signal. It was only with larger surveys that the significance of the BAO signal became sufficient to be able to extract cosmological constraints, and this was first done by \citet{EisensteinZehavi2005}, who reported a convincing BAO detection in the 2-point correlation function of the SDSS \citep[SDSS]{YorkAdelmanAnderson2000} DR3 Luminous Red Galaxy (LRG) sample at $z = 0.35$. In this section, I will introduce some modern analyses of the BAO signal in different surveys, and detail their model creation process.\\

Whilst increased target counts is possible by using photometric redshifts instead of spectroscopic redshifts \citep[see][for analysis of the BAO signal from the SDSS Luminous Red Galaxies (LRGs) catalogue for further examples]{BlakeCollister2007,Padmanabhan2007,HoCuesta2012}, the increased uncertainty associated with photometric analysis makes BAO analysis only possible with tomographic projection, and has not been achieved yet in cosmological surveys. As such the papers investigated in this section will be limited to those utilising spectroscopic data.\\

As discussed in \S\ref{sec:wigglez} the WiggleZ dataset has had the BAO signal analysed in previous studies. The signal has in fact been analysed using the complete dataset, where \citet{BlakeKazin2011} measured the BAO feature at $z=0.6$, making a distance measurement accurate to 4\%. The measurement was refined by  \citet{BlakeDavis2011} by breaking the analysis into separate redshift bins, which respectively provided distance measurements of accuracy 7.2\%, 4.5\% and 5.0\% in three bins centred at redshifts $z=0.44,0.60,0.73$. \citet{BeutlerBlake2011} made a distance measurement at $z=0.106$ with 6dF Galaxy Redshift Survey \cite[6dFGRS:][]{JonesRead2009} accurate to 4.5\%. The Sloan Digital Sky Survey (SDSS) has also had multiple BAO analyses carried out after their data releases. One example is that of \citet{PercivalReid2010}, who did power-spectrum analysis of SDSS DR7 and achieved a 2.7\% accurate measurement of distance-redshift relation centred at redshift $z=0.275$. The SDSS dataset is rich enough that many different analyses of galaxy distribution have been carried out, using analyses of the power spectrum \citep{TegmarkBlanton2004,Huetsi2005,BlakeCollister2007,Padmanabhan2007, PercivalCole2007,PercivalReid2010, ReidPercival2010}, or analyses of the correlation function \citep{EisensteinZehavi2005, Sanchez2009, OkumuraMatsubara2008, CabreGaztanaga2009, Martinez2009,KazinBlanton2010,ChuangWangHemantha2012}. Other studies using SDSS LRG sample include \citet{Huetsi2006, PercivalNichol2007,Sanchez2009, KazinBlanton2010}, but shall not be investigated in detail in this document.\\

Given the finalisation of the WiggleZ dataset, the main challenge performing the 2D BAO analysis involves creating an accurate cosmological model that can be compared to the dataset. I have therefore selected several relevant prior studies that span multiple methodologies for both 1D and 2D analysis, and have investigated their model construction methodologies. From the chosen studies, \citet{CabreGaztanaga2009} measured the linear redshift space distortion parameter $\beta$, galaxy bias $b$ and mean density $\sigma_8$ from SDSS DR6 LRGs. \citet{Gaztanaga2009} obtained measurement of $H(z)$ by measuring the shape of the two point correlation function along line of sight. \citet{KazinBlanton2010} determines $D_V$ from analysis of the 1D BAO signal in SDSS DR7 LRGs. \citet{ChuangWangHemantha2012} gives a method to obtain constraints without assuming dark energy model of flat universe. \citet{Kazin2010} and \citet{SanchezKazinBeutler2013} extract cosmological constraints from the 2D BAO signal using the BOSS dataset. Combined, these analyses provide multiple methodologies for constructing a 1D BAO model, and then adding anisotropic features to generate a 2D model.

\section{Correlation function and Covariance Matrix}

A cosmological survey starts with a collection of angular positions on the sky and redshift measurements, and these observations need to be converted into a three dimensional galaxy distribution function.
In all analyses investigated, the observed correlation function was determined from observational data using the \citet{LandySzalay1993} estimator,
\begin{equation}
\xi(s) = \frac{DD(s) - DR(s) + RR(s)}{RR(s)},
\end{equation}
where $D$ is used to denote the observed distribution and $R$ a random distribution, where the density of the random distribution used is denser than the observed distribution (by a factor of 20 for \citet{Gaztanaga2009} and a factor of 50 for \citet{SanchezScoccola2012}), and the random distribution follows the same selection function as used for the observed distribution. The small angle approximation is used in this estimator up to scales of approximately 10 degrees, to which it remains accurate \citep{Szapudi2004, Matsubara2000Correlation}. Alternative estimators were compared by \citet{Gaztanaga2009}, such as the estimator based on pixel density fluctuations \citep{BarrigaGaztanaga2002}, and no significant changes in results were observed. Several studies utilised the \citet{LandySzalay1993} estimator to produce an angle independent correlation function $\xi(s)$ \citep{BlakeDavis2011, ChuangWang2012}, whilst other studies produce a two dimensional cross correlation function $\xi(s,\mu)$ due to survey geometry introducing angular dependence in the random distributions \citep{SanchezScoccola2012, SamushiaPercivalGuzzo2011, KazinSanchezBlanton2012}.\\

Successive data points in a galaxy correlation function are highly correlated, and as such accurate estimation of the covariance between points is critical in being able to generate correct results. A popular methodolody is to estimate covariance through the utilisation of simulations created to replicate survey conditions and geometry. As with \citet{SanchezScoccola2012}, \citet{AndersonAubourg2012} states that the dataset covariance for the BOSS data was recovered from 600 galaxy mock catalogues, as detailed in \citet{ManeraScoccimarro2013}. For more detail, the mocks were generated using a method similar to PTHalos \citep{ScoccimarroSheth2002}, in which second order perturbation theory (2LPT) was used to generate the matter fields corresponding to the fiducial cosmology, and these fields were calibrated using suite of $N$-body simulations from LasDamas \citep{McBride2011}. The halos were populated with galaxies using a halo occupation distribution as described by \citet{ZhengCoilZehavi2007}. Mocks were then reshaped to fit the survey geometry and modified so as to include redshift-space distortions, follow sky completeness and downsampled to match the radial number density of observed data. Covariance was calculated for the LRG dataset of SDSS with the use of 216 mock catalogues \citep[MICEL7860; see][for details]{FosalbaGaztanaga2008, CrocceFosalbaCastander2010}, and \citet{Gaztanaga2009} compared this covariance to Jack-knife error and analytic error estimation. Agreement between comparisons validated the analytic error model, which is used in rest of their analysis.\\

\citet{BlakeDavis2011} utilised a series of lognormal realisations to estimate uncertainty in the binned data points. Lognormal realisations are reasonably accurate whilst the data remains in the linear and quasi-linear regimes, which is generally sufficient for analysis of large scale structure such as the BAO \citep{ColesJones1991}. The method of generating these realisations is detailed \citet{BlakeGlazebrook2003} and \citet{GlazebrookBlake2005}. In contrast to this, I will be using uncertainties derived from the WizCOLA simulations \citep{KodaBlake2015}. These use a technique to create fast simulations that are more accurate than lognormal realisations in the non-linear regime. Although WizCOLA simulations are not as accurate as a full $N$-body simulations, they are much faster to generate. That means they are ideal in situation where many realisations of a survey are required, from which one can calculate correlation function covariance.

\section{Model Creation}

Whilst the physical considerations taken into account when modelling the correlation function are fairly consistent across studies, the methodology used varies significantly. The greatest difference in model creation depends on whether the analysis seeks to take the broad shape of the correlation function into account, or whether they simply seek to marginalise over this broad structure and fit a pseudo-Gaussian peak. As this analysis will utilise the full correlation function and not just the peak, I will only describe in detail models that also use the full shape of the correlation function. It is important to note that many of the steps discussed in this section can be applied onto both the power spectrum and correlation function representations of the cosmological model, and that, whilst we shall see a trend of starting with a power spectrum and finishing with a correlation function, different methodologies transform them at different points. 

\subsection{Base Model}
All studies investigated begin with a linear power spectrum $P_{\rm{lin}}(k)$, which is commonly generated using the \camb{} software created by \citet{Lewis2000}. $P_{\rm{lin}}(k)$ is calculated by treating overdensities and underdensities as small perturbations in a homogeneous background. These density fluctuations are then evolved using linear perturbation theory, and as such are valid only for small fluctuations $\delta$ such that $\delta \equiv (\rho - \bar{\rho})/\bar{\rho} \ll 1$. Density fluctuations on the scale of galactic structures are firmly in the non-linear regime, and thus the linear model only forms the beginning of our model. Methods exist for including non-linear growth, the most accurate of which are $N$-body simulations. I discuss how I take non-linear growth into account in \S\ref{sec:nonlinear}.\\

The default \camb{} software, utilised by \citet{ChuangWang2012,BlakeDavis2011} test the default Flat $\Lambda$CDM cosmology. Modifications to \camb{} to support other cosmologies is possible \citep{FangHuLewis2008, KeislerReichardt2011, ConleyGuySullivan2011, SanchezScoccola2012}, however my analysis does not extend to that scope and as such I shall utilise the base version of \camb{} to generate a linear power spectrum.

\subsection{BAO damping}

\begin{figure}[h!]
  \begin{center}
    \includegraphics[width=\textwidth]{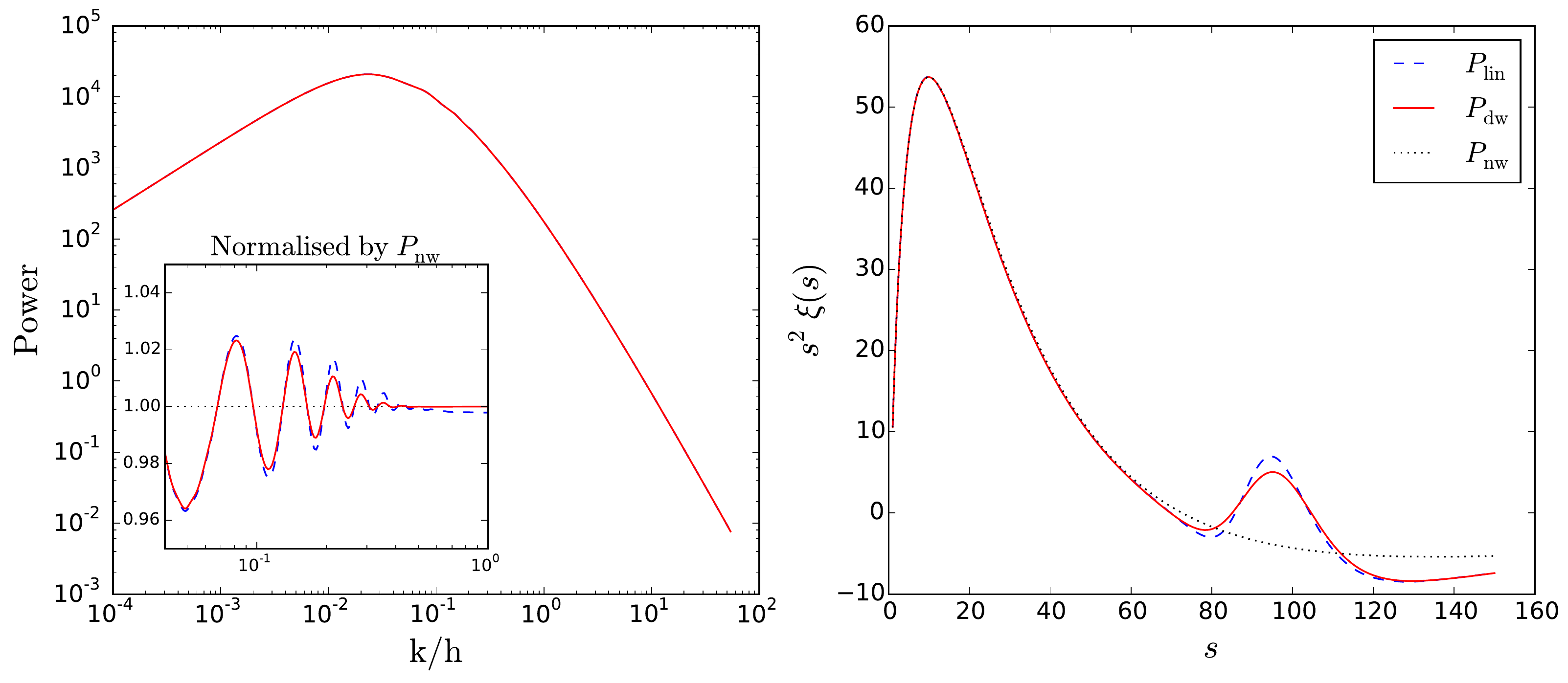}
  \end{center}
  \caption{An illustration on the effect of dampening the BAO peak in both the power spectrum (left) and resultant correlation function (right). The signal without wiggles, $P_{\mathrm{nw}}$ is shown dotted, the original linear power spectrum $P_{\mathrm{lin}}$ is shown dashed, and the damped power spectrum $P_{\mathrm{dw}}$ is shown in red.}
  \label{fig:dwExample}
\end{figure}

One of the common quasilinear effects taken into account by all studies is that of BAO peak smoothing caused by displacement of matter due to bulk flows \citep{CrocceScoccimarro2006, EisensteinSeoWhite2007,CrocceScoccimarro2008,Matsubara2008}. The degradation in the acoustic peak can be modelled with a smoothing parameter \citep{CrocceScoccimarro2008}, which was tested by \citet{SanchezBaughAngulo2008} against $N$-body simulations and subsequently used in many analyses \citep{Sanchez2009, BlakeDavis2011,BeutlerBlake2011}. This smoothing parameter takes the form of a Gaussian dampening term which reduces the amplitude of the BAO signal as a function of $k$:
\begin{align} \label{{eq:blake1}}
P_{\rm{dw}}(k) = \exp(-k^2 \sigma_v^2) P_{\rm{lin}}(k) + (1 - \exp(-k^2 \sigma_v^2)) P_{\rm{nw}}(k),
\end{align}
where $P_{\rm{nw}}(k)$ is a power spectrum without the BAO signal (the BAO peak is visible as a wiggle in the power spectrum, so `nw' denotes `no wiggles'), and $\sigma_v$ is the smoothing parameter. \citet{ChuangWang2012} utilise the same method, but call their smoothing parameter $k_*$, such that $\sigma_v = 1/(\sqrt{2} k_*)$. An identical method is utilised by \citep{AndersonAubourg2012} and \citet{XuPadmanabhan2012}, who follow \citet{EisensteinSeoWhite2007} and smooth their linear power spectrum as
\begin{align}
P_{\rm{dw}}(k) = \left[ P_{\rm{lin}}(k) - P_{\rm{nw}}(k) \right] e^{-k^2 \Sigma_{\rm{NL}}^2 / 2} + P_{\rm{nw}}(k),
\end{align}
where we can see that we have different notation for the smoothing parameter, giving $\sigma_v = \Sigma_{\rm{NL}} / \sqrt{2}$. Analogous approaches are also utilised by \citet{MontesanoSanchezPhelps2012} and \citet{SanchezScoccola2012}, and the effect of damping the BAO signal is shown in Figure \ref{fig:dwExample}.\\

Whilst advances in renormalization perturbation theory (RPT) \citep{CrocceScoccimarro2008} allow a theoretical determination of $\sigma_v$ as
\begin{align} \label{eq:sigmav}
\sigma_v^2 = \frac{1}{6\pi^2} \int P_{\rm{lin}}(k)\, dk,
\end{align}
this requires knowledge of the power of the spectrum, which is also marginalised over in all examined models. As such $\sigma_v$ (or equivalent variable) is often set as a free parameter. However, as the smoothing parameter $\sigma_v$ does not provide substantial impact to cosmological fitting \citep{ReidPercival2010, XuPadmanabhan2012}, it can also been fixed to a specific value, where \citet{XuPadmanabhan2012} (and companion papers) fix $\Sigma_{\rm{NL}}$ to the value corresponding with maximum likelihood when the parameter was initially allowed to vary.\\

The power spectrum without the BAO signal present is generated using the \verb;tffit; algorithm given by \citet{EisensteinHu1998} in the majority of studies. \citet{ReidPercival2010} investigated an alternate method of generating a no-wiggle power spectrum from the linear \camb{} power spectrum in which an 8 node b-spline was fitted to the linear power spectrum, concluding the likelihood surfaces generated when fitting using splines and the algorithm from \citet{EisensteinHu1998} agree well. For my work I attain a no-wiggle power spectrum $P_{\mathrm{nw}}(k)$ utilising polynomial subtraction. For a comparison of this methodology against the \verb;tffit; algorithm supplied by \citet{EisensteinHu1998} or spline fitting, please see Appendix \ref{app:dewiggle}.

\subsection{Non-linear growth}\label{sec:nonlinear}.

\begin{figure}[h!]
  \begin{center}
    \includegraphics[width=\textwidth]{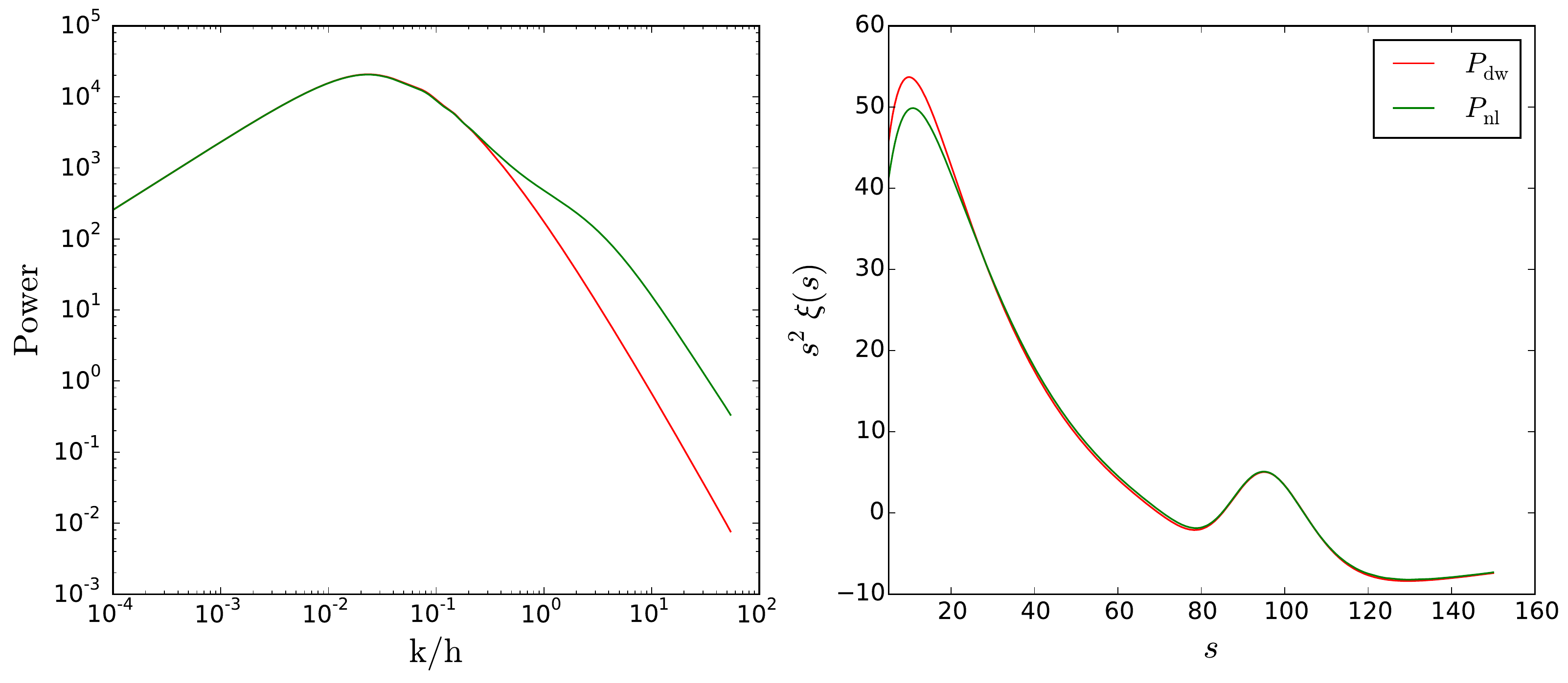}
  \end{center}
  \caption{An illustration on the effect of incorporating non-linear growth into the power spectrum (left), and the resultant changes in the correlation function (right). Notice the effect of the increase in power of rapid oscillations (high k/h) lead to dampening of the correlation function at small scales.}
  \label{fig:nlExample}
\end{figure}

The non-linear effects of gravitational growth require model corrections to account for the enhancement of small scale structure growth that was not modelled with linear perturbation theory. The software package \halofit{} from \citet{Smith2003} is utilised by many studies to generate a power ratio $r_{\rm{halo}}$ as a function of $k$ \citep{ReidPercival2010, BlakeDavis2011, ChuangWang2012}, which is applied onto the model:
\begin{align}
P_{\text{nl}} = P_{\text{dw}} r_{\text{halo}}.
\end{align}
Figure \ref{fig:nlExample} has been created to help visualise the effects from non-linear growth. \citet{XuPadmanabhan2012} does not detail this step in their model creation, but instead, after converting to the power spectrum $P_{\rm{dw}}(k)$ to a correlation function $\xi(s)$, add a nuisance function $A(s)$, such that
\begin{align} \label{eq:as}
A(s) = \frac{a_1}{s^2} + \frac{a_2}{s} + a_3,
\end{align}
which acts to marginalise over any changes in broad correlation shape which the non-linear correction would give (where moving from power spectra to correlation functions is discussed in \S\ref{sec:prior:cor}). In their analysis, \citet{XuPadmanabhan2012} compared the effects of $A(s) = 0$, $A(s) = a_1 / s^2$, $A(s) = a_1 / s^2 + a_2 / s$ and the $A(s)$ detailed in equation \eqref{eq:as}, which motivated their final selection of equation \eqref{eq:as}. The original function $A(s)$ was picked due to transformation simplicity, as in Fourier space it becomes $A^\prime(k) = \frac{a^\prime_1}{k} + \frac{a^\prime_2}{k^2} + \frac{a^\prime_3}{k^3}$.

\begin{figure}[h!]
  \begin{center}
    \includegraphics[width=0.5\textwidth]{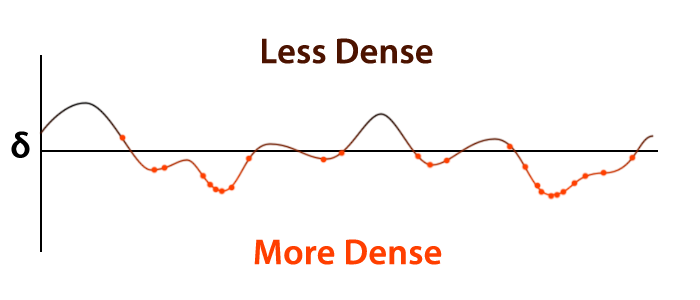}
  \end{center}
  \caption{As galaxies are more likely to form in regions of higher density, the distribution of galaxies (shown as dots on the density fluctuations) does not trace out the underlying matter distribution fully, instead is found to be biased towards regions of increased density.}
  \label{fig:bias}
\end{figure}

As illustrated in Figure \ref{fig:bias}, we observe galaxy density and not the true underlying total matter (dark matter and standard matter) distribution, and as such a biasing term $b$ is required to move from a matter power spectrum to a galaxy power spectrum. This is often incorporated into the model as a factor $b^2$ in the correlation function, such that $\xi_g(s) = b^2 \xi(s)$, or equivalently into the power spectrum, giving $P_g(k) = b^2 P_{\rm{nl}}(k)$ \citep{BlakeDavis2011, AndersonAubourg2012, ChuangWang2012, MontesanoSanchezPhelps2012, XuPadmanabhan2012}. In case of any dependence between the cosmological model and non-linear corrections added using \halofit{}, \citet{ReidPercival2010} expands upon the bias factor of $b^2$, instead utilising a more complex model that introduces scale dependent bias:
\begin{align}
F(k) = b^2\left(1 + a_1 \left(\frac{k}{k_*}\right) a_2 \left( \frac{k}{k_*} \right)^2 \right),
\end{align}
where their non-linear power spectrum is given by 
\begin{align}
P_{\text{nl}} = P_{\text{dw}} r_{\text{halo}} + r_{\text{halo}} F(k).
\end{align}

\citet{BlakeDavis2011} also incorporate scale dependent bias into to their model. The scale dependent bias, denoted $B(s)$ is included via
\begin{align}
\xi_{\rm{galaxy}}(s) = B(s) \xi(s),
\end{align}
where $B(s) = 1 + (s/s_0)^\gamma$, with $s_0 = 0.32 h^{-1}\,$Mpc and $\gamma = -1.36$. This fit was determined using halo catalogues extracted from the GiggleZ dark matter simulation. The magnitude of this correction ($\sim 1\%$) is far less than that found in more biased galaxy samples such as the SDSS LRG sample, which has corrections on the order of $\sim 10\%$ \citep{EisensteinZehavi2005}. The decreased bias in the WiggleZ survey is due to targeting bright blue galaxies, as opposed to SDSS targeting luminous red galaxies. A correction of the same form was also included in the SDSS BAO analysis undertaken in \citet{VeropalumboMarulliMoscardini2014}.

\subsection{Magnification Bias}

One model adjustment not found in the majority of papers was the consideration of magnification bias, which was only investigated by  \citet{Gaztanaga2009}. Two main effects were discussed: gravitational lensing increasing brightness via magnification, and lensing increasing apparent area (corresponding to a decrease in number density of galaxies). The net effect of these two factors is called magnification bias, and can be accounted for by determining the slope of the number counts over galaxy magnitude \citep{TurnerOstriker1984,  WebsterHewett1988, Fugmann1988, Narayan1989, Schneider1989, BroadhurstTaylor1995, MoessnerJain1998}:
\begin{align}
s = \frac{d \log_{10} N(<m)}{dm},
\end{align}
where $N(<m)$ refers to the number of galaxies in the survey with apparent magnitude brighter than $m$. \citet{Gaztanaga2009} utilise the photometric dataset from SDSS DR6 \citep[DR6:][]{Adelman2008} to estimate this slope within and beyond the spectroscopic limit, and from this applied corrections for magnification bias. For a more detailed derivation of the magnification bias effects, see \citet[\S 2.2]{Gaztanaga2009}.\\

I do not investigate magnification bias in my work because it is expected to be negligible for the WiggleZ analysis.

\subsection{Anisotropies}

At this point, various anisotropic effects can be incorporated into the model to take it from the one dimensional model produced so far to a two dimensional model which can extract cosmological information from these anisotropies.

\subsubsection{Kaiser effect}

A prominent source of anisotropy in cosmological models is due to the Kaiser effect, where the Doppler shift from coherent infall of galaxies in a cluster produces anisotropic distortions that appear to flatten the two dimensional cross correlation function. These distortions can be modelled simply in Fourier space \citep{Kaiser1987}:
\begin{align} \label{eq:gaztanga1}
P_{\rm{nl}}(k, \mu) = (1 + \beta \mu^2)^2 P_{g}(k),
\end{align}
where $P_{g}$ is the power spectrum of galaxy density fluctuations $\delta_g$, $\mu$ is the cosine of the angle to line of sight, and $\beta$ is the growth rate of growing modes in linear theory. Given that galaxy over-density is linearly biased by a factor of $b$ \citep{ReidSpergelBode2009}, we can relate $P_{\text{nl}}$ and $P_{g}$ as proportional. We can also approximate $\beta$ as \citep{Linder2005}
\begin{align}
\beta \approx \frac{\Omega_m^{0.55}}{b}.
\end{align}
This correction for the Kaiser effect has been utilised by \citet{Gaztanaga2009, ChuangWang2012, XuPadmanabhan2012} for identifying the unreconstructed BAO signal in survey data. When reconstructing the BAO signal \citep[see][for details]{PadmanabhanXuEisenstein2012,KazinKoda2014}, the Kaiser effect is corrected for and thus does not have to be inserted into the cosmological model.

\subsubsection{Fingers of God}

\begin{figure}[h!]
  \begin{center}
    \includegraphics[width=0.8\textwidth]{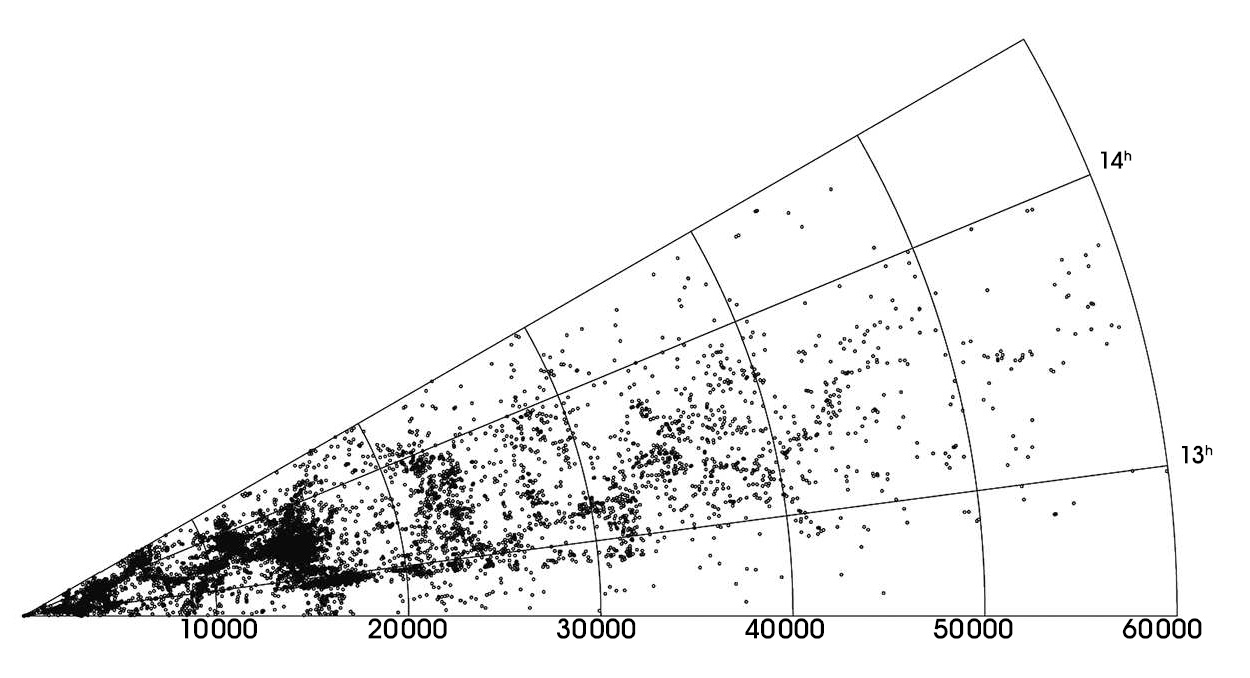}
  \end{center}
  \caption{A cone diagram (right ascension) around the Shapley supercluster from \citet{ProustQuintana2006}. Finger's of God are present in distinctly noticeable in the denser section of galaxy distribution, as elongated shapes pointing towards the origin.}
  \label{fig:fingersOfGod}
\end{figure}

Peculiar velocity does not have to be coherent to effect observational cosmology, and the random peculiar velocities of galaxies in clusters, which are related to the cluster mass via the virial theorem, create artefacts known as Fingers of God. Fingers of God elongate the observed position of galaxies along the line of sight, as illustrated in Figure \ref{fig:fingersOfGod}. \citet{SanchezKazinBeutler2013} incorporates this effect via an additional exponential prefactor:
\begin{align}
P_{\rm{gal}} = \left( \frac{1}{1 + (k f \sigma_v \mu )^2} \right)^2  P_{\rm{nl}}(k, \mu), \label{eq:lorentzian}
\end{align}
where $f = \frac{d \ln D(a)}{d \ln a}$, $D(a)$ is the growth factor, and $\sigma_v$ is the pairwise peculiar velocity dispersion. Notational differences aside, the same prefactor is used by \citet{XuPadmanabhan2012}, who also tested an alternate Gaussian form prefactor, finding little difference between results. In the investigation of growth rate with WiggleZ data, \citet{BlakeBroughColless2011} adopts a Lorentzian model of velocity dispersion with prefactor $[1 + (k \sigma_v \mu)^2]^{-1}$ due to the better fitting results found in \citet{HawkinsMaddoxCole2003} and \citet{CabreGaztanaga2009}.

This velocity dispersion is accounted for by \citet{ChuangWang2012} by convolving their 2D correlation function with a distribution of velocities. The convolution is given by
\begin{align}
\xi(\sigma, \pi) = \int_\infty^\infty \xi^* \left(\sigma, \pi - \frac{v}{H(z) a(z)} \right) f(v)\, dv,
\end{align} 
following \citet{Peebles1980}, where the random motions take exponential form \citep{RatcliffeShanks1998,Landy2002}
\begin{align}
f(v) = \frac{1}{\sigma_v \sqrt{2}} \exp\left(- \frac{\sqrt{2}\abs{v}}{\sigma_v} \right)
\end{align}
where $\sigma_v$ is the pairwise peculiar velocity dispersion, and not to be confused with the $\sigma_v$ denoted in the Gaussian dampening used by \citet{BlakeDavis2011}. In all of these analyses, the distribution itself is marginalised over, where $\sigma_v$ is often set as a free parameter. In my analysis, I will follow the Lorentzian methodology utilised by \citet{BlakeBroughColless2011}.


\subsection{Moving to a correlation function} \label{sec:prior:cor}

The power spectrum and correlation functions are related to each other via Fourier transform. One dimensional BAO analyses generally look at the angle-averaged correlation function, which is simply the monopole moment. A power function can be decomposed into its multipole components via 
\begin{align}
P_{\ell}(k) = \frac{2\ell + 1}{2} \int_{-1}^1 P_{\rm{gal}}(k, \mu) \ \mathcal{L}_\ell \  d\mu
\end{align}
where $\mathcal{L}_\ell$ represents the $\ell$'th Legendre polynomial. These multipole components can be turned into correlation functions by Fourier transforming them, giving
\begin{align}
\xi_\ell(s) = \frac{1}{(2\pi)^3} \int 4\pi k^2 \ P_\ell(k) \ j_\ell(ks)
\end{align}
where $j_\ell(ks)$ are spherical Bessel functions of the first kind. As the increased power of small scale oscillations from the non-linear corrections decreases convergence of this function,  \citet{AndersonAubourg2012} add a Gaussian factor $\exp(-k^2 a^2)$ to improve convergence, where they have set $a= 1\, h\, \rm{Mpc}$ (and found cosmology insensitive to changes in $a$). This is in contrast to other methods that may be used to increase convergence, such as truncating the numerical integral after a specific number of periods in the spherical Bessel function. Interestingly, whilst the \citet{BlakeDavis2011} WiggleZ analysis does not contain the Gaussian dampening term seen in \citet{AndersonAubourg2012}, it is present in the correlation function model used in \citet{BlakeKazin2011}.

\subsubsection{Wedges} \label{sec:prior:cor:wedge}

Having obtained the multipole expansion of the correlation function, a 2D analysis can reconstruct the parallel to line-of-sight and perpendicular to line-of-sight correlation functions \citep{KazinSanchezBlanton2012, SanchezKazinBeutler2013}, where the correlation functions are given as 
\begin{align}
\xi_\perp(s) &= \xi_0(s) - \frac{3}{8} \xi_2(s) + \frac{15}{128} \xi_4(s), \\
\xi_\parallel(s) &= \xi_0(s) + \frac{3}{8} \xi_2(s) - \frac{15}{128} \xi_4(s).
\end{align}
The distance scale is transformed via
\begin{align}
s_\perp &\rightarrow \alpha_\perp s_\perp \\
s_\parallel &\rightarrow \alpha_\parallel s_\parallel,
\end{align}
which is respectively used to constrain
\begin{align}
\alpha_\perp &= \frac{D_A(z)}{D_A^{\mathcal{D}}(z)}, \\
\alpha_\parallel &= \frac{H^{\mathcal{D}}(z)}{H(z)},
\end{align}
where the superscript $\mathcal{D}$ is used to denote the value given from fiducial cosmology. Having defined $\alpha_\perp$ and $\alpha_\parallel$, they are used to define transformation functions:
\begin{align}
s(\mu^\prime, s^\prime) &= s^\prime \sqrt{\alpha_\parallel^2(\mu^\prime)^2 + \alpha_\perp^2 (1 - (\mu^\prime)^2)}, \\
\mu(\mu^\prime) &= \frac{\alpha_\parallel \mu^\prime}{\sqrt{\alpha_\parallel^2(\mu^\prime)^2 + \alpha_\perp^2 (1 - (\mu^\prime)^2)}},
\end{align}
where data wedges can be extracted via
\begin{align}
\xi^\prime_{\Delta \mu}(s^\prime) = \frac{1}{\Delta \mu^\prime} \int_{\mu^\prime_{\rm{min}}}^{\mu^\prime_{\rm{max}}} \xi(\mu(\mu^\prime), s(\mu^\prime, s^\prime)) d\mu^\prime.
\end{align}
In both the analyses by \citet{KazinSanchezBlanton2012} and \citet{SanchezKazinBeutler2013}, the observational data is found in two wedges, corresponding to $0 < \mu < 0.5$ and $0.5 < \mu < 1$ respectively. When considering only a one dimensional BAO analysis, such as in \citet{BlakeDavis2011}, the monopole moment has its $s$ transformed to $\alpha s$ (as opposed to scaling $s_\perp$ and $s_\parallel$ as seen above), where we can now provide constraints on $D_V(z)$ via
\begin{align}
\alpha = \frac{D_V(z)}{D_V^{\mathcal{D}}(z)}.
\end{align}

\subsubsection{Multipoles} \label{sec:prior:cor:mp}

Other analyses utilise fitting to the multipole expansion directly, instead of reconstructing the parallel and tangential to line-of-sight wedges. As detailed in \citet{PadmanabhanWhite2008}, \citet{KazinSanchezBlanton2012} and \citet{XuCuesta2013}, one can introduce isotropic scaling factor $\alpha$ and warping parameter $\epsilon$, where we transform the distance scales such that
\begin{align}
s_\parallel &\rightarrow s_\parallel \alpha (1 + \epsilon)^2 \\
s_\perp &\rightarrow s_\perp (1 + \epsilon)^{-1},
\end{align}
where $\alpha$ gives constraints on $D_V$, as mentioned previously:
\begin{align}
\alpha = \left( \frac{H^{\mathcal{D}}(z)}{H(z)}\right) ^{1/3} \left( \frac{D_A(z)}{D_A^{\mathcal{D}}(z)} \right)^{2/3},
\end{align}
where a superscript $\mathcal{D}$ is used again to indicate the value from fiducial cosmology. Similarly, $\epsilon$ gives
\begin{align}
1 + \epsilon = \left( \frac{H^{\mathcal{D}}(z) D_A^{\mathcal{D}}(z)}{H(z) D_A(z)} \right) ^{1/3}.
\end{align}
From these transformations, we have that
\begin{align}
s^\mathcal{D} &= \alpha (1 + 2\epsilon P_2(\mu)) s + \mathcal{O}(\epsilon^2) \\
\mu^\mathcal{D} &= \mu^2 + 6 \epsilon (\mu^2 - \mu^4) + \mathcal{O}(\epsilon^2)
\end{align}
By combining this with the definition of the multipole expansion,
\begin{align}
\xi(s, \mu) = \sum\limits_{\rm{even }\ \ell} P_\ell(\mu) \xi_\ell(s),
\end{align}
we can conclude, to first order in $\epsilon$ and discarding hexadecapole contribution, that the multipole moments should be transformed as \citep{KazinSanchezBlanton2012}
\begin{align}
\xi_0(s) &= \xi_0(\alpha s) + \frac{2}{5}\epsilon \left[ 3 \xi_2(\alpha s) + \frac{d \xi_2(\alpha s)}{d \log(s)}\right], \label{eq:ximp}\\
\xi_2(s) &= 2\epsilon \frac{d \xi_0(\alpha s)}{d\log(s)} + \left( 1 + \frac{6}{7}\epsilon\right) \xi_2(\alpha s) + \frac{4}{7} \epsilon \frac{d \xi_2(\alpha s)}{d \log(s)}. \label{eq:multipoles} 
\end{align}
If we choose to include the hexadecapole terms as done by \citet{XuCuesta2013}, we would need to add to the $\xi_2(s)$ 
\begin{align}
\frac{4}{7}\epsilon \left[ 5 \xi_4 (\alpha s) + \frac{d \xi_4(\alpha s)}{d \log(s)} \right].
\end{align}
It is interesting to note some disagreement in prior literature about the form of the derivative terms and whether the distance $s$ should be transformed or not. \citet{KazinSanchezBlanton2012} do not transform this scale, using for example $d\xi_2(s)/d\log(s)$, whilst \citet{XuCuesta2013} do include the scale factors, such that they use $d\xi_2(\alpha s) / d\log(s)$. Following the more detailed derivation from \citet{XuCuesta2013}, the scale factors will be included inside the derivative terms for my analysis. Finally, we can note that for the monopole moment $\xi_0(s)$ found in equation \eqref{eq:ximp}, the terms in the square bracket effectively cancel out, such that we can use the transformation $\xi_0(s) = \xi_0(\alpha s)$. This simplification is tested in \citet{Sanchez2009} and \citet{EisensteinZehavi2005}.

With these transformations in place, we can then utilise $\alpha$ and $\epsilon$ to fit for the multipole moments of the correlation function. For small $\epsilon$, we can can constrain $H(z)$ and $D_A$ via
\begin{align}
\alpha(1 + \epsilon)^2 \approx \alpha(1 + 2\epsilon) &\approx \frac{H^\mathcal{D}(z)}{H(z)} \label{eq:alpha1}\\
\alpha(1 + \epsilon)^{-1} \approx \alpha(1 - \epsilon) &\approx \frac{D_A}{D_A^\mathcal{D}} \label{eq:alpha2}
\end{align}

\newpage
\section{Fitting} \label{sec:trunc}

Having generated a model correlation function, the final step is to compare that against the correlation function computed with observational data and a fiducial cosmology. However, as the created model breaks down at low distances and all models are similar at high distances where sample variance increases data uncertainty, matching normally only occurs on a subsection of the data. Comparisons between different papers are shown in Table \ref{tab:truncation}. Given the wide range of dataset truncation values and lack of clear support for one cutoff over another, this represents a parameter that requires investigation. This investigation can be found in Appendix \ref{app:truncation}.

\begin{table}[h]
\centering
\caption{A comparison of data fitting ranges found in prior literature}
\label{tab:truncation}
\begin{tabular}{lll}
\specialrule{.1em}{.05em}{.05em} 
Study & Data Range $(h^{-1}$Mpc) & Comments \\
\specialrule{.1em}{.05em}{.05em} 
\citet{XuPadmanabhan2012}      &     $30 < s < 200$       &          \\
\citet{SanchezScoccola2012}      &    $40 < s < 200$        &          \\
\citet{Sanchez2009}     &       $40 < s < 200$     &        \\  
\citet{Gaztanaga2009}     &       $20 < s$     &        \\  
\citet{ChuangWang2012}     &       $40 < s < 120$     &     \specialcell{Low upper limit due to \\similarity of all models}   \\  
\citet{EisensteinZehavi2005}     &       $10 < s < 180$     &        \\  
\citet{BlakeDavis2011}     &       $10 < s < 180$     &        \\  
\citet{KazinSanchezBlanton2012}     &       $40 < s < 150$     &        \\  
\citet{BlakeDavis2011}     &       $30 < s < 180$     &   \specialcell{Insufficient to determine $\Omega_m h^2$ \\from clustering pattern alone.}     \\  
\citet{BlakeDavis2011}     &       $50 < s < 180$     &   \specialcell{Insufficient to determine $\Omega_m h^2$ \\from clustering pattern alone. }    \\  
\specialrule{.1em}{.05em}{.05em} 
\end{tabular}
\end{table}

\chapter{Cosmological Model}
\label{ch:model}

In this section I outline the steps I used to create a robust model for the BAO signal, and the consistency checks I put it through. I present the base, one dimensional model and check it against the prior WiggleZ analysis from \citet{BlakeKazin2011}. I then turn this base model into a two dimensional model by including anisotropies in the model, and utilise the WizCOLA simulations to check the consistency of the angular dependent model using both the wedges and multipole expansion methodologies.

\section{Confirming the base model}

In order to determine whether my model was consistent with prior literature, I attempt to recover the fits found by \citet{BlakeKazin2011} by utilising their model creation method. The underlying linear model is computed using the CAMB software \citep{Lewis2000}, following prior studies \citep{BlakeDavis2011, SanchezScoccola2012, ChuangWang2012}. The parameter $\Omega_c h^2$ is free, with $\Omega_b h^2$ set to $0.0226$\footnote{As $\Omega_b h^2$ is well constrained by CMB data and variations even up to $5\sigma$ are negligible to the BAO model, I have set it constant in my model. Note that this holds for Flat $\Lambda$CDM cosmology, and if my analysis extended beyond this cosmology I would not be able to fix $\Omega_b h^2$.} and $h=0.705$ following the WizCOLA fiducial model and fiducial model adopted by \citet{BlakeKazin2011}. The quasi-linear correction due to matter flow displacement is incorporated into the model via blending between the linear model and a model without the BAO feature - denoted $P_{\rm{nw}}$ - with a Gaussian dampening term:
\begin{align}
P_{\text{dw}}(k) = P_{\text{lin}}(k) \exp\left(-\frac{k^2}{2k_*^2}\right)  + P_{\text{nw}}(k) \left(1 - \exp\left(-\frac{k^2}{2k_*^2}\right)\right),
\end{align}
where $k_*$ is set as a free parameter. Due to the lack of strong constraints on $k_*$ given by the analysis in \citet{BlakeKazin2011}, instead fit over the space $\log(k_*)$ to check that the parameter is bounded in both limits. The no-wiggle power spectrum is calculated in \citet{BlakeKazin2011} using the formula from \citet{EisensteinHu1998}, whilst I have utilised weighted polynomial subtraction as detailed in Appendix \ref{app:dewiggle}. The non-linear growth of structure is accounted for by utilising the \textsc{halofit} algorithm from \citet{Smith2003} which gives growth ratio $r_{\rm{halo}}$ as a function of $k$, such that we have
\begin{align}
P_{\text{nl}} = P_{\text{dw}} r_{\text{halo}}
\end{align}
The Spherical Hankel transformation (a special form of the Fourier transform) is used to move from power spectrum to correlation function,
\begin{align}
\xi_\ell(s) = \frac{1}{(2\pi)^3} \int 4\pi k^2 \ P_\ell(k) \ j_\ell(ks) e^{-k^2 a^2},
\end{align}
where we have introduced the gaussian dampening term following \citet{AndersonAubourg2012}. Scale dependent growth calibrated from the GiggleZ simulation is also applied onto the correlation functions following \citet{BlakeDavis2011}, such that $\xi_B(s) = B(s) \xi(s)$, where $B = 1 + (s/s_0)^\gamma$ with $s_0 = 0.32\, h^{-1}\,\rm{Mpc}$ and $\gamma = -1.36$. Bias factors $b^2$ and horizontal dilation $\alpha$ were applied to the model, giving a final correlation function of 
\begin{align}
\xi^{\rm{fin}}(s) = b^2 \xi_B(\alpha_0 s)
\end{align}
Fits were created utilising this model and the WiggleZ unreconstructed dataset over the same data range utilised by \citet{BlakeDavis2011} and \citet{BlakeKazin2011}: $10<s<180 h^{-1}\rm{Mpc}$. \citet{BlakeKazin2011} employed a grid search due to the low number of parameters, whilst I employ an MCMC based fitting analysis. The fitting values found by \citet{BlakeKazin2011} compared to this analysis are detailed in Table \ref{tab:blakekazintable}. If we take the uncertainty from \citet{BlakeKazin2011} and use that to determine the difference in units of $\sigma$ for both $\Omega_m h^2$ and $\alpha$, we find the difference in recovered $\Omega_m h^2$ to be $(0.25\sigma, 0.31\sigma, 0.46\sigma)$ for the effective redshift bins $0.44, 0.6, 0.73$ respectively. We also find $\alpha$ to be recovered at a shift of $(0.15\sigma, 0.25\sigma, 0.26\sigma)$ for the same respective effective redshift bins as before. The consistent sign of both the $\Omega_m h^2$ and $\alpha$ recovery values may be indicative of a systematic bias in our model when compared to the model used by \citet{BlakeKazin2011}, or simply a product of different fitting methodologies.

\begin{table}[h]
\centering
\caption{A comparison between the fits found in this analysis and those found in \citet{BlakeKazin2011}.}
\begin{tabular}{cc|ccc|ccc}
\specialrule{.1em}{.05em}{.05em} 
Sample & $z_{\rm{eff}}$ & \multicolumn{3}{c}{\citet{BlakeKazin2011}}  & \multicolumn{3}{c}{This analysis}\\
&  & $\chi^2$ & $\Omega_m h^2$ &$\alpha$ & $\chi^2$ & $\Omega_m h^2$ & $\alpha$ \\
\specialrule{.1em}{.05em}{.05em} 
$0.2 < z < 0.6$ & $0.44$ & $11.4$ & $0.143\pm0.020$ &$1.024\pm0.093$ & $13.9$ & $0.138\pm 0.016$ & $1.038\pm 0.098$ \\
$0.4 < z < 0.8$ & $0.60$ & $10.1$ & $0.147\pm0.016$ &$1.003\pm0.065$ & $11.7$ & $0.142\pm 0.014$ & $1.019\pm 0.082$ \\
$0.6 < z < 1.0$ & $0.73$ & $13.7$ & $0.120\pm0.013$ &$1.113\pm0.071$ & $15.1$ & $0.114\pm 0.012$ & $1.132\pm 0.074$ \\
\specialrule{.1em}{.05em}{.05em} 
\end{tabular} \label{tab:blakekazintable}
\end{table}

\begin{figure}[h]
  \begin{center}
    \includegraphics[width=\textwidth]{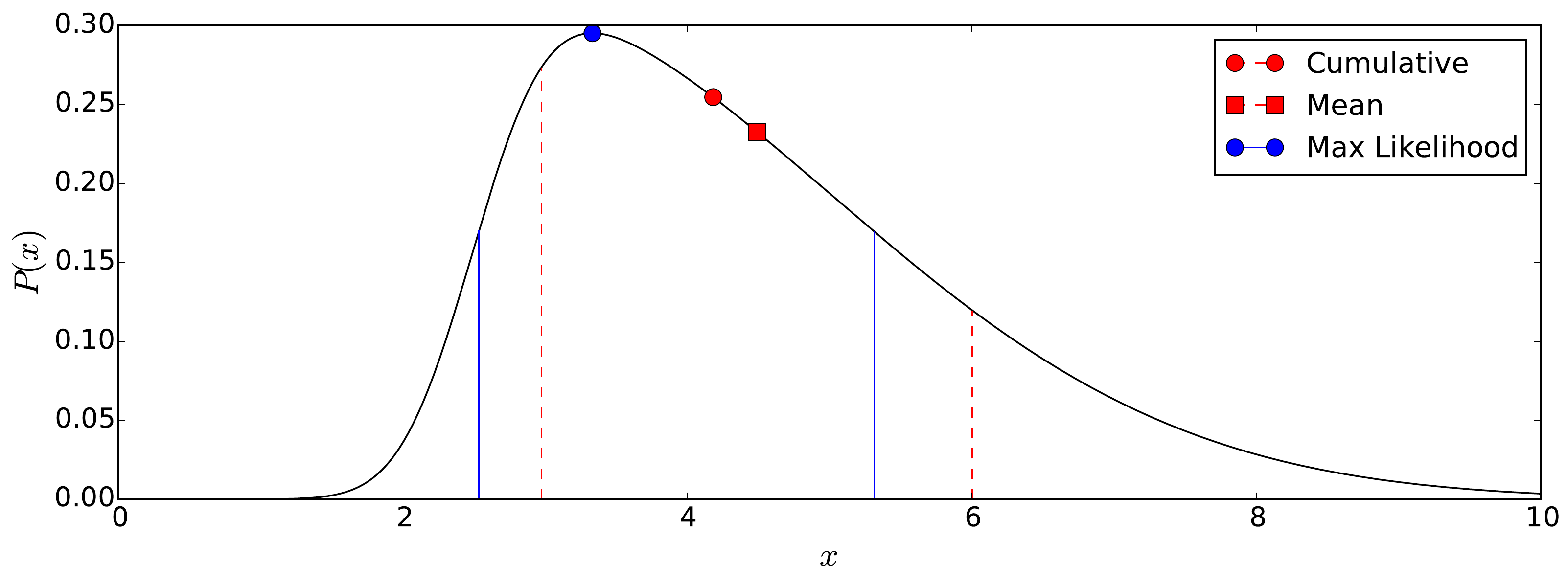}
  \end{center}
  \caption{Three different statistics are presented in the above figure. The red uncertainty bounds and circular marker indicated as ``cumulative'' represent statistics that are determined directly from the cumulative probability distribution function, where the lower $1\sigma$ confidence bound, mean value, and upper bound are respectively given by the $x$ value of the cumulative function at points $P(x) = 0.15865$, $0.5$, and $0.84135$. The second statistic displayed, denoted ``mean'', is similar to the prior method in determining parameter bounds, but achieves a symmetric error by taking the mean value as simply the mean of the $1\sigma$ parameter bounds. The final statistic, ``max likelihood,'' is based upon the point of maximum likelihood, which is given as the quoted parameter value. Uncertainty in this value is then given by the parameter values which have equal probability and together enclose an area of $68.27\%$. All three statistics converge to the same outcome for Gaussian distributions. The statistics given in \citet{BlakeKazin2011} follow a method similar to the mean methodology to achieve symmetric upper and lower error bounds. I have instead chosen to use the maximum likelihood based statistics for my own results, and only quote ``mean'' statistics in Table \ref{tab:blakekazintable} to try and match their output. As such, it is useful to compare differences in final parameter bounds for the values quoted in Table \ref{tab:blakekazintable}. This comparison is found in Table \ref{tab:stats}.}
  \label{fig:statistics}
\end{figure}

It is interesting to note that the choice of statistics - the process of moving from likelihood surface to numeric constraints - used to extract parameter bounds can have a significant effect on the final constraints achieved. Three methods of extracting constraints from distributions are contrasted in Figure \ref{fig:statistics}, where a skewed Gaussian distribution has been used as the underlying probability distribution. All three methodologies investigate converge to the same numeric constraints for a Gaussian distribution, however our probability distributions do not always take this ideal shape. To capture the important information contained in the point of maximum likelihood in my distributions, I utilise statistics that preserve this information and give asymmetric error bars, shown as ``Max likelihood'' statistics in Figure \ref{fig:statistics}.

\begin{table}[h]
\centering
\caption{A comparison between the parameter outcomes from Table \ref{tab:blakekazintable} when using different statistics. We should note that no one particular choice of statistics dramatically improves the level of agreement between my results and those of \citet{BlakeKazin2011}.}
\begin{tabular}{c|cc|cc|cc} 
\specialrule{.1em}{.05em}{.05em} 
Sample  & \multicolumn{2}{c}{Cumulative}  & \multicolumn{2}{c}{Mean} & \multicolumn{2}{c}{Max likelihood} \\
 $z_{\rm{eff}}$ &  $\Omega_c h^2$ & $\alpha$ &  $\Omega_c h^2$ & $\alpha$ &  $\Omega_c h^2$ & $\alpha$ \\
\specialrule{.1em}{.05em}{.05em} 
$0.44$ & $0.136^{+0.018}_{-0.014}$ & $1.026^{+0.110}_{-0.086}$ & $0.138\pm 0.016$ & $1.038\pm 0.098$ & $0.133^{+0.019}_{-0.013}$ & $1.038\pm0.098$ \\
0.6 & $0.140^{+0.016}_{-0.013}$ & $1.011^{+0.090}_{-0.074}$ & $0.142\pm 0.014$ & $1.019\pm0.082$ & $0.139^{+0.016}_{-0.013}$ & $1.003^{+0.089}_{-0.072}$ \\
0.73 & $0.113^{+0.013}_{-0.011}$ & $1.134^{+0.072}_{-0.077}$ & $0.114\pm 0.012$ & $1.132\pm 0.074$ & $0.111^{+0.013}_{-0.010}$ & $1.146^{+0.072}_{-0.075}$ \\
\specialrule{.1em}{.05em}{.05em} 
\end{tabular}\label{tab:stats}
\end{table}

In all cases in the comparison from Table \ref{tab:stats}, we can see that the change in mean value is within the uncertainty of all statistical methods, indicating that the symmetry of all recovered distributions is great enough that the choice of statistical methodology does not significantly impact final parametrisations. Whilst deviations from the results of \citet{BlakeKazin2011} are larger than expected, as they are well within a $1\sigma$ limit, I will move on to constructed the 2D BAO model.

\pagebreak
\section{Confirming the angle dependent BAO model with WizCOLA}

In moving to a 2D analysis, anisotropies must now be taken into account. I use the same technique as the 1D analysis to calculate up to the non-linear power spectrum $P_{\rm{nl}}(k)$. Considering anisotropic effects, distortions due to coherent infall are corrected via an angle dependent factor:
\begin{align}
P_{\rm{nl}}(k, \mu) = \left(1 + \beta \mu^2\right)^2 P_{\text{nl}}(k),
\end{align}
where $\mu$ is the cosine of the angle with line-of-sight, and $\beta$ is the growth rate. The growth rate is marginalised over in this study, and can be checked for consistency by comparing it to the approximate value
\begin{align}
\beta \approx \frac{\Omega_m^{0.55}}{b},
\end{align}
where $b$ is galaxy bias. The effect of galaxy bias on the power of the spectrum is marginalised over with free parameter $b$, and the pairwise velocity dispersion of galaxies is reflected in the Lorentz distribution factor (see discussion after equation \ref{eq:lorentzian}), such that we get
\begin{align}
P_{\rm{gal}}(k, \mu) = \frac{b^2 P_{\rm{nl}}(k, \mu)}{1 + \left(\sigma_v H(z) k \mu\right)^2},
\end{align} 
where $\sigma_v$ is the velocity dispersion, and $\sigma_v H(z)$ is marginalised over. The multipole expansion of this power spectrum is given by
\begin{align}
P_{\ell}(k) = \frac{2\ell + 1}{2} \int_{-1}^1 P_{\rm{gal}}(k, \mu) \ \mathcal{L}_\ell \  d\mu = (2\ell + 1) \int_{0}^1 P_{\rm{gal}}(k, \mu) \ \mathcal{L}_\ell \  d\mu
\end{align}
where $\mathcal{L}_\ell$ represents the $\ell$'th Legendre polynomial. For the monopole moment, this gives
\begin{align}
P_0(k) = \int_0^1 P_{\rm{gal}}(k, \mu) \  d\mu,
\end{align}
 and similarly for the quadrupole we get
\begin{align}
P_2(k) = 5 \int_0^1 \frac{3\mu^2 - 1}{2}\  P_{\rm{gal}}(k, \mu)\  d\mu.
\end{align}
The monopole and quadrupole moments of the power spectrum are then Fourier transformed to give the moments of the correlation function $\xi(s)$, such that
\begin{align}
\xi_\ell(s) = \frac{b^2 B(s)}{(2\pi)^3} \int 4\pi k^2 \ P_\ell(k) \ j_\ell(ks) e^{-k^2 a^2},
\end{align}
where $j_\ell(ks)$ are spherical Bessel functions of the first kind, and the Gaussian dampening term has been added following \citet{AndersonAubourg2012} to improve convergence, with $a= 0.5\, h/\rm{Mpc}$. My chosen value of $a$ differs to that of \citet{AndersonAubourg2012} as I found a lower value of $a$ gave more numerically accurate results for low $s$, as can be seen in Appendix \ref{app:pk2xi}. The multiplicative factors $b^2 B(s)$ represent the linear bias and scale dependent growth seen previously in the 1D model. From these multipole moments of the correlation function, I can construct both the multipole data and wedge data representations of my 2D model.

\subsection{Testing WizCOLA multipoles}

I now need to transform the basic multipole correlation functions such that they match the multipole data expressions as outlined in \S\ref{sec:prior:cor:mp} to introduced a quantised anisotropic warping. Following equation \eqref{eq:multipoles}, the multipoles of the correlation function were transformed such that
\begin{align}
\xi_0(s) &= \xi_0(\alpha s) + \frac{2}{5}\epsilon \left[ 3 \xi_2(\alpha s) + \frac{d \xi_2(\alpha s)}{d \log(s)}\right] \\
\xi_2(s) &= 2\epsilon \frac{d \xi_0(\alpha s)}{d\log(s)} + \left( 1 + \frac{6}{7}\epsilon\right) \xi_2(\alpha s) + \frac{4}{7} \epsilon \frac{d \xi_2(\alpha s)}{d \log(s)} + \frac{4}{7}\epsilon \left[ 5 \xi_4 (\alpha s) + \frac{d \xi_4(\alpha s)}{d \log(s)} \right] ,
\end{align}
where I have not utilised the approximation to simplify $\xi_0(s)$, nor discarded the hexadecapole term in $\xi_2(s)$. For small $\epsilon$, cosmological parameters can be extracted via
\begin{align}
\alpha(1 + \epsilon)^2 &\approx \frac{H^\mathcal{D}(z)}{H(z)} \\
\alpha(1 + \epsilon)^{-1} &\approx \frac{D_A}{D_A^\mathcal{D}}.
\end{align}
As the fiducial cosmology used in the WizCOLA simulations is identical to the simulation cosmology, we do not expect to observe anisotropic warping when fitting to the simulation realisations. As such, I can validate my model by checking if it can recover $\alpha = 1.0$ and $\epsilon = 0.0$ when fitting the WizCOLA correlation functions.\\

\begin{figure}[h!]
  \begin{center}
    \includegraphics[width=\textwidth]{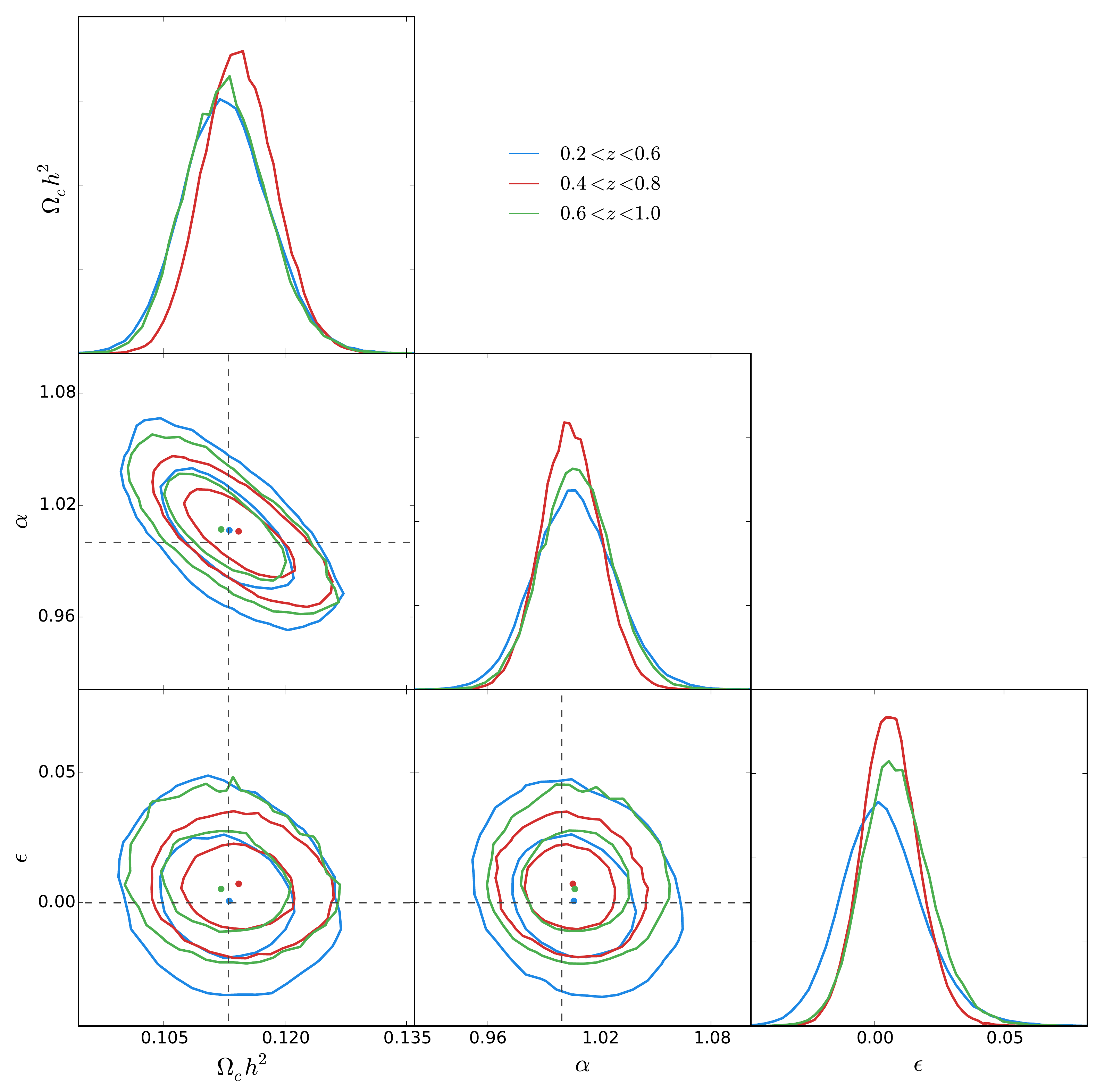}
  \end{center}
  \caption{Model fitting using mean data from the 600 WizCOLA realisations, where we are fitting to the multipole expansion. The dashed line represents the desired parameter recoveries of $\Omega_c h^2 = 0.113$, $\alpha=1.0$, $\epsilon=0.0$. Increasing the range of data points included in the matching served to draw the recovered $\Omega_c h^2$ further away from the desired recovery value.}
  \label{fig:wizmp}
\end{figure}

The WizCOLA simulations were configured with cosmology $\Omega_m = 0.273$, $\Omega_\Lambda = 0.727$, $\Omega_b = 0.0456$, $h=0.705$, $\sigma_8 = 0.812$ and $n_s = 0.961$, following WMAP5 cosmology \citep{Komatsu2009}. Putting this in terms of $\Omega_c h^2$, we have $\Omega_c h^2 = 0.113$. The WizCOLA data is presented both in multipole expansion and in data wedges, and in this section I will fit to the multipole expansion. To reduce statistical uncertainty as much as possible, the input data was created using the mean of all 600 realisations of the WizCOLA simulations, and the variance was thus reduced by a factor of  $\sqrt{600}$ as each realisation should be independent. Data was fit in the range $25<s<180$, and a comparison of the effects of dataset truncation can be found in Appendix \ref{app:truncation}. As with the one dimensional fitting performed previously, fit parameters $\Omega_c h^2$, $b^2$, $\log(k_*)$ were allowed to vary, with additional angular dependent marginalisation parameters $\beta$ and $\sigma H(z)$ also allowed to vary. The main parameters of interest for cosmology are the dilation and warp parameters $\alpha$ and $\epsilon$, since these relate directly to $H(z)$ and $D_A(z)$. Likelihood surfaces and marginalised parameter distributions for fits to all three redshift bins are shown in Figure \ref{fig:wizmp}, and final parameter constraints are detailed in Table \ref{tab:wizmp}.\\

For all redshift bins, all desired recovery parameters were recovered well within the $1\sigma$ uncertainty limit. We can also see that, looking at the mean value of the determined values for $\log(k_*) = -2.10$, this gives a $\sigma_v = 5.77 h^{-1}$ Mpc, which  is in the magnitude expected by the theory given in equation \eqref{eq:sigmav} and the values found found in \citet{BlakeKazin2011} and \citet{BlakeDavis2011}. It should also be noted that within the range $\sigma_v \in [0,10]$, no significant difference is found in determined $\chi^2$ values, indicating that setting a specific $\sigma_v$ when refitting is sensitive to the magnitude, but is not tightly constrained within theoretically predicted ranges. Fixing $\sigma_v = 5 h^{-1}$ Mpc, I recover fits also consistent with a $1\sigma$ recovery of simulation cosmology.

\begin{table}[h]
\centering
\caption{Recovered parameter constraints when fitting to the combined 600 realisations of the WizCOLA simulation data multipoles.}
\begin{tabular}{cc|ccccc}
\specialrule{.1em}{.05em}{.05em} 
Sample & $z_{\rm{eff}}$ & min $\chi^2$ & $\Omega_c h^2$ &$\alpha$ & $\epsilon$ & $\log(k_*)$\\
\specialrule{.1em}{.05em}{.05em} 
$0.2 < z < 0.6$ & $0.44$ & $12.0$ & $0.112^{+0.005}_{-0.006}$ &$1.006^{+0.023}_{-0.022}$ & $0.002^{+0.016}_{-0.016}$ & $-2.18^{+0.22}_{-0.20}$\\
$0.4 < z < 0.8$ & $0.60$ & $8.6$  & $0.114^{+0.005}_{-0.004}$ &$1.004^{+0.017}_{-0.016}$ & $0.005^{+0.012}_{-0.010}$ & $-2.05^{+0.20}_{-0.17}$\\
$0.6 < z < 1.0$ & $0.73$ & $10.8$ & $0.113^{+0.005}_{-0.005}$ &$1.006^{+0.021}_{-0.018}$ & $0.007^{+0.015}_{-0.012}$ & $-2.07^{+0.26}_{-0.23}$\\
\specialrule{.1em}{.05em}{.05em} 
Input & & & 0.113 & 1.0 & 0.0 & \\
\specialrule{.1em}{.05em}{.05em} 
\end{tabular}\label{tab:wizmp}
\end{table}

\begin{figure}[h!]
  \begin{center}
    \includegraphics[width=\textwidth]{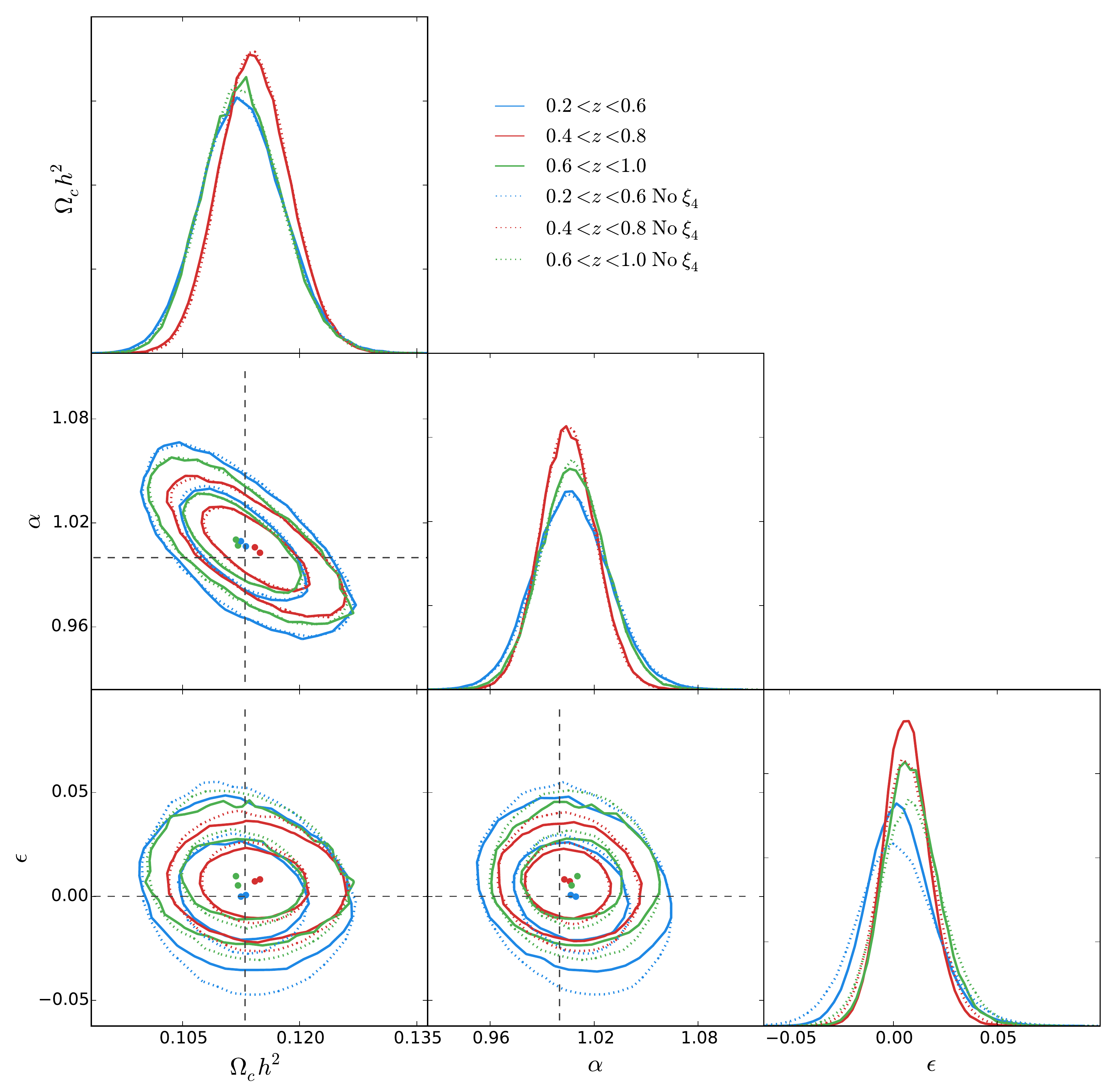}
  \end{center}
  \caption{A comparison of likelihood surfaces for the WizCOLA data when including the contribution from the hexadecapole term and when discarding it. Any change in likelihood surfaces and final distributions is negligible when compared to the statistical uncertainty in the final results.}
  \label{fig:wizmpfastComp}
\end{figure}

As shown in \S\ref{sec:prior:cor:mp}, whether or not the hexadecapole terms are included in the multipole analysis changes depending on which study one is examining. In order to test the significance of the hexadecapole term, the analysis shown in Figure \ref{fig:wizmp}, which includes the hexadecapole term, was rerun with the term removed. The comparison likelihood surfaces are shown in Figure \ref{fig:wizmpfastComp}, and we can see from this that the statistical uncertainty dominates any loss of information contained in the hexadecapole signal. Due to computational constraints and the low impact of the term, the hexadecapole contribution was left out of the final model.\\

\begin{figure}[h!]
  \begin{center}
    \includegraphics[width=\textwidth]{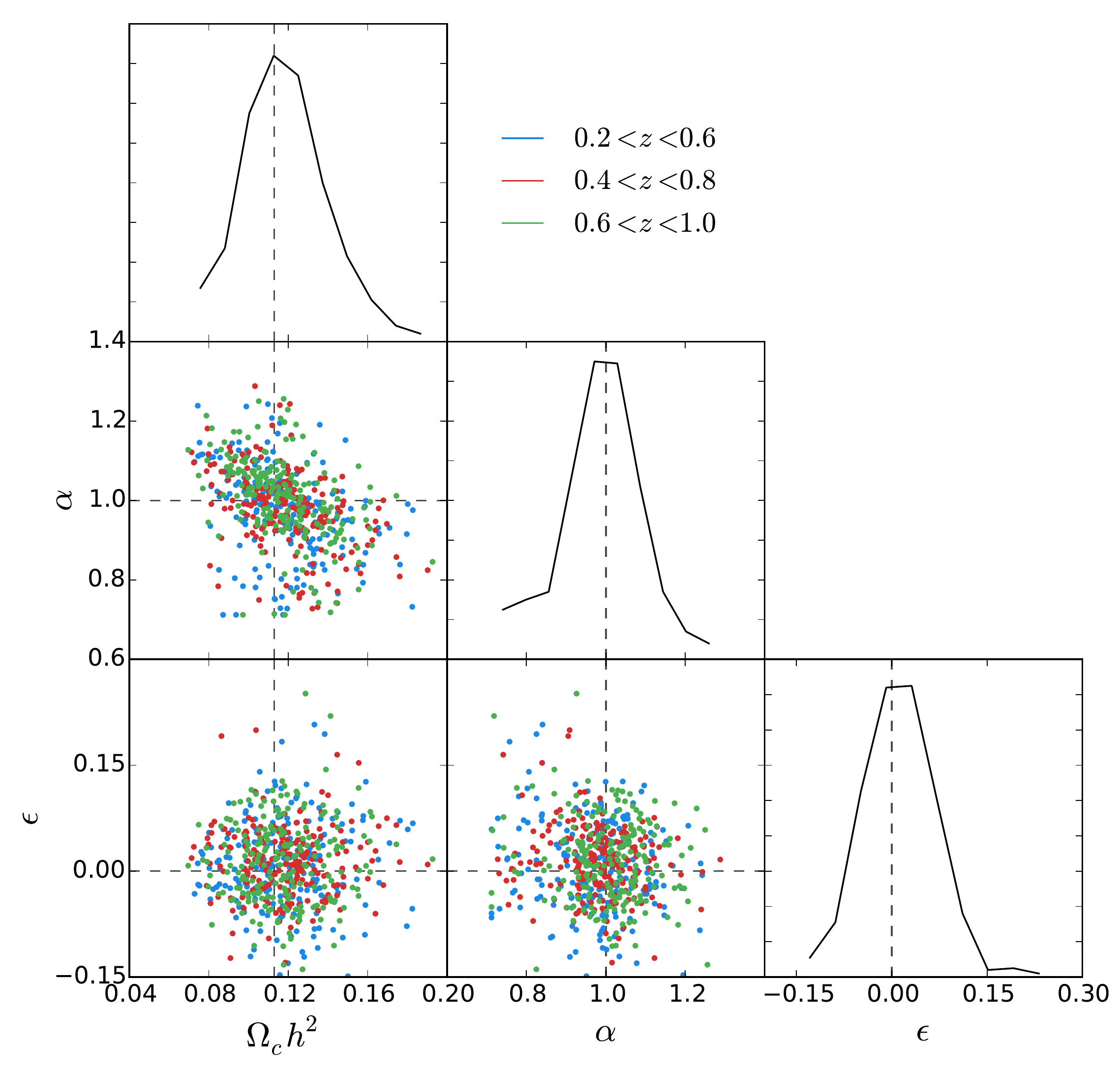}
  \end{center}
  \caption{Maximum likelihood $\Omega_c h^2$, $\alpha$ and $\epsilon$ values from a total of 623 realised bins are shown in the bottom left corner plots. Dashed black lines indicate simulations parameters, and the solid black distributions in the diagonal subplots represent the final distribution across all bins for the specific parameter.}
  \label{fig:mpDist2}
\end{figure}

Finally, we can perform a final validation of the multipole methodology by fitting to individual realisations of the WizCOLA simulation instead of the mean data set. Due to the decreased data strength, I move from allowing the parameter $\log(k_*)$ to vary, to fixing the parameter $\sigma_v = 5\,$h$^{-1}$ Mpc, which was validated using the mean simulation data. Fixing the parameter was not found to significantly modify output parameters, and a reduction in the number of free parameters results in tighter constraints. Figure \ref{fig:mpDist2} confirms that there is no significant offset in the recovered parameter distribution from the simulation parametrisation.

\newpage \phantom{blabla} \clearpage

\section{Testing against WizCOLA wedges}

\begin{figure}[h!]
  \begin{center}
    \includegraphics[width=\textwidth]{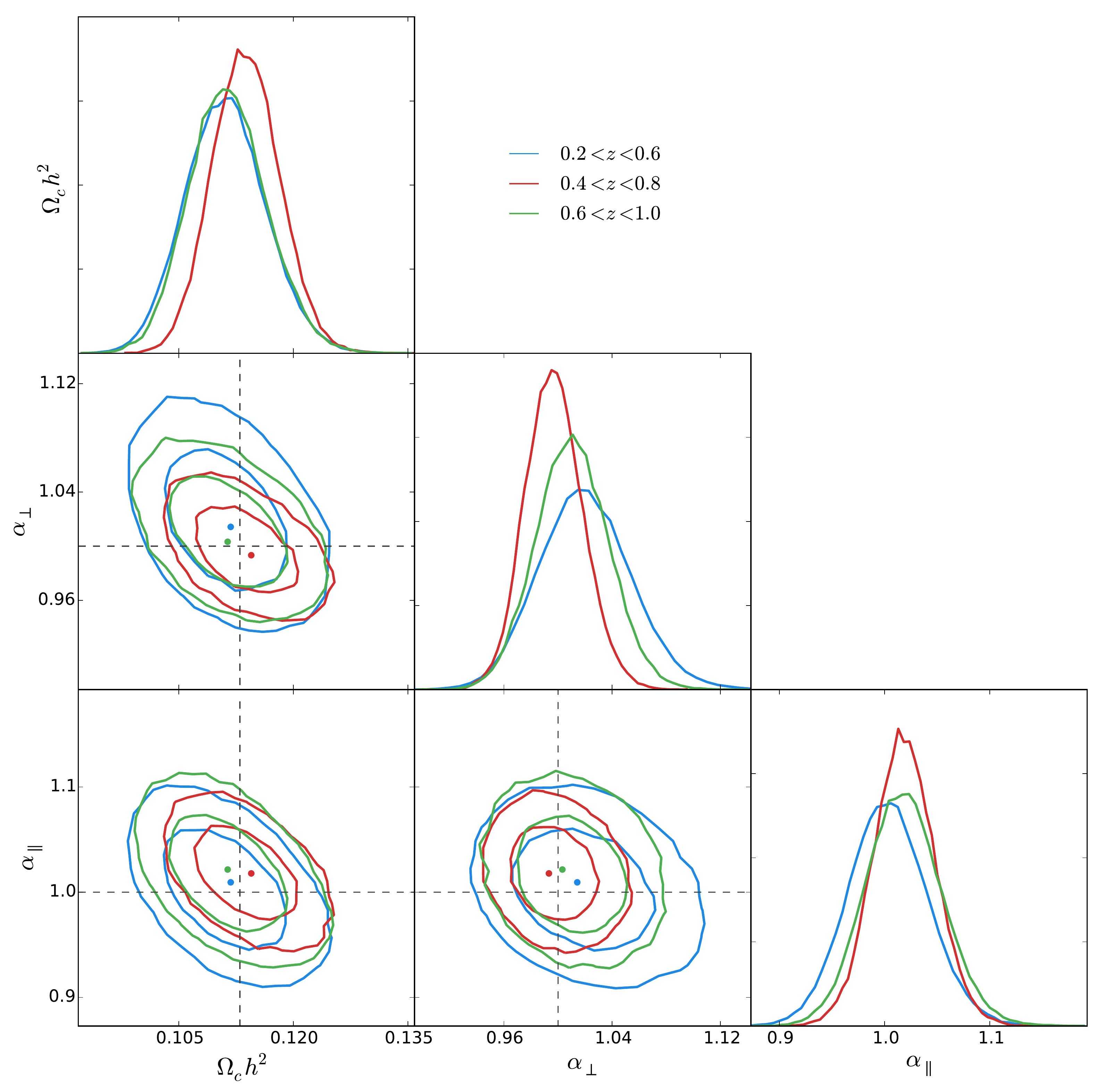}
  \end{center}
  \caption{Model fitting using mean data from the 600 WizCOLA realisations, where we are fitting to the wedged data. The dashed line represents the desired parameter recovery of $\Omega_c h^2 = 0.113$, $\alpha_\perp=1.0$, $\alpha_\parallel=1.0$. }
  \label{fig:wizwedge}
\end{figure}

I also test the wedge methodology used by \citet{KazinSanchezCuesta2013} and \citet{SanchezKazinBeutler2013}, whereby one transforms the scale of the tangential and parallel to line-of-sight directions separately in the form of $\alpha_\perp$ and $\alpha_\parallel$. From these transformations we define
\begin{align}
s(\mu^\prime, s^\prime) &= s^\prime \sqrt{\alpha_\parallel^2(\mu^\prime)^2 + \alpha_\perp^2 (1 - (\mu^\prime)^2)}, \\
\mu(\mu^\prime) &= \frac{\alpha_\parallel \mu^\prime}{\sqrt{\alpha_\parallel^2(\mu^\prime)^2 + \alpha_\perp^2 (1 - (\mu^\prime)^2)}},
\end{align}
where data wedges can be extracted via
\begin{align}
\xi^\prime_{\Delta \mu}(s^\prime) = \frac{1}{\Delta \mu^\prime} \int_{\mu^\prime_{\rm{min}}}^{\mu^\prime_{\rm{max}}} \xi(\mu(\mu^\prime), s(\mu^\prime, s^\prime)) d\mu^\prime.
\end{align}
The WizCOLA data provides two wedges, $0 < \mu < 0.5$ and $0.5 < \mu < 1$, which I fit against. As in the previous section, we intend to recover simulation parameters $\Omega_c h^2 = 0.113$ and - due to identical fiducial cosmology and simulation cosmology, expect to recover $\alpha_\perp = \alpha_\parallel = 1.0$. Fit parametrisation is illustrated in Figure \ref{fig:wizwedge}, and recovered parameters are detailed in Table \ref{tab:wizwedge}. As with the multipole fitting, the recovered $\log(k_*)$ is consistent with prior literature and theory, and the parameter recovery is consistent with simulation cosmology. Similarly to the multipole analysis, we recover simulation cosmology after fixing $\sigma_v = 5 h^{-1}$ Mpc. In this way, the mean data for all WizCOLA simulations is used to provide a $\sigma_v$ value for individual realisations, which individually are unable to provide tight constraints on $\sigma_v$.

\begin{table}[h]
\centering
\caption{Recovered parameter constraints when fitting to the combined 600 realisations of the WizCOLA simulation data wedges.}
\begin{tabular}{cc|ccccc}
\specialrule{.1em}{.05em}{.05em} 
Sample & $z_{\rm{eff}}$ & min $\chi^2$ & $\Omega_c h^2$ &$\alpha_\perp$ & $\alpha_\parallel$ & $\log(k_*)$\\
\specialrule{.1em}{.05em}{.05em} 
$0.2 < z < 0.6$ & $0.44$ & $9.8$ & $0.111^{+0.005}_{-0.006}$ & $1.018^{+0.036}_{-0.033} $ & $1.002^{+0.038}_{-0.038} $ & $ -2.16^{+0.24}_{-0.21}$\\
$0.4 < z < 0.8$ & $0.60$ & $9.1$  & $0.113^{+0.005}_{-0.004} $ &$ 0.996^{+0.022}_{-0.022} $ & $ 1.017^{+0.032}_{-0.029}$ & $ -2.02^{+0.21}_{-0.20}$\\
$0.6 < z < 1.0$ & $0.73$ & $10.9$ & $ 0.111^{+0.006}_{-0.005}$ &$1.012^{+0.026}_{-0.030} $ & $ 1.014^{+0.040}_{-0.035}$ & $ -2.04^{+0.28}_{-0.26} $\\
\specialrule{.1em}{.05em}{.05em} 
Input &  & & 0.113 & 1.0 & 1.0 & \\
\specialrule{.1em}{.05em}{.05em} 
\end{tabular}\label{tab:wizwedge}
\end{table}

\begin{figure}[h!]
  \begin{center}
    \includegraphics[width=\textwidth]{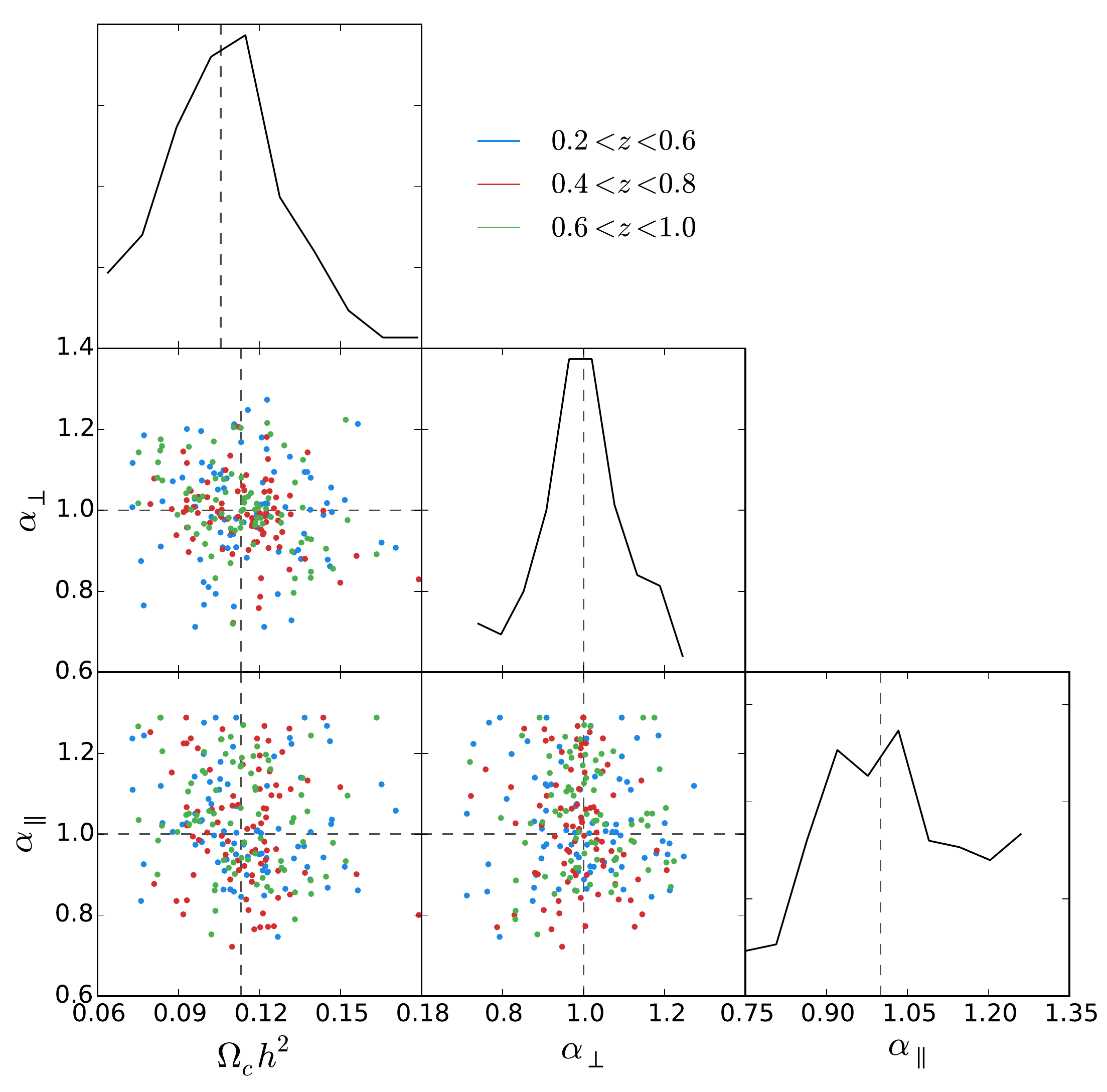}
  \end{center}
  \caption{Maximum likelihood $\Omega_c h^2$, $\alpha_\perp$ and $\alpha_\parallel$ values from a total of 245 realisation bins are shown in the bottom left corner plots. Dashed black lines indicate simulations parameters, and the solid black distributions in the diagonal subplots represent the final distribution across all bins for the specific parameter. The clear physical boundaries of the fits are visible in the $\alpha_\parallel$ constraints, where the imposed $\alpha_\parallel < 1.3$ does not appear sufficient to constrain the parameter, raising a potential problem with the wedge analysis. Increasing $\alpha_\parallel$ is not only $>10\sigma$ from Planck results, but introduces cyclic patterns into the MCMC chain, worsening the fits for $\Omega_c h^2$ and $\alpha_\perp$.}
  \label{fig:wdgDist2}
\end{figure}

As with the multipole expansion, we can perform a final validation of the wedge methodology by fitting to individual realisations of the WizCOLA simulation. This has been done in Figure \ref{fig:wdgDist2}, and confirms that there is no significant offset in the recovered parameter distribution from the simulation parametrisation. Unlike the multipole expansion, there is some cause for concern with the wedge fitting, as the $\alpha_\parallel$ distribution does not converge to zero as we depart from $\alpha_\parallel = 1.0$. Comparing the $\alpha_\perp$ and $\alpha_\parallel$ distribution, we can see that the ability to constrain $\alpha_\parallel$ is considerably less than the ability to constrain $\alpha_\perp$. In terms of data, this is not unexpected, as the $\alpha_\parallel$ correlation function comes from one dimension on the two point correlation function (line of sight), whilst the $\alpha_\perp$ contribution is the sum of the two transverse dimensions. Given this concern, potential contention between the mutlipole expansion and wedge analyses of the data should defer to the multipole analysis.

\newpage \phantom{blabla} \clearpage
\section{Combining data bins}

The data present in the WizCOLA simulations and the final WiggleZ dataset is available in three redshift bins, $0.2 < z < 0.6$, $0.4 < z < 0.8$ and $0.6 < z < 1.0$. In the circumstance where these bins were independent, final parameter constraints could simply be obtained by combining the results for each individual bin. However, the data bins that we have to work with overlap and are thus correlated. \\

There are two methods we can use to combined the binned data, and both methods will be utilised in my analysis so that I can check they give consistent results. The first method investigated is using the correlation between final parameter values, and the second method I investigate is to calculate the covariance between data points across all bins and run a separate fit.

\subsection{First Method: Parameter Covariance} \label{sec:parameterCov}

\begin{figure}[h!]
  \begin{center}
    \includegraphics[width=\textwidth]{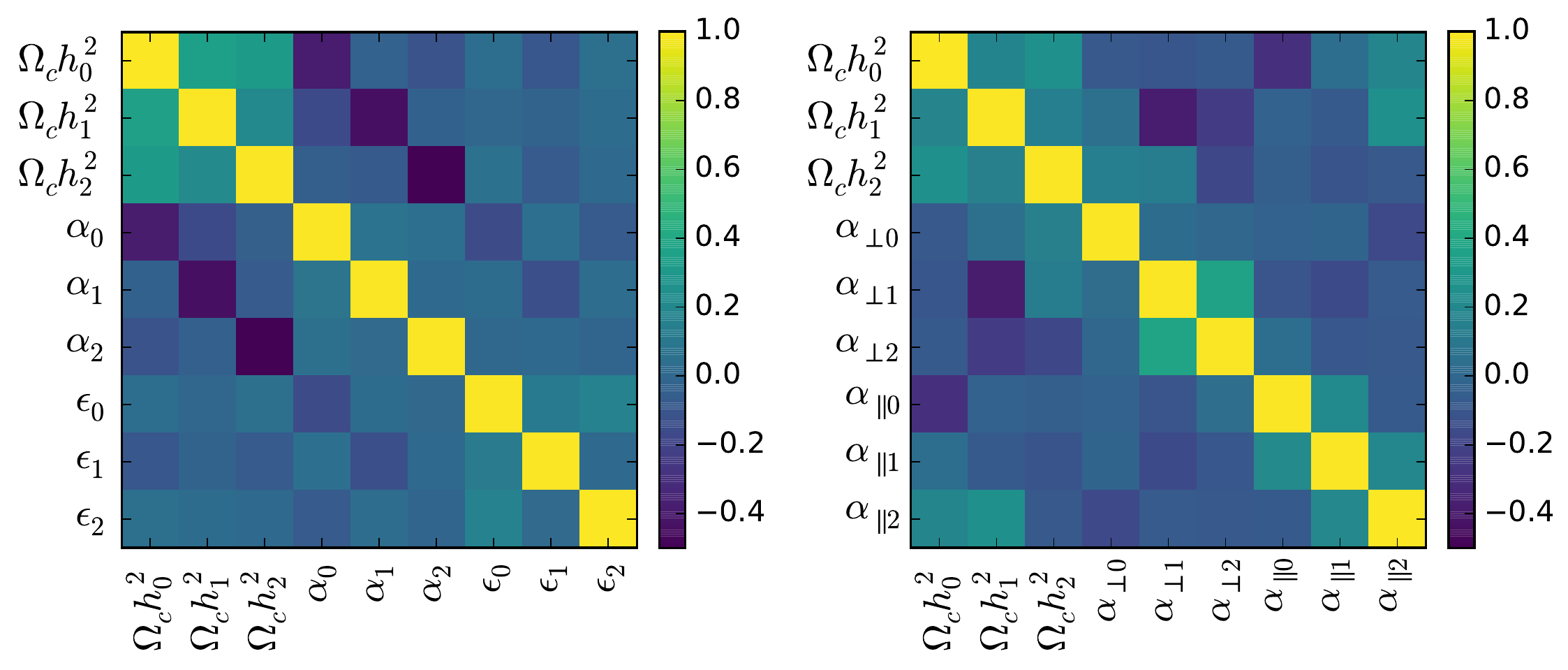}
  \end{center}
  \caption{Correlations between final cosmological parameters when fitting to the three redshift bins of each WizCOLA simulation for both the multipole data and wedge data formats. The subscript numbers after each parameter are used to denote the redshift bin, with $0$, $1$ and $2$ respectively denoting the $0.2 < z < 0.6$, $0.4 < z < 0.8$ and $0.6 < z < 1.0$ bins. }
  \label{fig:correlations}
\end{figure}

In order to determine final parametrisations across all redshift bins, the correlation between fit parameters from individual redshift bins needs to be quantified and accounted for. To do this, I fit individual realisations of the WizCOLA simulation, and construct a $9\times 9$ covariance matrix from the peak likelihood fit values for parameters ($\Omega_c h^2$, $\alpha$ and $\epsilon$ for a multipole analysis and $\Omega_c h^2$, $\alpha_\perp$ and $\alpha_\parallel$ for the wedge analysis) were utilised to construct a covariance matrix, such that we construct
\begin{align}
C_{ij} = \frac{1}{N} \sum\limits_{n=1}^{N} (P_i - \bar{P}_i)(P_j - \bar{P}_j),
\end{align}
where $P_i$ represents the list of parameters 
\begin{align*}
\left\lbrace \Omega_c h^2 (z = 0.44), \Omega_c h^2 (z = 0.60),  \Omega_c h^2 (z = 0.73), \right. \\ 
\alpha (z = 0.44), \alpha (z = 0.60),  \alpha (z = 0.73), \\
\epsilon (z = 0.44), \left. \epsilon (z = 0.60),  \epsilon (z = 0.73) \right\rbrace
\end{align*}
Similarly to the covariance matrix, we can also calculate the correlation matrix, defined similarly as
\begin{align}
R_{ij} = \frac{1}{N} \sum\limits_{n=1}^{N} \frac{(P_i - \bar{P}_i)(P_j - \bar{P}_j)}{\sigma_i \sigma_j},
\end{align}
where $\sigma_i$ represents the standard deviation of the $i$th parameter. The correlation matrix determined from analysis of $N$ realisations is shown in Figure \ref{fig:correlations}. Ideally, this matrix would be constructed using all $N=600$ realisations, however computational limitations have reduced this to $N \approx 200$, which may not be sufficient to produce smooth and accurate covariance matrices. I intend to continue to run fits for further realisations, and the results of these fits will be presented in a future paper.

\begin{figure}[h!]
  \begin{center}
    \includegraphics[width=0.8\textwidth]{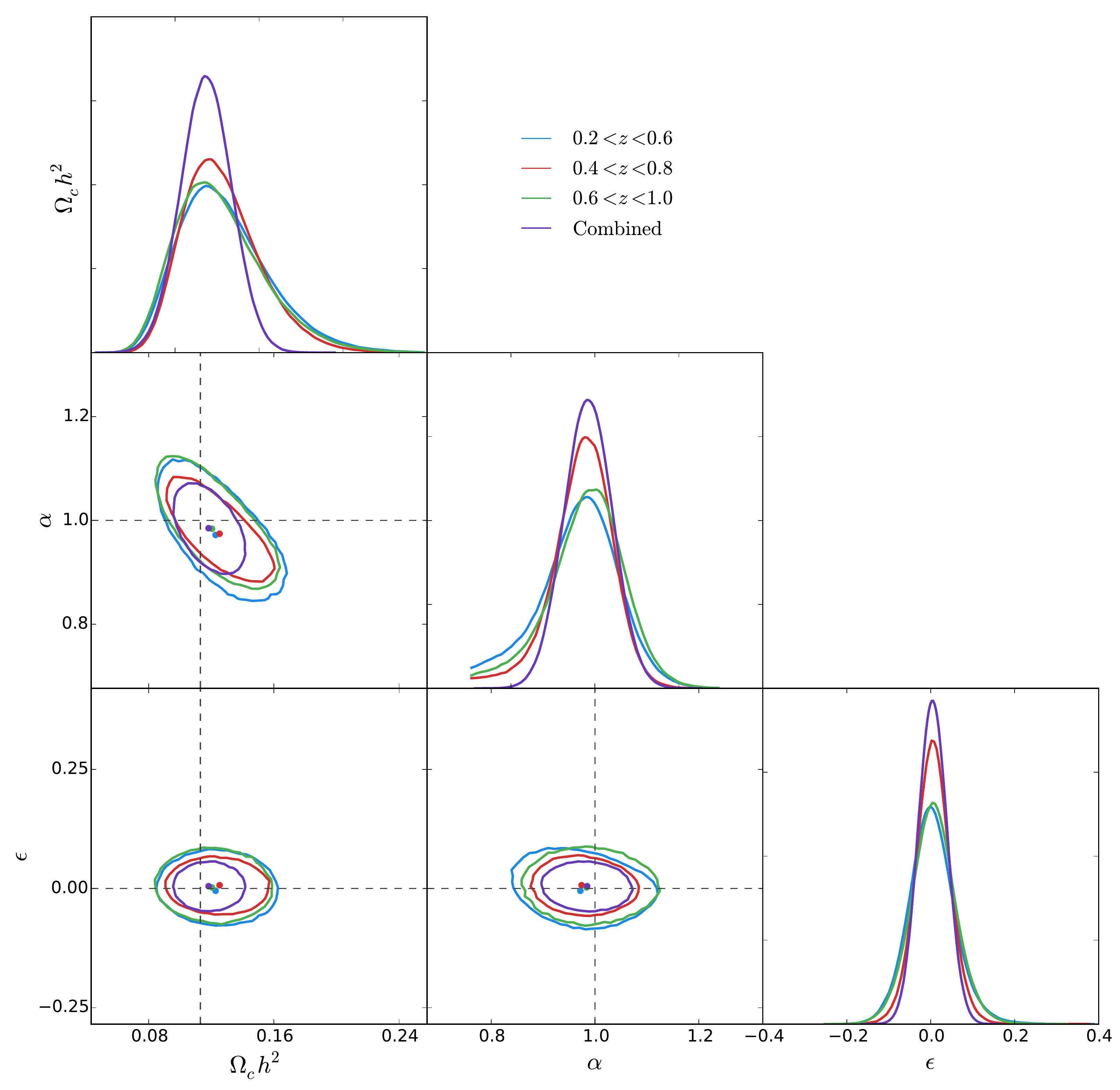}
  \end{center}
  \caption{Likelihood surfaces and marginalised distributions of $\Omega_ch^2$, $\alpha$ and $\epsilon$ for the mean data across all 600 WizCOLA simulations.}
  \label{fig:corCombinedMPWiz2}
\end{figure}

\begin{figure}[h!]
  \begin{center}
    \includegraphics[width=0.8\textwidth]{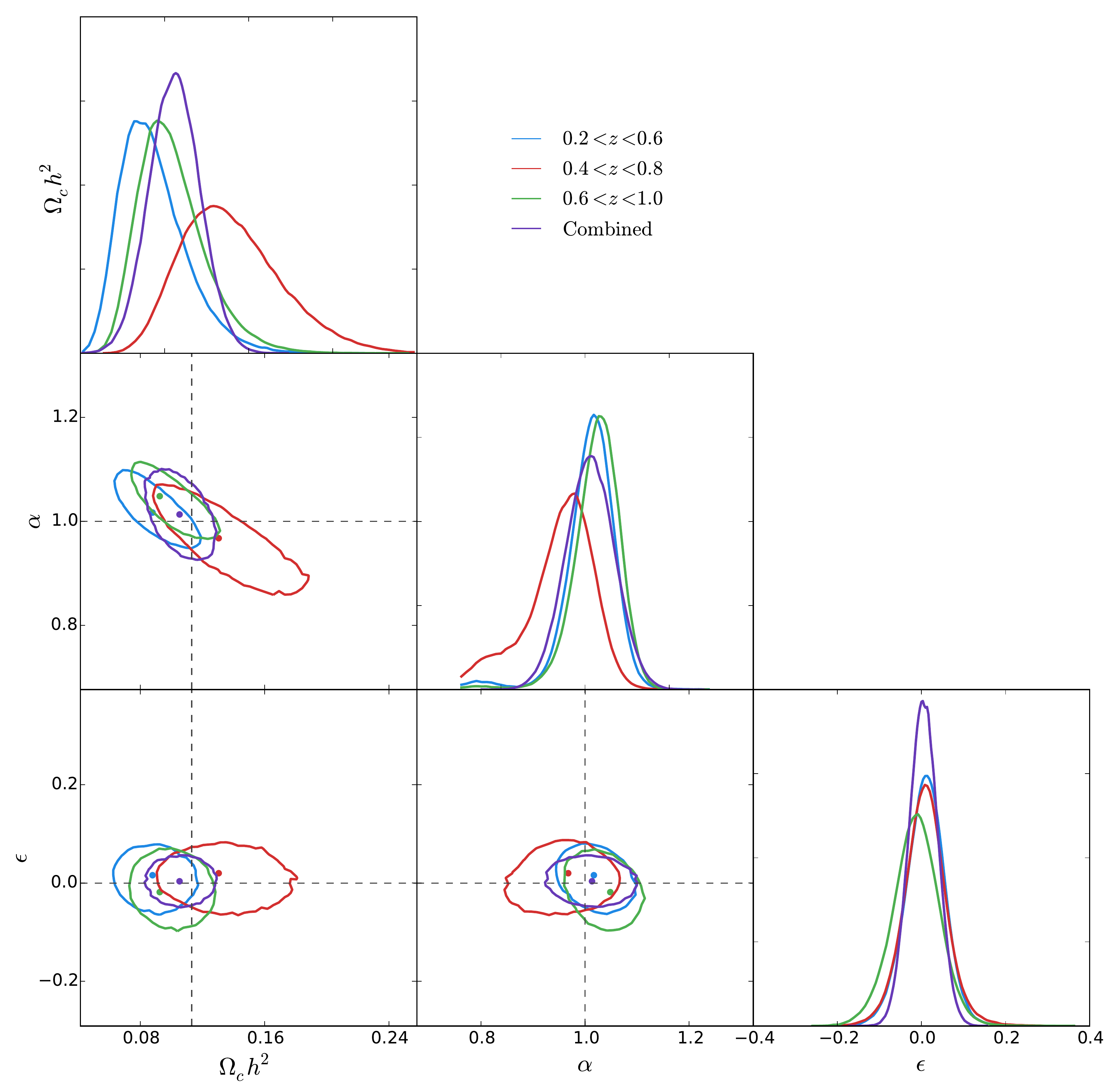}
  \end{center}
  \caption{Likelihood surfaces and marginalised distributions of $\Omega_ch^2$, $\alpha$ and $\epsilon$ for the $N=5$ realisation of the WizCOLA simulation. Simulation parametrisation is recovered within the $1-sigma$ limit.}
  \label{fig:corCombined_2}
\end{figure}

This covariance matrix can now be used to fit for a final $\Omega_c h^2$, $\alpha$ and $\epsilon$ (for a multipole analysis), or a final $\Omega_c h^2$, $\alpha_\perp$ and $\alpha_\parallel$ (for a wedge analysis). Treating only multipoles hereonin to simplify the text, this is done by  minimising the $\chi^2$ statistic, given as
\begin{align} \label{eq:covchi}
\chi^2(\Omega_c h^2, \alpha, \epsilon) = (\Omega_c h^2 - \Omega_c h^2_0, \Omega_c h^2 - \Omega_c h^2_1, ...,\  \epsilon - \epsilon_2)^T C_{ij}^{-1}(\Omega_c h^2 - \Omega_c h^2_0, \Omega_c h^2 - \Omega_c h^2_1, ...,\  \epsilon - \epsilon_2),
\end{align}
where again the subscript indices on the $\Omega_c h^2$, $\alpha$ and $\epsilon$ refer to the redshift bins. In essence, we utilise the parameters fitted to each bin as datapoints in a secondary model, which we minimise with respect to the final parameters $\Omega_c h^2$, $\alpha$ and $\epsilon$. To test this methodology, I first utilise it on a fit to the mean WizCOLA simulation values, illustrated in Figure \ref{fig:corCombinedMPWiz2}. With this test showing that tighter constraints can be achieved by combining bins, it has been applied to the $N=5$ WizCOLA realisation to more accurately test the effect on the final WiggleZ dataset.  The fitting results for the $N=5$ realisation are described in Table \ref{tab:n2Wizcola}. We can see from the results that combining the redshift bins recovers tighter parameter constraints that are closer to recovering simulation parameters than any individual redshift bin for this realisation. The likelihood surfaces and marginalised parameter distributions for the three bins and the combined data is shown in Figure \ref{fig:corCombined_2}. In addition to this test, once further realisations are fitted, I intend to calculate a distribution of final parameter values (by combining the fits from all 600 realisations) to ensure that this methodology is without systematic bias.

\begin{table}[h]
\centering
\caption{Recovered parameter constraints when fitting the $N=5$ realisation of the WizCOLA simulation. The parameter values after combination are shown at the bottom of the table, where the $\chi^2$ value for the combined dataset indicates the $\chi^2$ from equation \eqref{eq:covchi} instead of the $\chi^2$ when fitting the cosmological model to the WizCOLA data.}
\begin{tabular}{cc|ccccc}
\specialrule{.1em}{.05em}{.05em} 
Sample & $z_{\rm{eff}}$ & min $\chi^2$ & $\Omega_c h^2$ &$\alpha$ & $\epsilon$\\
\specialrule{.1em}{.05em}{.05em} 
$0.2 < z < 0.6$ & $0.44$ & $52.8$ & $0.084^{+0.022}_{-0.014}$ &$1.02^{+0.05}_{-0.05}$ & $  0.01^{+0.04}_{-0.05} $ \\
$0.4 < z < 0.8$ & $0.60$ & $28.2$ & $0.130^{+0.032}_{-0.025}$ &$0.97^{+0.06}_{-0.08}$ & $0.01^{+0.05}_{-0.05}$ \\
$0.6 < z < 1.0$ & $0.73$ & $37.8$ & $0.097^{+0.022}_{-0.015}$ &$1.04^{+0.05}_{-0.05}$ & $-0.01^{+0.05}_{-0.05}$ \\
\specialrule{.05em}{.05em}{.05em} 
Combined & $0.60$ & $3.7$ & $0.107^{+0.014}_{-0.016}$ &$1.01^{+0.06}_{-0.06}$ & $0.00^{+0.04}_{-0.03}$ \\
\specialrule{.1em}{.05em}{.05em} 
\end{tabular}\label{tab:n2Wizcola}
\end{table}

\clearpage
\subsection{Second method: All data covariance} \label{sec:allData}

\begin{figure}[h!]
  \begin{center}
    \includegraphics[width=\textwidth]{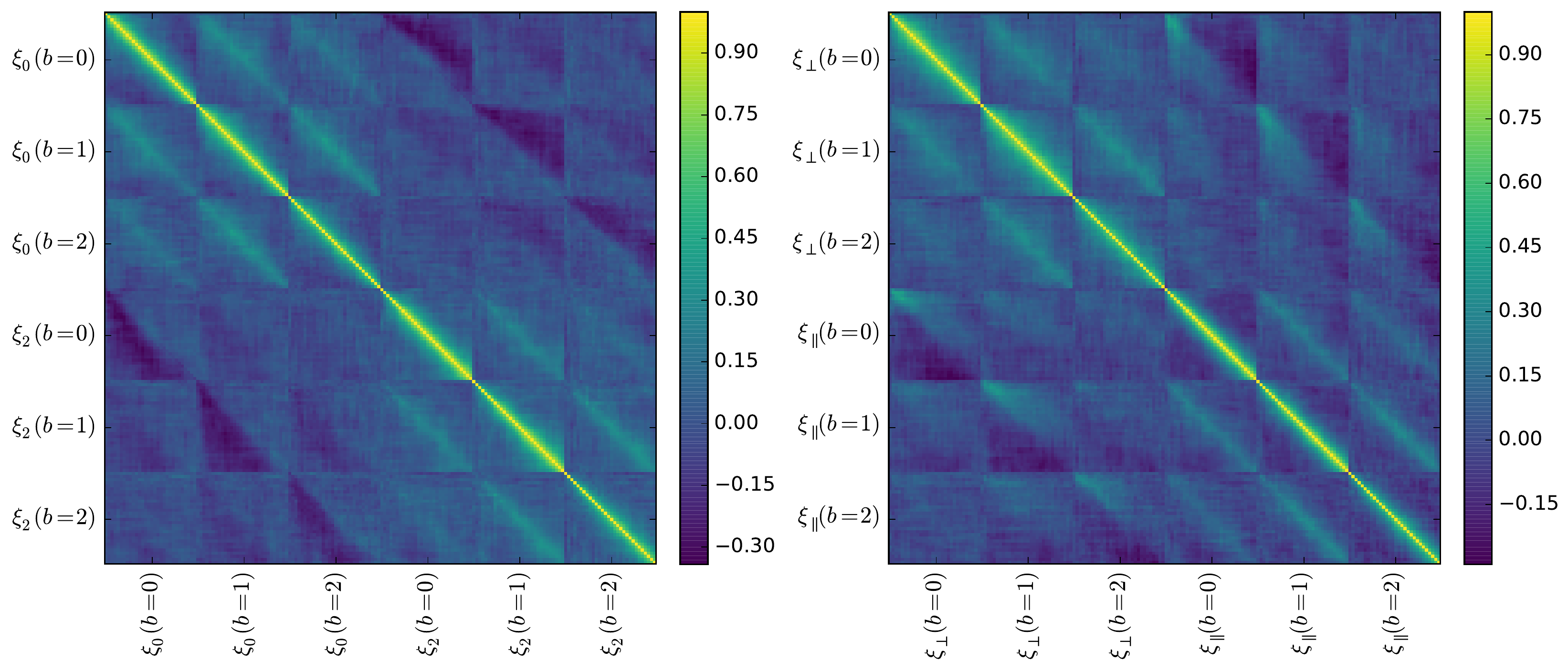}
  \end{center}
  \caption{Full data correlation matrices constructed for both the multipole (left) and wedge (right) expressions of the WizCOLA data. The $b=0$, $b=1$ and $b=2$ labels respectively refer to the redshift bins $0.2 < z < 0.6$, $0.4 < z < 0.8$ and $0.6 < z < 1.0$. We can see that, even though the $b=0$ and $b=2$ bins do not overlap, some faint correlation still persists. This is expected, as long modes in the simulation would span both bins.}
  \label{fig:fullCorrelations}
\end{figure}

The covariance matrices utilised so far in my analysis have been supplied from the WizCOLA simulations, and give data covariance inside each redshift bin. However, also having the 600 WizCOLA realisations, I can reconstruct a full covariance matrix to give me the covariance between values of the correlation function across redshift bins. The correlation matrix for the multipole and wedge data formats are shown in Figure \ref{fig:fullCorrelations}.\\

It is interesting to note that the computed covariance differs from the supplied WizCOLA covariance in off-diagonal cells, and an analysis of the significance of this difference can be found in Appendix \ref{app:wizDiff}. When using the full data covariance to simultaneously fit all three redshift bins, a further question becomes whether marginalisation parameters $b^2$, $\beta$, $\sigma H$ and $k_*$ should be free between redshift bins, or consistent across them. \\

From a physical motivation, we expect the bias parameter $b^2$ to be dependent on redshift bin. This is that the further out we look, the fainter galaxies appear. As such, at higher redshifts, we are more likely to successfully observe more massive, luminous galaxies, which increases bias. We can also see that the $\sigma H(z)$ nuisance parameter is explicitly a function of redshift, and so it is expected to change between redshift bins. However, when performing fits, $b^2$ and $\beta$ are well constrained, whilst $k_*$ and $\sigma H(z)$ are not. As this implies that those two parameters do not significantly contribute to the likelihood calculations, it is unknown if setting $\sigma H(z)$ free between bins will have a noticeable benefit. \\

\begin{figure}[h!]
  \begin{center}
    \includegraphics[width=\textwidth]{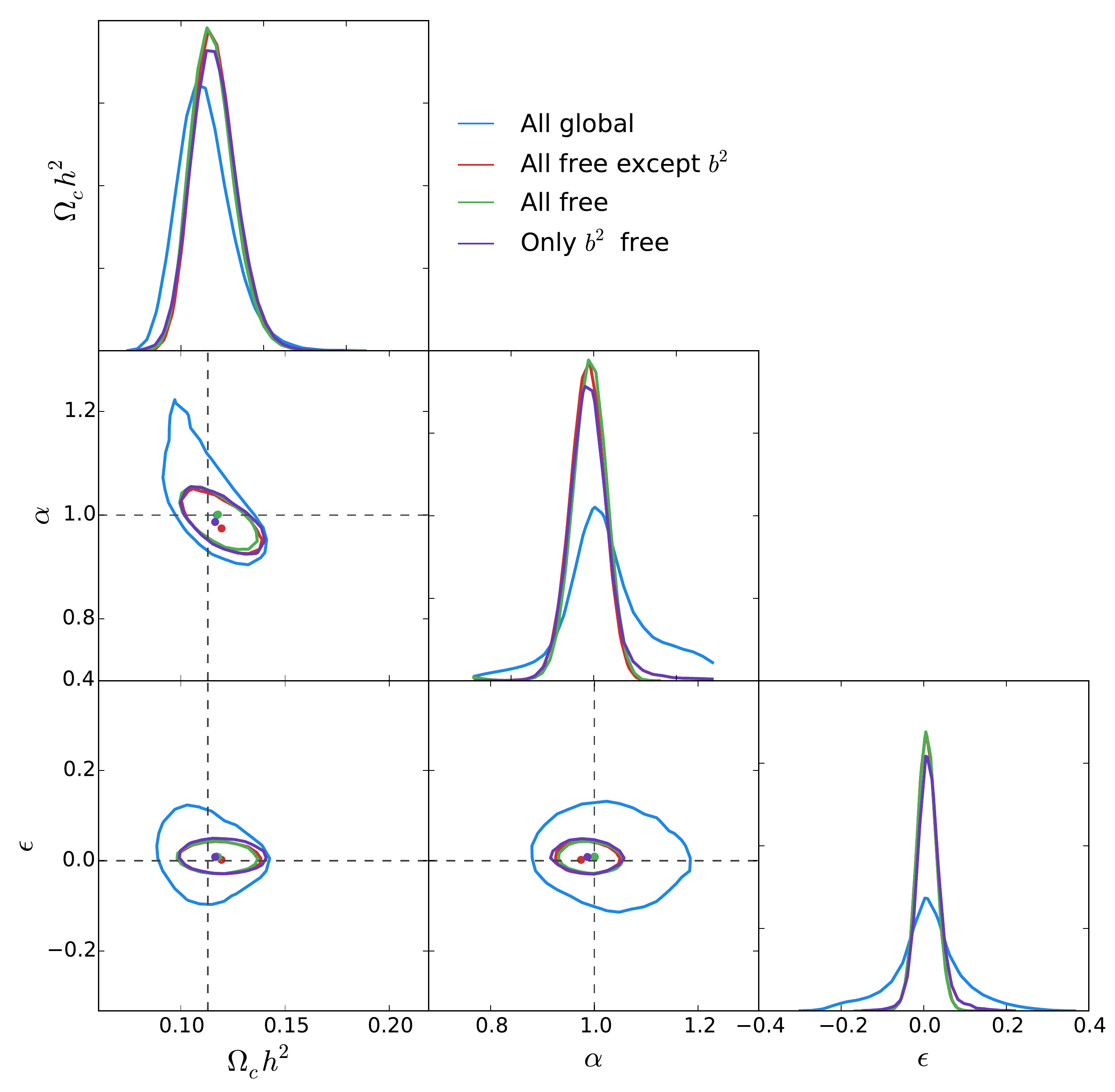}
  \end{center}
  \caption{Fits to the mean WizCOLA data using the computed covariance matrix. Marginalisation parameters for the fit are $b^2$, $\beta$, $\sigma_v$ and $\sigma H(z)$. There are four fits shown, where the word `free' has been used to indicate that the parameter in question has been made unique for each redshift bin. We can see that setting more than just $b^2$ as a nuisance parameter per bin does not increase the quality of obtained results.}
  \label{fig:wizcolaAllNormalCovCombined}
\end{figure}

To investigate this, I ran fits to the combined WizCOLA data where I set no nuisance parameters free between redshift bins, when I only set $b^2$ free, when I set all \textit{but} $b^2$ free, and then when I set all four nuisance parameters free. A comparison of their parameter distributions is shown in Figure \ref{fig:wizcolaAllNormalCovCombined}. These fits indicate a strong preference for fitting with separate marginalisation parameters across redshift bins due to tighter constraints achieved, however once can see that setting more parameters free than $b^2$ has negligible benefits (it neither increases fit strength or removes introduced bias), and adds computational time in the form of delayed chain convergence.\\

Based on these results, I will utilise independent $b^2$ values, whilst fixing $\beta$, $\sigma_v$ and $\sigma H(z)$ between bins when fitting with the full data set and full data covariance.\\

\clearpage
\section{Model testing conclusions}

In this chapter I have sought to validate my constructed BAO model. The one dimensional base model was first validated by successfully reproducing fits to the one dimensional WiggleZ survey as found in \citet{BlakeDavis2011} and \citet{BlakeKazin2011} to within a $1\sigma$ limit. Both methodologies for incorporating angular dependence to create a 2D BAO model, by modelling data wedges or multipole expansion, were then successfully tested using the combined WizCOLA simulation dataset by recovering the simulation parameters. The significance of the hexadecapole term was investigated, and was found to be small enough that discarding the term for computational benefit would have a negligible impact on parameter recovery when compared to statistical error.\\

\begin{figure}[h!]
  \begin{center}
    \includegraphics[width=\textwidth]{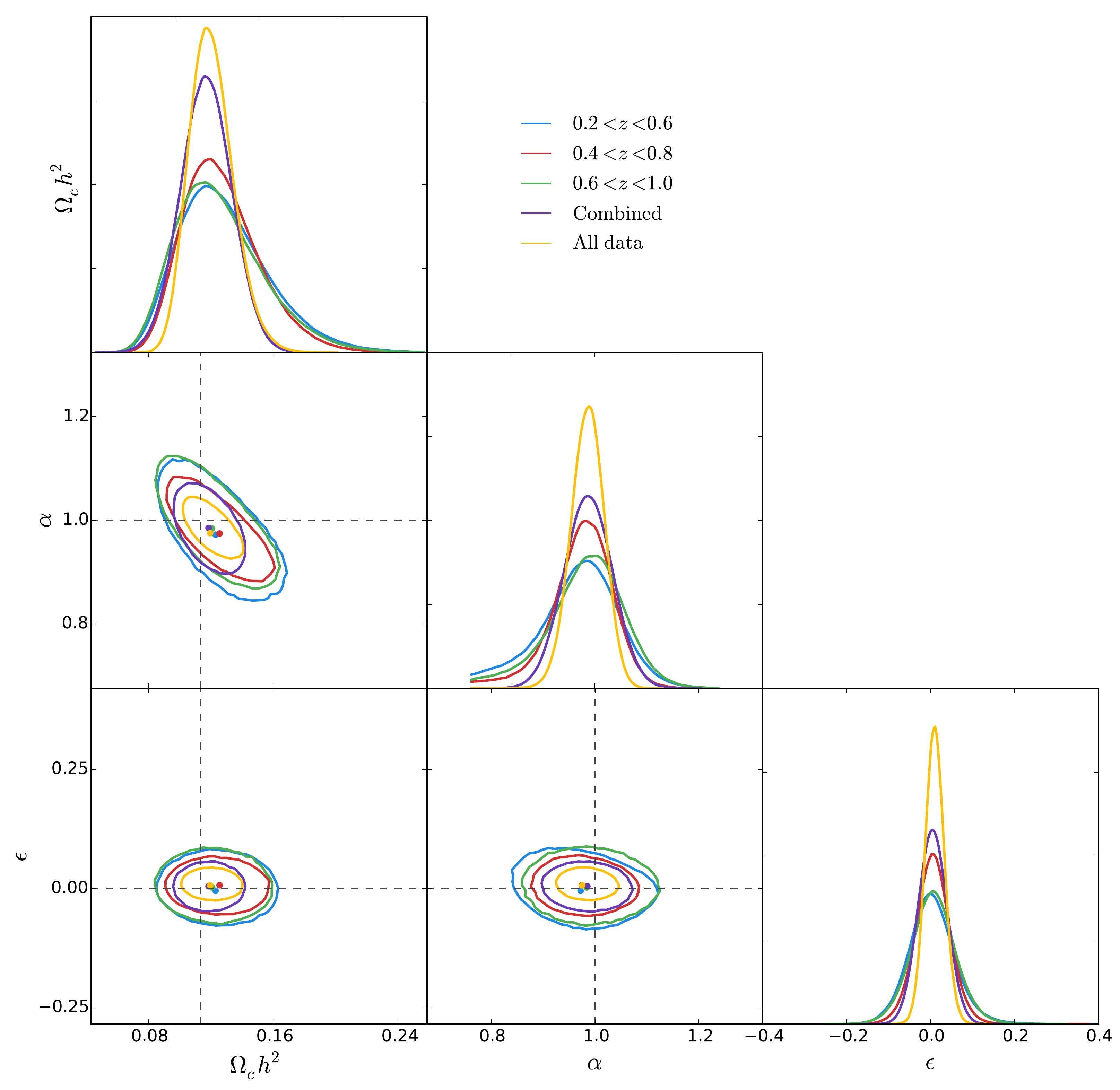}
  \end{center}
  \caption{Fits to the mean data of all 600 WizCOLA realisations (without reducing data uncertainty) for the multipole expansion expression of the data. Fits using all three bins simultaneously are shown as the ``All data'' fits, and the combination of maximum likelihood parameters using parameter covariance is shown as the ``Combined'' likelihood surfaces. In all cases we recover simulation cosmology well within $1\sigma$ limits.}
  \label{fig:corCombinedMPWiz}
\end{figure}

\begin{figure}[h!]
  \begin{center}
    \includegraphics[width=\textwidth]{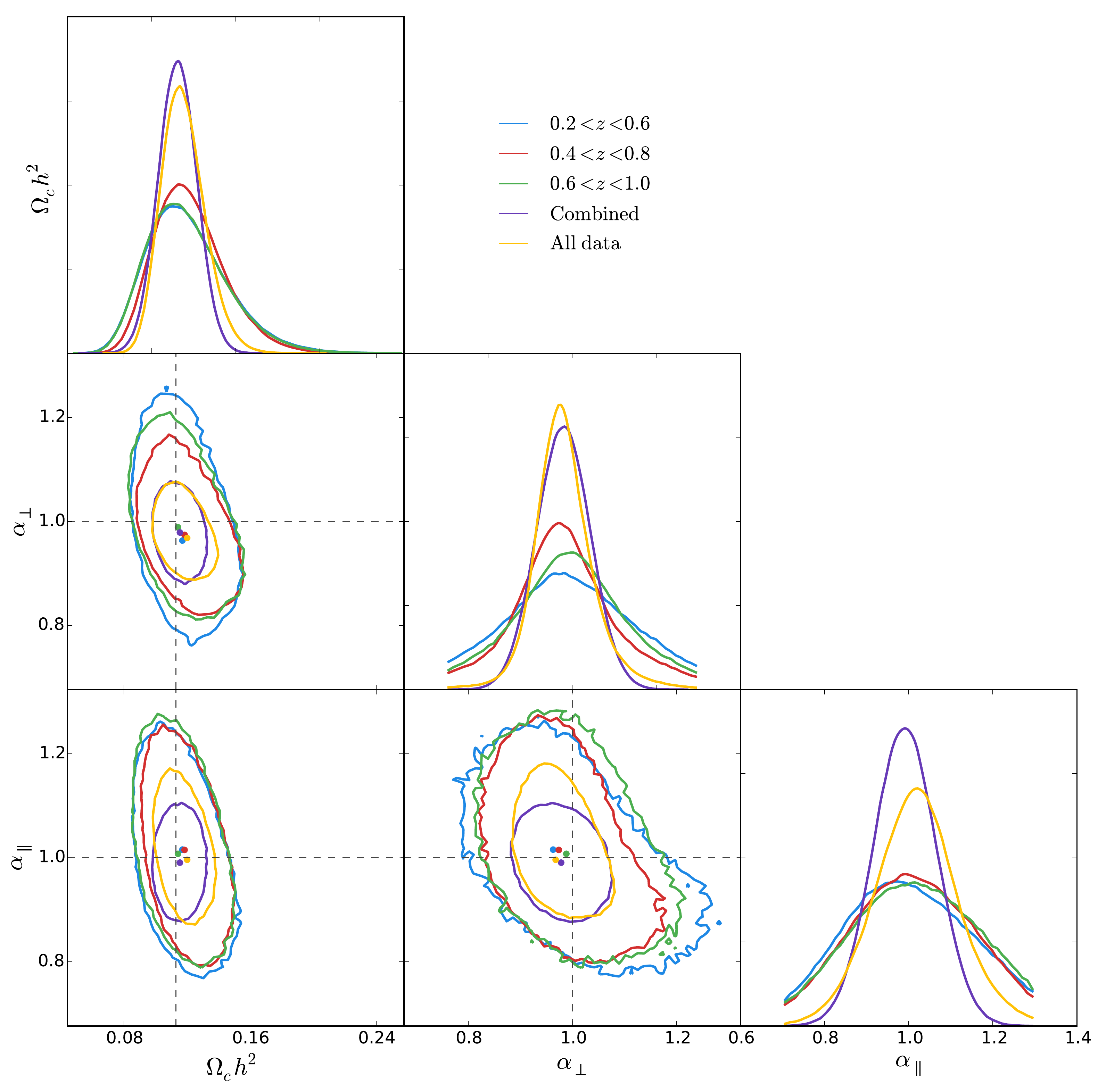}
  \end{center}
  \caption{Fits to the mean data of all 600 WizCOLA realisations (without reducing data uncertainty) for the wedge expansion expression of the data. Fits using all three bins simultaneously are shown as the ``All data'' fits, and the combination of maximum likelihood parameters using parameter covariance is shown as the ``Combined'' likelihood surfaces. In all cases we recover simulation cosmology well within $1\sigma$ limits, although we should note that the data strength is not significant enough to constrain $\alpha_\parallel$ to the $2\sigma$ limits even in this optimal example (where the data input is the mean of all 600 realisations).}
  \label{fig:corCombinedWedgeWiz}
\end{figure}

Due to the overlapping redshift bins in the analyses, covariance between the final cosmological parameters was determined by comparing individual realisations of the WizCOLA simulations, and this will be used in conjunction with an analysis using the all data across redshift bins to get global parameters. A graphical comparison of fits to the mean WizCOLA data for individual redshift bins, all data fits, and combining bin parameters can be found for the multipole data format in Figure \ref{fig:corCombinedMPWiz} and for the wedge data format in Figure \ref{fig:corCombinedWedgeWiz}. \\

Having performed various tests to determine that my constructed models is free of significant bias and is able to correctly recover cosmological values, I will now apply my model to the WiggleZ data. Due to potential issues constraining $\alpha_\parallel$ in the wedge analyses, the multipole analysis will form the primary analysis.

\chapter{Results}

In this section I present the cosmological results obtained when fitting to the final WiggleZ data. I utilise the three different fitting methods outlined in the previous section; fitting individual bins, fitting for all data, and combining binned fits, for both the multipole and wedge data representation.\\

\section{Fitting Results}

\begin{figure}[h!]
  \begin{center}
    \includegraphics[width=\textwidth]{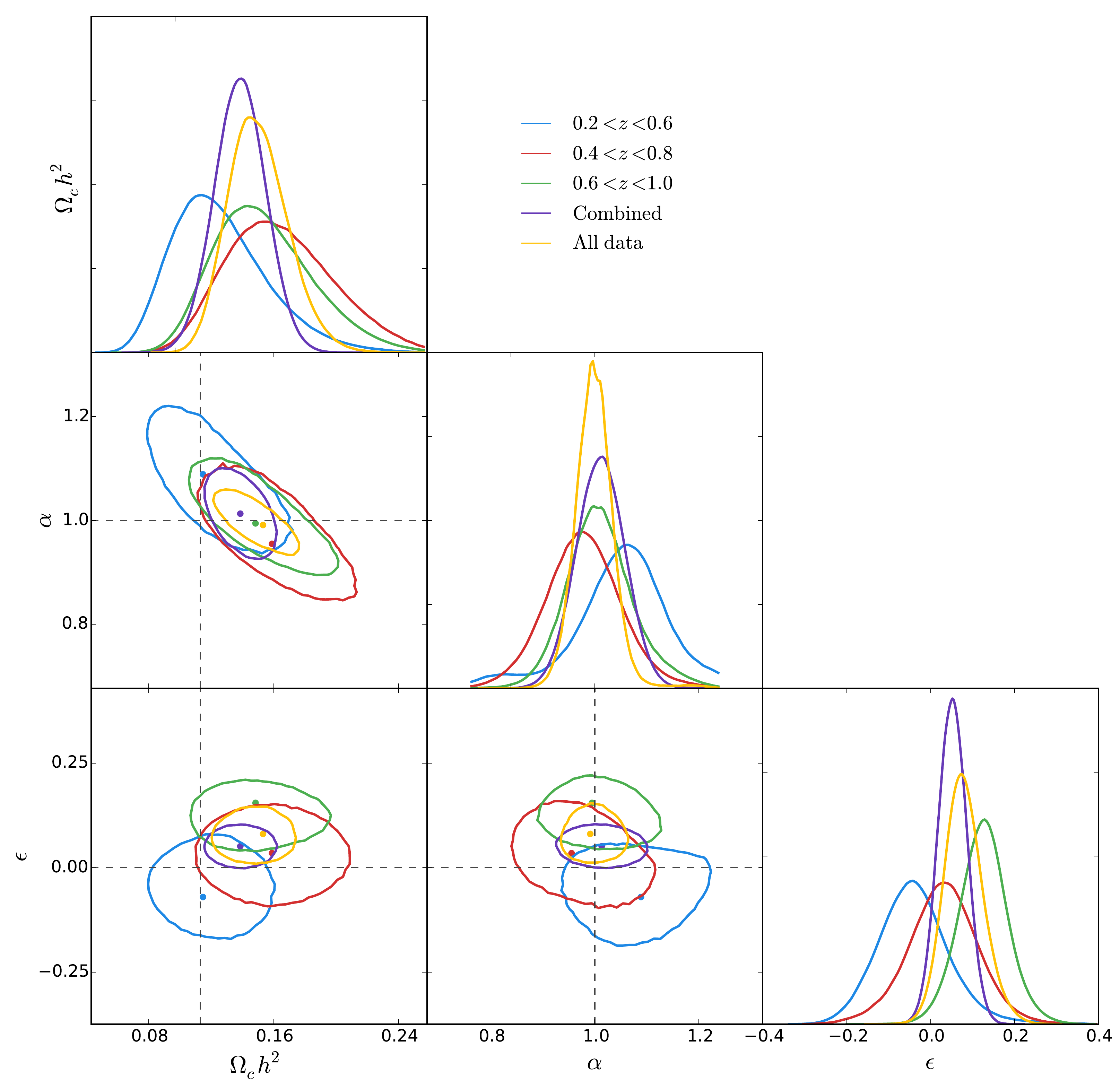}
  \end{center}
  \caption{Likelihood surfaces and marginalised distributions of $\Omega_ch^2$, $\alpha$ and $\epsilon$ for the WiggleZ multipole expression of the data. }
  \label{fig:wigglezBinsMP}
\end{figure}

\begin{figure}[h!]
  \begin{center}
    \includegraphics[width=\textwidth]{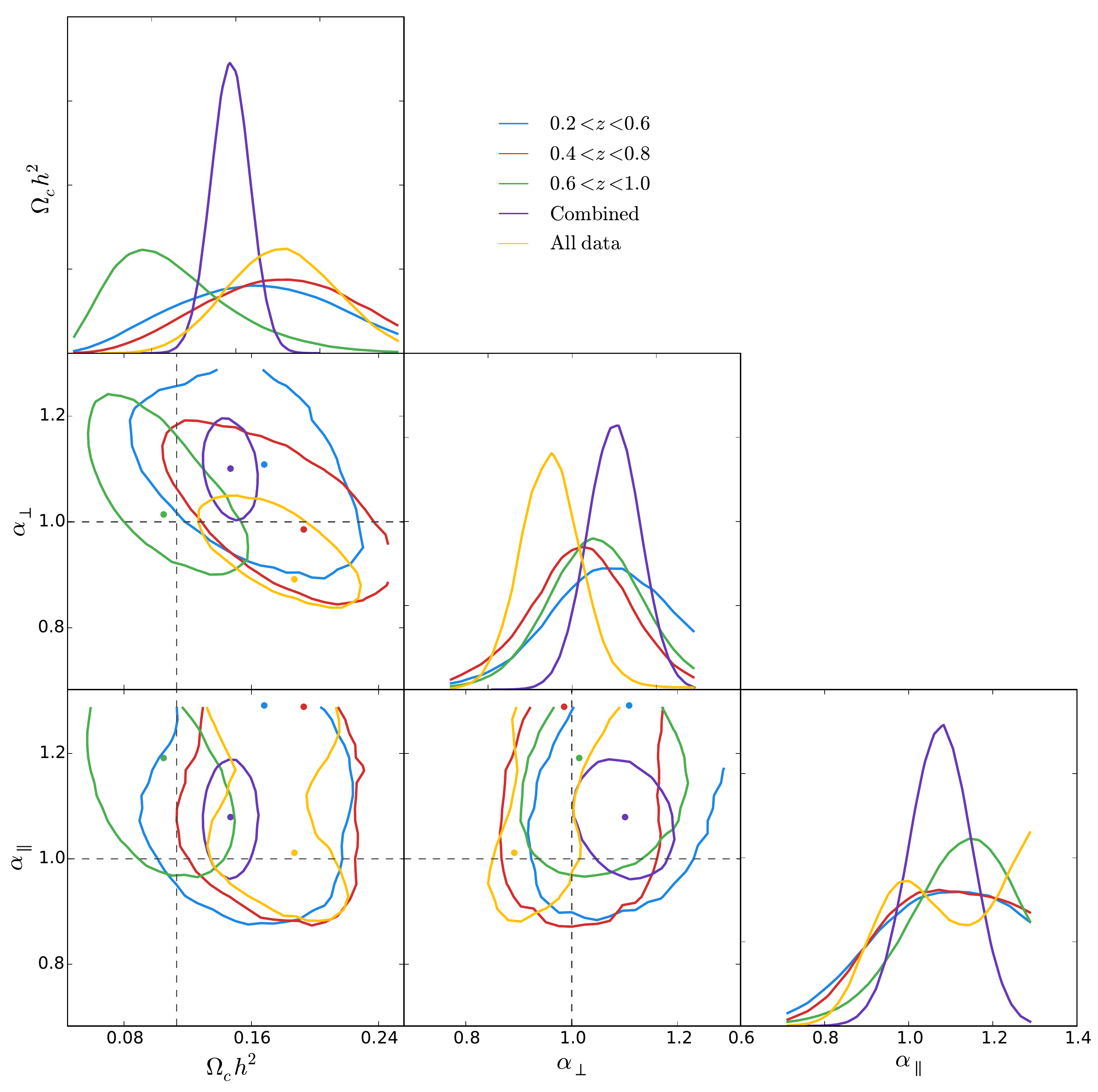}
  \end{center}
  \caption{Likelihood surfaces and marginalised distributions of $\Omega_ch^2$, $\alpha$ and $\epsilon$ for the WiggleZ wedge expression of the data. }
  \label{fig:wigglezBinsWdg}
\end{figure}

For the fits to individual redshift bins, I fit to the final unreconstructed WiggleZ dataset from \citet{KazinKoda2014}. I then combined these output parameters using the methodology outlined in \S\ref{sec:parameterCov}, and check the consistency of the combined parameters against the parameters recovered when fitting all redshift bins simultaneously as outlined in \S\ref{sec:allData}. The final distributions obtained are given in Table \ref{tab:wigglezBins} and are illustrated in Figures \ref{fig:wigglezBinsMP} and \ref{fig:wigglezBinsWdg} for mulitpole and wedge data respectively.\\

\begin{table}[h]
\centering
\caption{Recovered parameter constraints when fitting the WiggleZ multipole and wedge data data. The $\chi^2$ column represents the minimum $\chi^2$ attained in the fit. The parameter values after combination are shown at the bottom of the table, where the $\chi^2$ value for the combined dataset indicates the $\chi^2$ from equation \eqref{eq:covchi} instead of the $\chi^2$ when fitting the cosmological model to the WiggleZ data. The multipole data fits are sufficiently constrained that statistics are given from maximum likelihood points (referred to as Method C in Figure \ref{fig:statistics}). The wedge data format did not produce well behaved distributions which would give proper uncertainty using max likelihood statistics, and so have been presented using the mean statistics. The effective redshift for the data in the ``All'' and ``Combined'' rows is $z=0.6$.}
\resizebox{\columnwidth}{!}{
\begin{tabular}{c|cccc|cccc}
\specialrule{.1em}{.05em}{.05em} 
$z_{\rm{eff}}$ & \multicolumn{4}{c}{Multipole}  & \multicolumn{4}{c}{Wedge}\\
 & $\chi^2$ & $\Omega_c h^2$ &$\alpha$ & $\epsilon$ & $\chi^2$ & $\Omega_c h^2$ & $\alpha_\perp$ & $\alpha_\parallel$\\
\specialrule{.1em}{.05em}{.05em} 
$0.44$ & $51.8$ & $0.119^{+0.029}_{-0.026}$ & $1.07^{+0.10}_{-0.10}$ & $-0.03^{+0.07}_{-0.10}$ & $55.0$ & $0.156\pm 0.048$ & $1.08\pm0.14$ & $1.054\pm 0.15 $ \\
$0.60$ & $69.3$ & $0.151^{+0.038}_{-0.025}$ & $0.98^{+0.08}_{-0.10}$ & $0.05^{+0.07}_{-0.10}$  & $39.9$ & $0.169\pm 0.044$ & $1.02\pm 0.13 $ & $1.07\pm 0.15$ \\
$0.73$ & $59.1$ & $0.140^{+0.036}_{-0.022}$ & $1.00^{+0.08}_{-0.07}$ & $0.12^{+0.06}_{-0.05}$  & $42.6$ & $0.111\pm 0.036$ & $1.06\pm 0.12$ & $1.10\pm 0.12$ \\
\specialrule{.05em}{.05em}{.05em} 
All & $255.0$   & $0.147^{+0.018}_{-0.018}$ & $1.00^{+0.05}_{-0.04}$ & $0.08^{+0.04}_{-0.05}$  & $204.0$ & $0.176\pm 0.032$ & $0.95\pm 0.07$ & $1.09\pm 0.14$ \\
\specialrule{.05em}{.05em}{.05em} 
Combined & $5.6$ & $0.140^{+0.014}_{-0.017}$ & $1.00^{+0.07}_{-0.05}$ & $0.06^{+0.03}_{-0.04}$ & $20.3$ & $0.146\pm 0.013$ & $1.10\pm 0.07$ & $1.08\pm 0.07$ \\
\specialrule{.1em}{.05em}{.05em} 
\end{tabular}\label{tab:wigglezBins}
}
\end{table}

Whilst the process to move from the parameters $\Omega_c h^2$, $\alpha_\perp$ and $\alpha_\parallel$ to cosmological information is trivial, the parameters $\alpha$ and $\epsilon$ combine to give cosmological constraints as per equations \eqref{eq:alpha1} and \eqref{eq:alpha2}. To estimate final parameter variance we can utilise Fisher matrices. From
\begin{align}
\begin{pmatrix}
\sigma_{D_A}^2 & \sigma_{D_A H} \\ \sigma_{D_A H} & \sigma_H^2 \end{pmatrix}
 = 
\begin{pmatrix}
\frac{\partial D_A}{\partial \alpha} & \frac{\partial H}{\partial \alpha} \\ \frac{\partial D_A}{\partial \epsilon} & \frac{\partial H}{\partial \epsilon} \end{pmatrix} 
\begin{pmatrix}
\sigma_\alpha^2 & \sigma_{\alpha \epsilon} \\ \sigma_{\alpha \epsilon} & \sigma_\epsilon^2
\end{pmatrix}
\begin{pmatrix}
\frac{\partial D_A}{\partial \alpha} & \frac{\partial H}{\partial \alpha} \\ \frac{\partial D_A}{\partial \epsilon} & \frac{\partial H}{\partial \epsilon} \end{pmatrix} ^T,
\end{align}
we have
\begin{align}
\frac{\sigma_{D_A}^2}{D_A^2} &= \alpha^{-2} \sigma_\alpha^2 + \left(1 + \epsilon \right)^{-2} \sigma_\epsilon^2 - 2\alpha^{-1}\left(1 + \epsilon \right)^{-1} \sigma_{\alpha \epsilon} \\
\frac{\sigma_H^2}{H^2} &= \alpha^{-2} \sigma_\alpha^2 + 4(1+\epsilon)^{-2} \sigma_\epsilon^2 + 4 \alpha^{-1} (1 + \epsilon)^{-1} \sigma_{\alpha \epsilon}.
\end{align}

Using these relationships, I formulate parameter constraints shown in Table \ref{tab:wigglezBinsParams2}. We notice that, whilst most parameter recoveries are $1\sigma$ consistent between methodologies, there are notable discrepancies in the wedge analysis. In addition, we see in Figure \ref{fig:wigglezBinsWdg} that the distributions for $\alpha_\parallel$ did not converge properly for any of the redshift bins. This raises questions about the validity of the wedge analysis - the data present in the WiggleZ survey is simply insufficient to constrain the wedge parameters well. In order to confirm that increasing data quality would improve the wedge parameter constraints to an acceptable level, it would be possible to determine the cross-realisation variance of the recovered $\alpha_\parallel$ parameter as a function of data strength, and ensure that this is a strictly minimising function. This would be simple to do with the WizCOLA mocks - we already possess the variance of $\alpha_\parallel$ when fitting for only one realisation. Data quality could be increased by summing an arbitrary number of realisations (as they are independent), where the more realisations one sums into a single dataset would give increased data quality. Whilst this methodology could be used to further confirm the wedge methodology outline in prior sections, the wedge data analysis would still show disagreement between the combined and all data analysis and could not be used for final analysis, and the desired consistency check in final parameter values between the multipole analysis and wedge analysis could not be utilised.\\

Due to the lack of internal consistency for the wedge analysis, I will present results based off the multipole analysis hereonin, as the multipole analysis passes the internal consistency check - the all data analysis and the combined bin parameter analysis differ on average by only $0.3\sigma$ and so are well within a $1\sigma$ level of agreement.\\

Using the multipole data analysis and fiducial cosmology, I present final constraints from the WiggleZ analysis. To determine the significance of the BAO peak detected in my analysis, I reran the multipole analysis with a model devoid of the BAO peak and converted the $\Delta \chi^2$ into a detection significance. These results are shown in Table \ref{tab:wigglezBinsParams}.

\begin{table}[h]
\centering
\caption{Relative cosmological constraints on $\Omega_c h^2$, $D_A(z)$ and $H(z)$ from Table \ref{tab:wigglezBins}. Note that the combined wedge constraints for $D_A / D_A^{\mathcal{D}}$ and $H / H^\mathcal{D}$ are inconsistent with the bin values, as the bin values have been quoted using mean statistics as stated in Table \ref{tab:wigglezBins}, whilst the combined parameter constraints are derived from maximum likelihood statistics. The wedge data utilises different statistics as the equal probability points from the maximum likelihood methodology do not make sense given the maximum likelihood point abuts the physical boundaries (see the $\alpha_\parallel$ marginalised distribution in Figure \ref{fig:wigglezBinsWdg}). }
\resizebox{\columnwidth}{!}{
\begin{tabular}{c|ccccc|ccccc}
\specialrule{.1em}{.05em}{.05em} 
Parameter & \multicolumn{5}{c}{Multipole}  & \multicolumn{5}{c}{Wedge}\\
 & $z_{\rm{eff}} = 0.44$ & $z_{\rm{eff}} = 0.60$ & $z_{\rm{eff}} = 0.73$ & All & Combined & $z_{\rm{eff}} = 0.44$ & $z_{\rm{eff}} = 0.60$ & $z_{\rm{eff}} = 0.73$ & All & Combined \\
\specialrule{.1em}{.05em}{.05em} 
$\Omega_c h^2$                &  $0.119^{+0.029}_{-0.026}$ &   $ 0.147^{+0.018}_{-0.018}$  &  $0.140^{+0.036}_{-0.022}$ & $0.155^{+0.023}_{-0.015}$ & $0.140^{+0.014}_{-0.017}$ & $0.156\pm 0.048$  & $0.169\pm 0.044$ & $0.111\pm 0.036 $ & $0.176\pm 0.032$ &  $ 0.146\pm 0.013$\\
$D_A / D_A^{\mathcal{D}}$     &  $1.11 \pm  0.15$          &   $0.93 \pm  0.12$           &  $0.89 \pm  0.10$          & $0.92 \pm  0.06$          & $ 0.95 \pm  0.07$           & $1.08\pm0.14$ & $1.02\pm 0.13 $ & $1.06\pm 0.12$  & $0.95\pm 0.07$ & $1.10\pm 0.07$  \\
$H / H^\mathcal{D}$           &  $1.00 \pm 0.18$           &   $0.94 \pm 0.15$           &  $0.80 \pm 0.10$          & $0.87 \pm 0.09$           & $0.89 \pm 0.08$          &                     $0.89 \pm 0.13$          & $0.94 \pm 0.14$ & $0.91 \pm 0.11$ & $0.91 \pm 0.13$ & $ 0.93 \pm 0.08$ \\
\specialrule{.1em}{.05em}{.05em} 
\end{tabular}\label{tab:wigglezBinsParams2}
}
\end{table}

\begin{table}[h]
\centering
\caption{Parameter constraints from the WiggleZ redshift bins. $D_A$ is presented in units of $[\rm{Mpc}]$, and $H(z)$ is presented in units of [\kmsmpc]. The low significance of the BAO peak is expected: the 1D BAO analysis from \citet{BlakeDavis2011} found a significance of $3.2\sigma$ when using all data in one combined bin, and as my analysis used the data divided over three bins, and includes extra parameters to model angular dependence, it is expected the statistical significance of the BAO peak would decrease. The analysis in \citet{BlakeKazin2011}, which utilised three redshift bins, the same as my analysis, found statistical significances between $1.9\sigma$ and $2.4\sigma$, consistent with the results shown below.}
\resizebox{\columnwidth}{!}{
\begin{tabular}{cc|cc|cccc}
\specialrule{.1em}{.05em}{.05em} 
Sample & $z_{\rm{eff}}$ & $D^\mathcal{D}_A(z)$ & $H^\mathcal{D}(z)$ &  $\Omega_c h^2$   & $D_A(z)$ & $H(z)$ & BAO peak significance\\
\specialrule{.1em}{.05em}{.05em} 
$0.2 < z < 0.6$ &  $0.44$ & 1175.5  & 87.4  & $0.119^{+0.029}_{-0.026}$ & $1302 \pm  160$ & $87.0 \pm 15.7$  & $2.2\sigma$\\
$0.4 < z < 0.8$ &  $0.6$  & 1386.2  & 95.5  & $0.151^{+0.038}_{-0.025}$ & $1295 \pm  184$ & $89.7 \pm 15.1$  & $2.1\sigma$\\
$0.6 < z < 1.0$ &  $0.73$ & 1509.4  & 102.8 & $0.140^{+0.036}_{-0.022}$ & $1348 \pm  161$ & $82.0 \pm 12.7$  & $2.3\sigma$\\
\specialrule{.1em}{.05em}{.05em} 
\end{tabular}\label{tab:wigglezBinsParams}
}
\end{table}

\clearpage
\section{External comparison}

We can now check consistency of the results obtained in the previous section with results from other BAO analyses. In Table \ref{tab:external} I source cosmological constraints on $D_A(z)$ and $H(z)$, and compare them to the Wigglez constraints I have derived in Table \ref{tab:wigglezBinsParams}. We can see that the derived $D_A(z)$ values diverge (non-significantly) from Planck cosmology in the higher redshift bins, due to the low $D_A/D^{\mathcal{D}}_A$ values seen in Table \ref{tab:wigglezBinsParams2}. We should further note that if we had instead utilised the wedge analysis, the $D_A$ values would consistently have been higher than Planck cosmology! For the determined values of $H(z)$, we notice a (not significantly) lower value of $H(z = 0.73)$ than may have been expected from the trend observed in external data sets.\\

The constraints included in Table \ref{tab:external} from \citet{BlakeBroughColless2012} determine $D_A(z)$ and $H(z)$ from the WiggleZ survey by combining measurements of the 1D BAO peak and the Alcock-Paczynski distortion. The 1D BAO signal analysed in prior studies provides constraints on $D_V$, which is a function of $D_A$, $H(z)$ and $z$, as per equation \eqref{eq:dv}. The  Alcock-Paczynski (AP) test can be used to measure $\approx (1+z)D_A(z) H(z)/c$, and can thus be used in conjunction with the angle-average BAO signal to constrain $D_A(z)$ and $H(z)$. The AP test was applied to the Wigglez data in \citet{BlakeGlazebrook2011}, and the angle average BAO peak was measured in \citet{BlakeKazin2011}. The resultant constraints on $D_A(z)$ and $H(z)$ by combining these results and a Gaussian prior on $\Omega_m h^2$ from CMB fits \citep{Komatsu2009} are shown in Table \ref{tab:external}.

\begin{table}[h]
\centering
\caption{External constraints on $D_A(z)$ and $H(z)$.}
\resizebox{\columnwidth}{!}{
\begin{tabular}{cc|c|cc}
\specialrule{.1em}{.05em}{.05em} 
Paper & Source &  $z_{\rm{eff}}$ & $D_A(z)$ [Mpc] & $H(z)$ [km/s/Mpc]\\
\specialrule{.1em}{.05em}{.05em} 
\citet{Gaztanaga2009} & SDSS LRG DR7 & 0.24 & & $79.69\pm2.32$ \\
 & & 0.34 & &  $83.80\pm2.96$ \\
 & & 0.43 & &  $86.45\pm3.27$ \\
 
\citet{BlakeBroughColless2012} & WiggleZ & 0.44 & $1205\pm114$ & $82.6\pm7.8$ \\
& & 0.6 & $1380\pm95$ & $87.9\pm6.1$ \\
& & 0.73 & $1534\pm107$ & $97.3\pm7.0$ \\
\cite{ChuangWang2012} & SDSS LRG DR7 & 0.35 & $1048^{+60}_{-58}$ & $82.1^{+4.8}_{-4.9}$ \\
\citet{AndersonAubourg2012} & SDSS-III BOSS DR11 & 0.57 & $(1421\pm20) (r_d/r_{d,\rm{fid}})$ & $(96.8\pm3.4) (r_{d,\rm{fid}}/r_d)$ \\
\citet{FontRiberaKirkby2014} & SDSS-III BOSS DR11 Quasars & 2.36 & $1590\pm60$ & $226\pm 8$ \\
\specialrule{.1em}{.05em}{.05em} 
\end{tabular}\label{tab:external}
}
\end{table}

\begin{figure}[h!]
  \begin{center}
    \includegraphics[width=\textwidth]{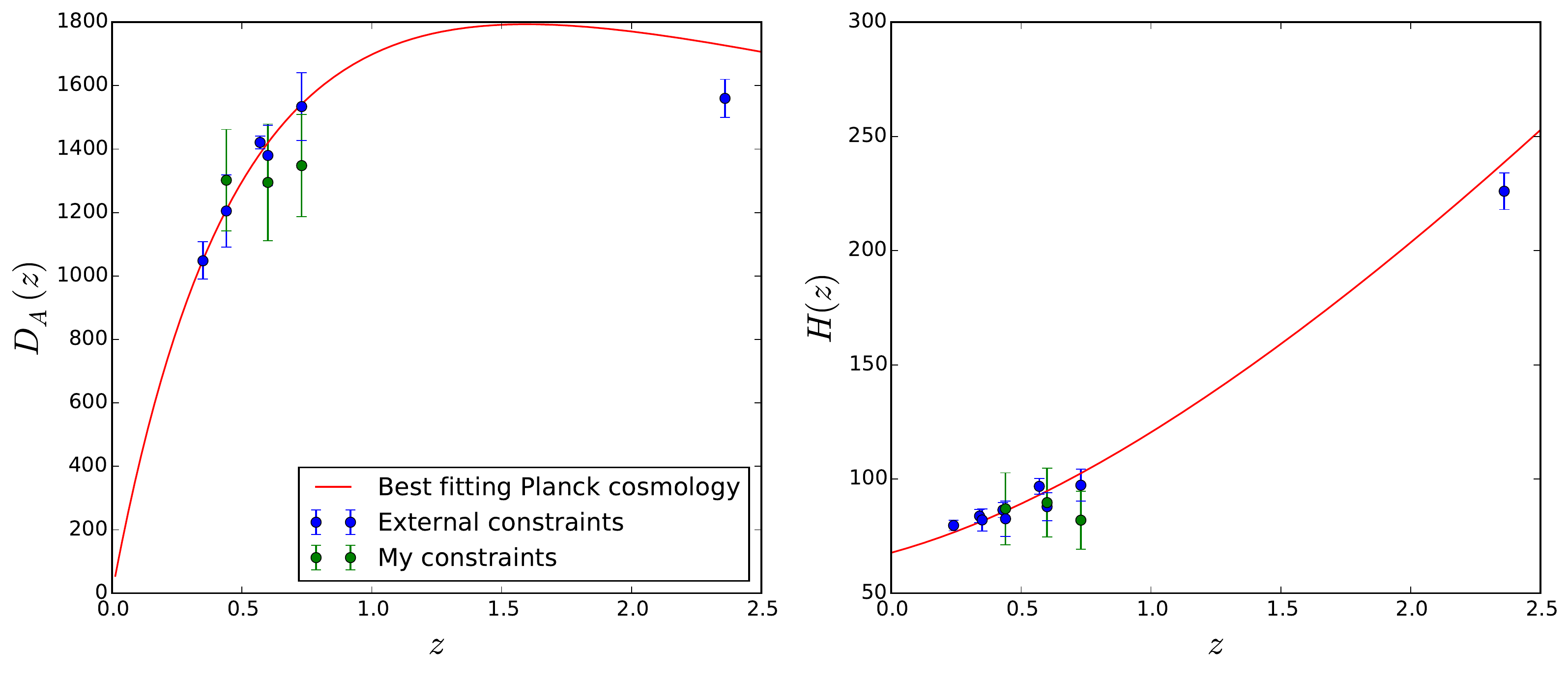}
  \end{center}
  \caption{A comparison with the constraints on $D_A(z)$ and $H(z)$ found in external papers, best fitting Planck cosmology for a Flat $\Lambda$CDM universe \citep{Planck2015Parameters}, and the constraints listed in Table \ref{tab:wigglezBinsParams}.}
  \label{fig:external}
\end{figure}

\chapter{Conclusion}

In the previous section I analysed the full WiggleZ dataset from \citet{KazinKoda2014} for the 2D BAO signal. My methodology follows that of prior studies, and has been validated against prior WiggleZ analyses of the 1D BAO signal, and I have checked for systematic biases in my model by fitting to the WizCOLA simulations \citep{KodaBlake2015}. The WizCOLA simulations also provide improved covariance estimates over previous analyses of the WiggleZ data which estimate covariance from lognormal realisations. Using the WizCOLA covariance and WiggleZ data, I fit for the cosmological parameters $\Omega_c h^2$, $D_A(z)$ and $H(z)$ for the three redshift bins $z \in \left[0.44, 0.60, 0.73\right]$ using both the wedge and multipole data expressions provided.\\

After showing the wedge data analysis was internally inconsistent and unable to properly constrain required parameters, I present the multipole data results as the final constraints from the 2D BAO analysis, given in Table \ref{tab:final}. As shown in Figure \ref{fig:external}, these results are consistent with the Flat $\Lambda$CDM cosmology derived from best-fitting Planck cosmological values.\\

\begin{table}[h]
\centering
\caption{Final constraint summary, from the WiggleZ multipole data analysis. $H(z)$ is given in units of \kmsmpc, and $D_A(z)$ is given in units of Mpc.}
\begin{tabular}{cc|cccc}
\specialrule{.1em}{.05em}{.05em} 
Sample & $z_{\rm{eff}}$   &  $\Omega_c h^2$   & $D_A(z)$ & $H(z)$ \\
\specialrule{.1em}{.05em}{.05em} 
$0.2 < z < 0.6$ &  $0.44$ & $0.119^{+0.029}_{-0.026}$ & $1300 \pm  160$ & $87 \pm 16$ \\
$0.4 < z < 0.8$ &  $0.60$  & $0.151^{+0.038}_{-0.025}$ & $1300 \pm  180$ & $90 \pm 15$ \\
$0.6 < z < 1.0$ &  $0.73$ & $0.140^{+0.036}_{-0.022}$ & $1350 \pm  160$ & $82 \pm 13$ \\
\specialrule{.1em}{.05em}{.05em} 
\end{tabular}\label{tab:final}
\end{table}

\chapter*{References}
\begingroup
\addcontentsline{toc}{chapter}{References}
\renewcommand{\addcontentsline}[3]{}
\renewcommand{\chapter}[2]{}
\bibliography{bibliography}
\endgroup

\begin{appendices}

\chapter{Dewiggling Process} \label{app:dewiggle}

In the literature review, we saw the prevalence for using the \verb;tffit; algorithm developed by \citet{EisensteinHu1998} to generate a power spectrum without the BAO feature. However, the use of this algorithm necessarily constrains an analysis to not only the precision of the algorithm, but also to the cosmologies considered when the algorithm was developed. Whilst most changes in cosmological models have been subtle in the past decade, a quick inspection of the changelog for CAMB\footnote{\url{http://camb.info/readme.html}} \citep{Lewis2000} shows over fifty software releases since the publication of the \verb;tffit; algorithm - representing a continual divergence between CAMB and \verb;tffit; as CAMB continues to become more accurate and consistent with modern cosmological models, whilst \verb;tffit; remains static. \\

Given these reasons, it was decided to develop an alternate method for generating a power spectrum without the BAO feature present. Given the regular updating of the CAMB software, a replacement algorithm would be most useful if it was capable of taking a standard linear power spectrum from CAMB and returning a filtered version, such that any changes in future cosmology would be reflected in the no wiggle power spectrum simply due to its presence in the original linear power spectrum from CAMB. To this end, several different methods of filtering power spectra were investigated, implemented, and tested, and these implementations are detailed in this chapter.

\section{Comparison of methods}

\begin{figure}[h]
  \begin{center}
    \includegraphics[width=0.8\textwidth]{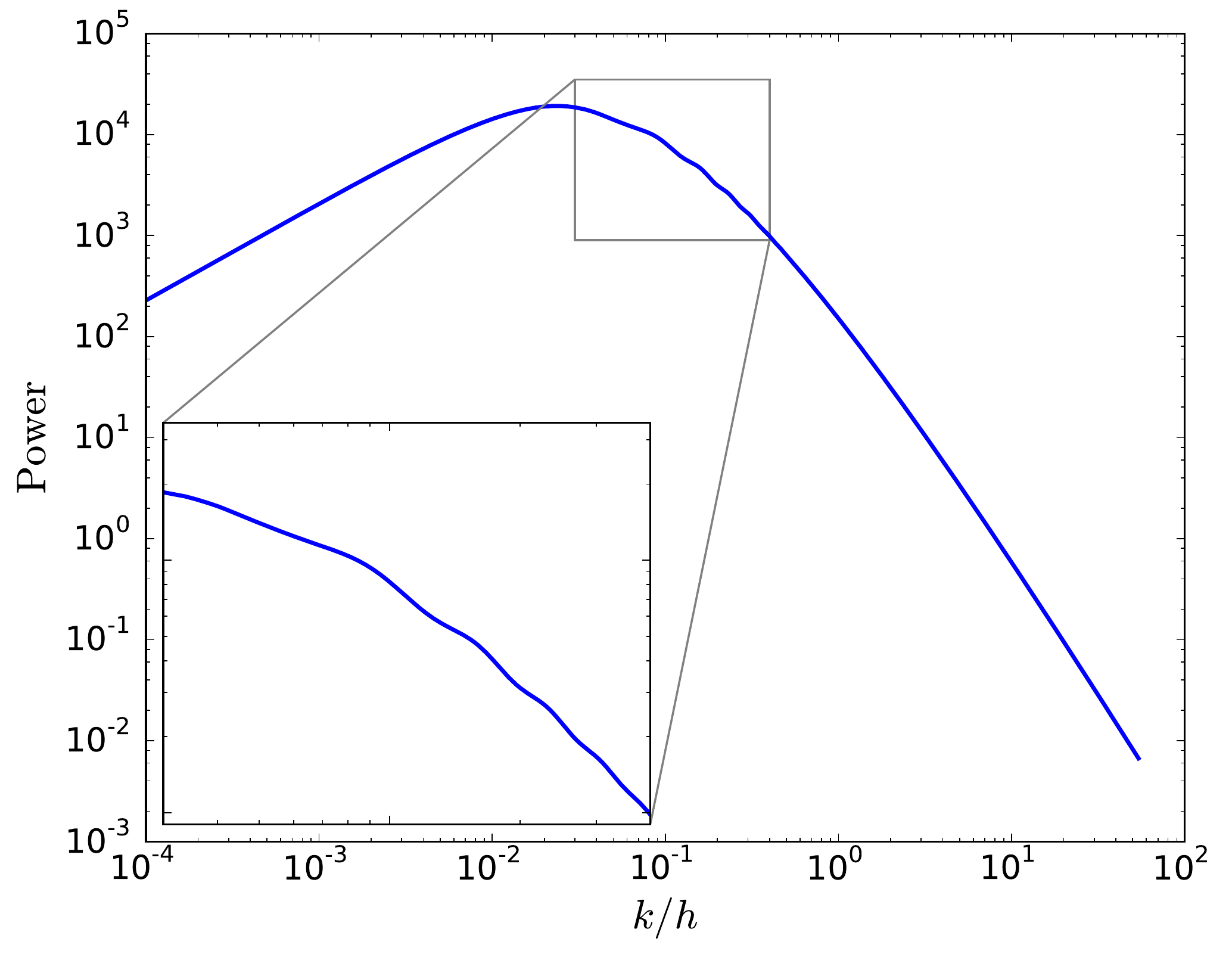}
  	\caption{A detailed look at the BAO signature present in the linear power spectrum. It presents as a set of wiggles at approximately $k/h = 0.1$.}
  	\label{fig:Alinear}
  \end{center}
\end{figure}

The BAO signal is present in the linear power spectrum generated by CAMB in the form of small scale oscillations after the main power peak, as illustrated in Figure \ref{fig:Alinear}.

Given the BAO signal is of small amplitude and restricted periodicity, both polynomial data fitting, low order spline interpolation and frequency based filtering are all viable candidates for investigation.

\subsection{Low Pass and Band Stop Filters}

\begin{figure}[h]
  \begin{center}
    \includegraphics[width=\textwidth]{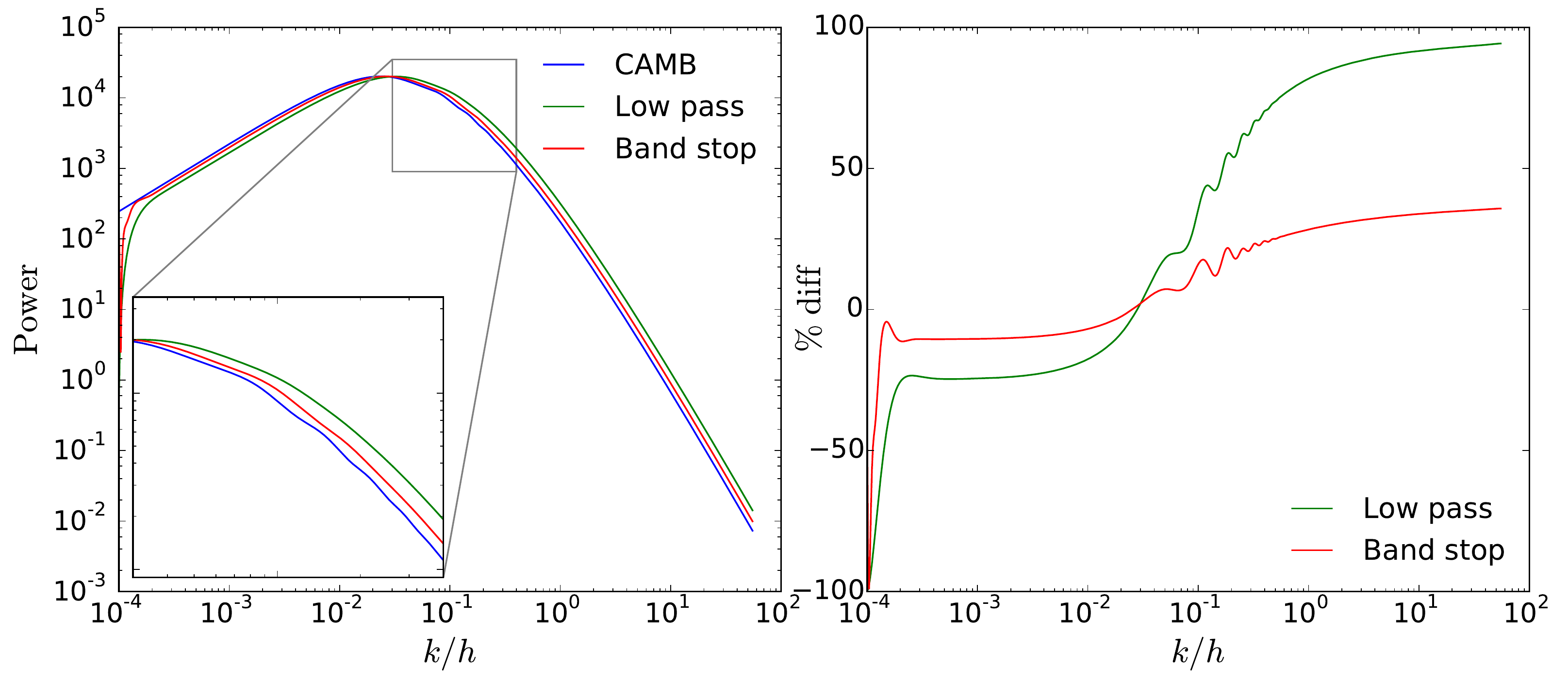}
  	\caption{Herein lies the failed attempts at using an easily available digital signal processing library to remove the BAO signal without changing the broad power spectrum.}
  	\label{fig:Alowpass}
  \end{center}
\end{figure}

It was hoped that, due to the characteristic periodicity observed in the BAO signal, it might be possible to remove it with either a low pass filter or a band stop filter. Unfortunately, the strong broad range signal present in the power spectrum (likened to a strong continuum) means that signal remains present at all frequencies, and thus there were no viable methods of extracting only the BAO signal. Figure \ref{fig:Alowpass} illustrates the difficulty of the low pass and band stop filters, namely that crushing sufficient frequencies to remove the BAO peak ends up distorting the entire shape of the power spectrum.

\subsection{Polynomial regression}

Polynomial regression are a tried and tested method for determining broad shape in a given spectrum \citep{baldry2014galaxy}. The higher order the polynomial fit becomes, the better the broad band shape extraction becomes, at the cost of eventually, as one keeps increasing the order, the polynomial model becomes detailed enough it begins to recover BAO signal. To counter this, one can introduce weights on the points, where the data points in the range of the BAO wiggle are down weighted. To make this method more viable, a specific $k/h$ is not chosen as the centre point (as this strongly removes our model independence), instead we can note that the wiggle will appear approximately at the data peak, and down weight this area using a Gaussian weighting function, such that the weights supplied to the polynomial regression take the form $w = 1 - \alpha \exp\left(-k^2/2 \sigma^2\right)$. Using this, we can construct an array of polynomial fits where the polynomial degree, Gaussian width and amount of down weighting are varied to determine the most effective construction to remove the BAO signal. In order to take advantage of the smooth shape of the power spectrum in the log domain, the polynomial regression is applied to the logarithm of the power spectrum.

\begin{figure}[h]
  \begin{center}
    \includegraphics[width=\textwidth]{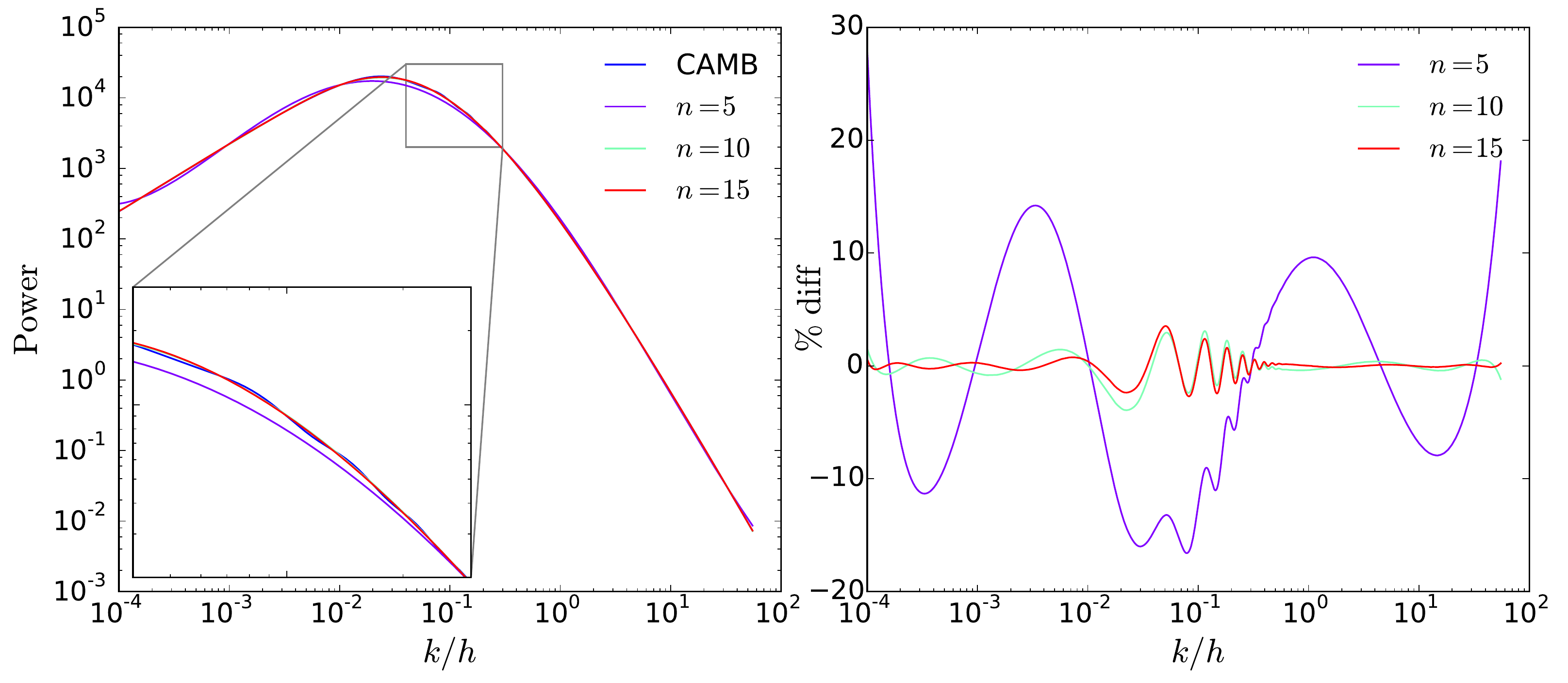}
  	\caption{A comparison of the effects of increasing polynomial weight. Due to the high number of data points in the linear CAMB model ($>600$), even a high degree polynomial such as the 15 degree polynomial displayed in red, does not attempt to recover the BAO signal. Given the range of $k_*$ values typically used in model fitting, the right hand side of the graph where $k/h > 0.1$ is most relevant. It is desired that the polynomial fit converge to the CAMB power spectrum at high $k/h$, as occurs with higher order polynomial fits.}
  	\label{fig:ApolyDegree}
  \end{center}
\end{figure}

\begin{figure}[h]
  \begin{center}
    \includegraphics[width=\textwidth]{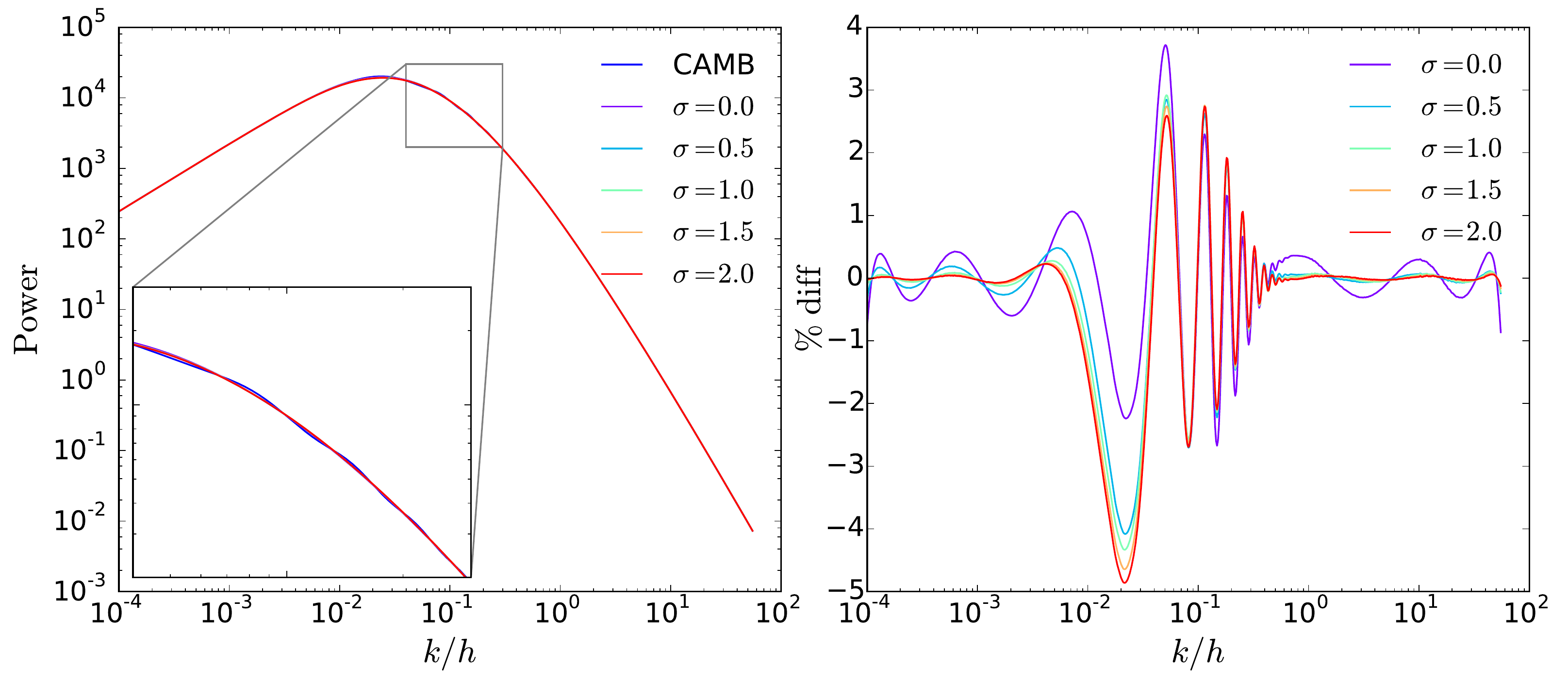}
  	\caption{With polynomial degree fixed to $n = 13$, the width of the Gaussian used to down weight the peak of the spectrum is compared in this plot. It can be seen that no Gaussian ($\sigma= 0.0$) results in oscillations at high $k/h$, whilst the increasing $\sigma$ initially leads to better convergence at high $k/h$, with continually increasing $\sigma$ reducing the completeness of the BAO signal subtraction.}
  	\label{fig:ApolySigma}
  \end{center}
\end{figure}

\begin{figure}[h]
  \begin{center}
    \includegraphics[width=\textwidth]{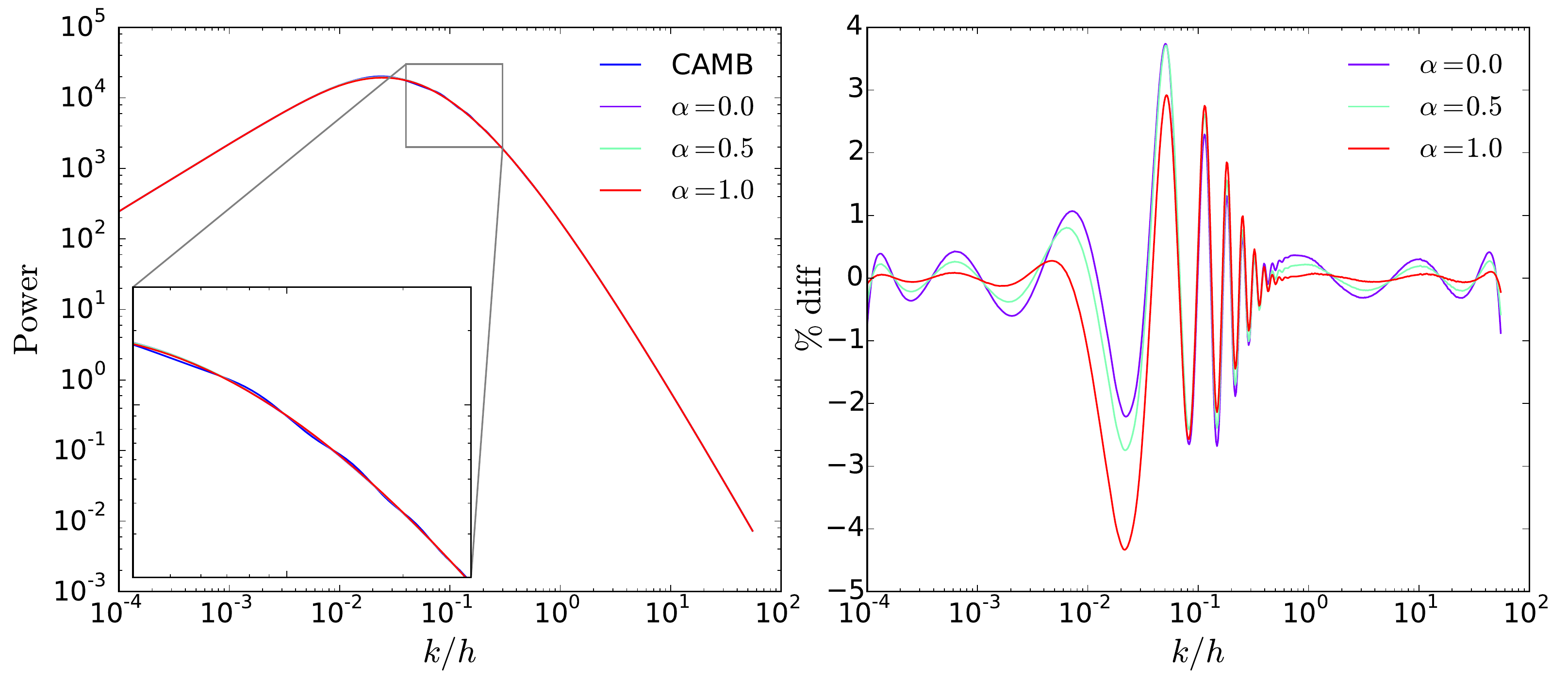}
  	\caption{Setting $\sigma = 1$, we can examine the effect of the weight $\alpha$ of the Gaussian down weighting. As expected, setting the weight to zero gives the oscillations at high $k/h$ found in Figure \ref{fig:ApolySigma}. Setting the subtraction to full strength with $\alpha = 1.0$, we see that there is a downward shift in the polynomial fit (as the peak which lifts the fit has effectively been removed). Thus a compromising value in between must be chosen.}
  	\label{fig:ApolyWeight}
  \end{center}
\end{figure}

By comparing a wide array of parametrisations of polynomial degree $n$, Gaussian width $\sigma$ and Gaussian weight $\alpha$, a final combination of $n=13, \sigma=1, \alpha=0.5$ we chosen to act as the best choice for both strong BAO signal subtraction and non distortion of the original linear power spectrum.

\clearpage
\subsection{Spline Interpolation}

The final method of removing the BAO signal from the linear power spectrum investigated was using spline interpolation. Similarly to the polynomial fits, it has the option of being supplied relevant weights for each data point, and thus a similar investigation as to weights was carried out for spline interpolation as was carried out for polynomial fitting. The spline fitting was found to be completely insensitive to modified weights, but highly sensitive to the positive smoothing factor $s$. A value of $s = 0.18$ compromises between BAO subtraction and low levels of distortion at high $k/h$, as determined by minimising the difference between the resultant spline model and the output of \verb;tffit;. Spline interpolation was similarly investigated in \citet{ReidPercival2010}, who found that use of a cubic b-spline with eight nodes fitted to $P_{\rm{lin}}(k) k^{1.5}$ produced likelihood surfaces in high agreement with formula from \citet{EisensteinHu1998}. In testing this methodology for potential use, no benefit was found to come from rotating the power spectrum via the $k^{1.5}$. This was found for both tests using a univariate spline and a b-spline, however the similarity between the results of the different splines was such that only the univariate spline is documented.

\begin{figure}[h!]
  \begin{center}
    \includegraphics[width=\textwidth]{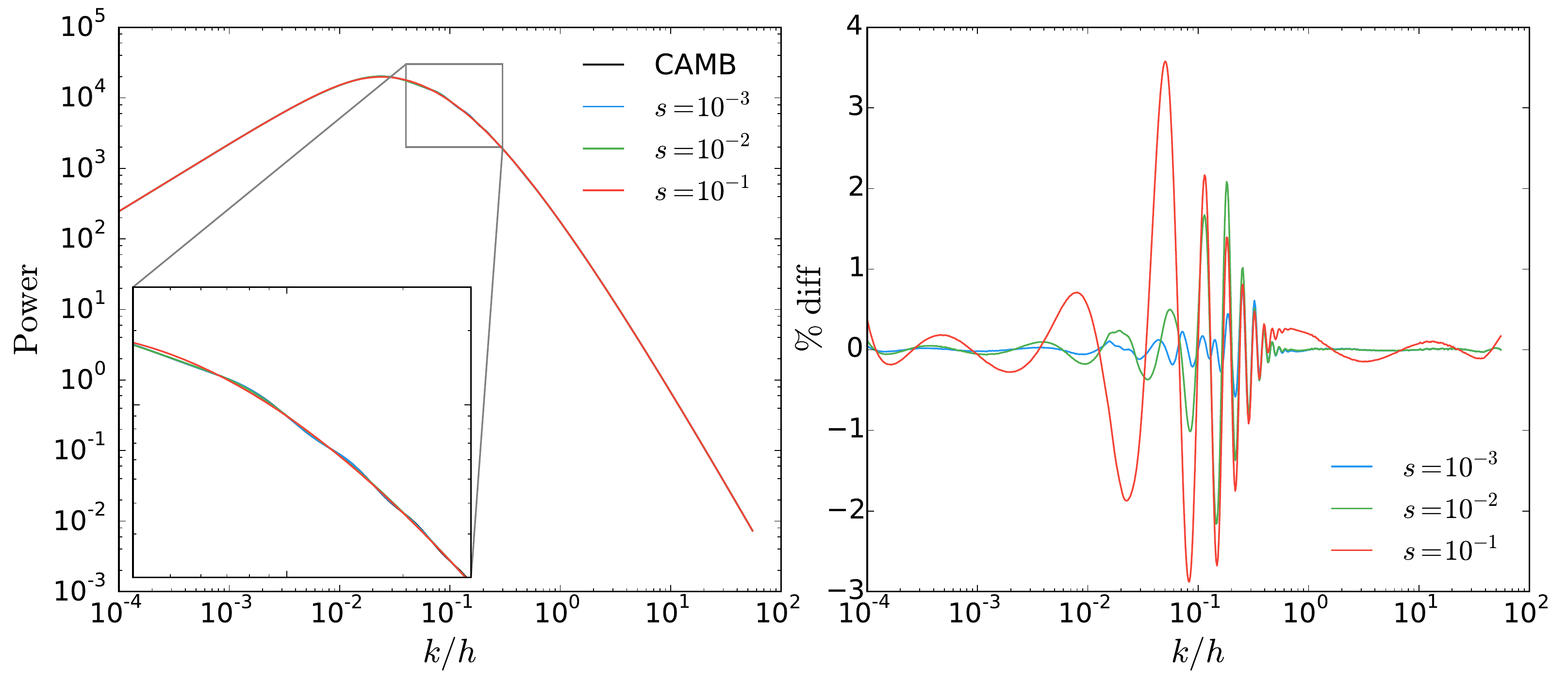}
  	\caption{Modifying the positive smoothing factor $s$ when computing a 5-point univariate spline has dramatic effects on the extraction of BAO signal. Setting $s < 0.01$ stops the spline from effectively removing the BAO signal, whilst setting it higher such that $s > 0.3$, the deviation from the linear power spectrum starts becoming significant at higher $k/h$ values.}
  	\label{fig:AsplineSmooth}
  \end{center}
\end{figure}

\section{Selection of final model}

\begin{figure}[h!]
  \begin{center}
    \includegraphics[width=0.7\textwidth]{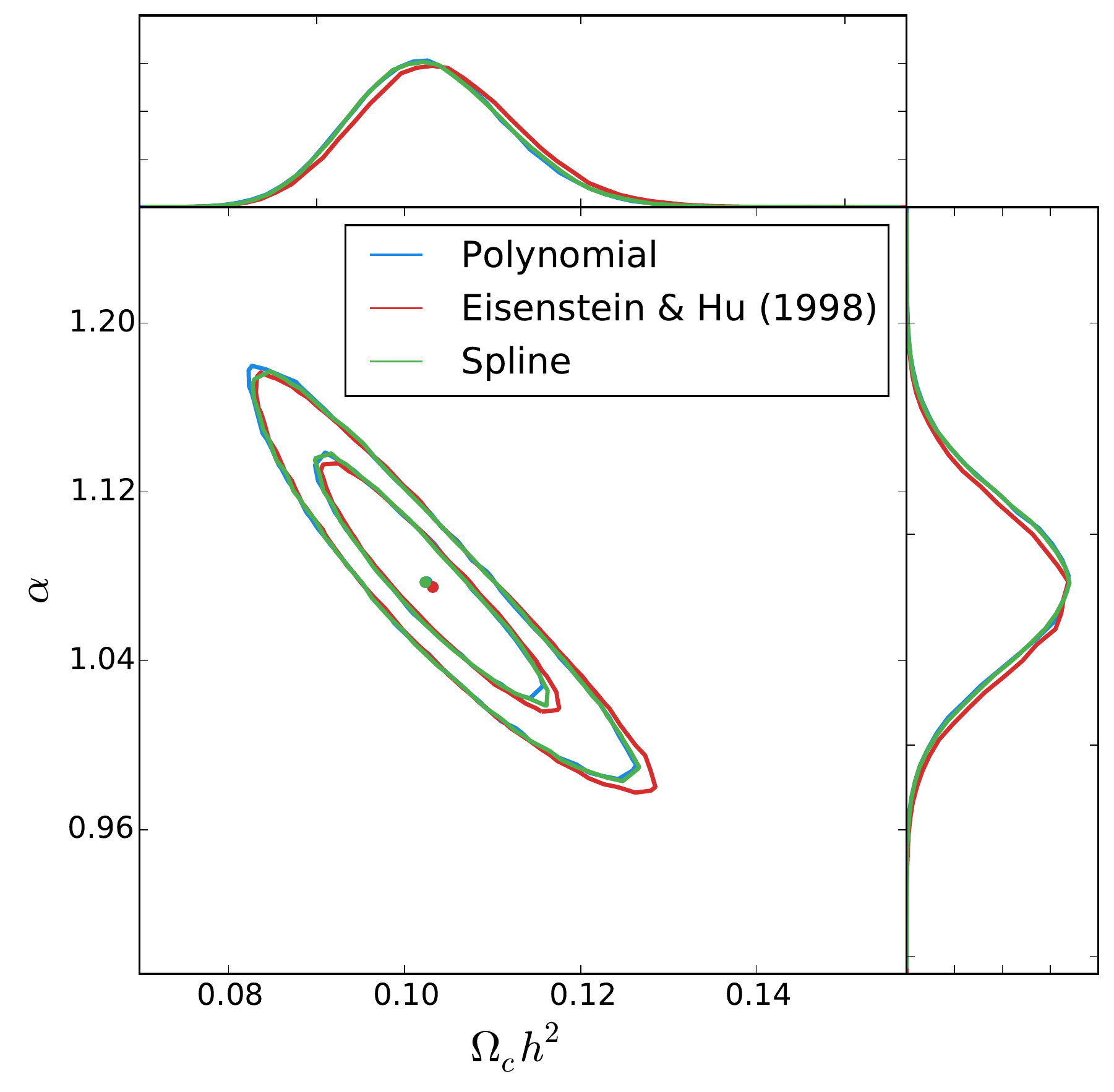}
  \end{center}
  \caption{A cosmological sensitivity test between the algorithm from \citet{EisensteinHu1998}, polynomial fitting and spline fitting. Likelihood surfaces and marginalised distributions were calculated using the WizCOLA simulation data at the $z=0.6$ redshift bin, where all 600 realisations have been used as input data, and $k_*$ fixed to $0.1$. With the low value of $k_*$ to increase the significance of the dewiggling algorithm and high data quality to reduce statistical uncertainty beyond the scope of the WiggleZ dataset, any deviation between the different methodologies should represented in the likelihood surfaces represents extremal values of diverge. However, as all likelihood surfaces agree to a high degree, we can conclude any difference in methodology is negligible in comparison to statistical uncertainty.}
  \label{fig:AcosmologyTest}
\end{figure}

Selecting the final method of dewiggling input spectra was done via looking explicitly at how the spectra are used in cosmological fitting: they are transformed into correlation functions and compared to observed data points. As such, the chosen optimal configurations for the polynomial and spline method were compared to \verb;tffit; by performing a cosmological sensitivity test wherein fits to WizCOLA data using the polynomial method, spline method and the algorithm given by \citet{EisensteinHu1998} are directly compared. To ensure this is robust, the value $k_*$ is fixed to 0.1, representing a fit with a very high level of dewiggling (hard thresholds are often limited to around this value, ie \citet{ChuangWang2012} have minimum $k_* = 0.09$), whilst still preserving some of the BAO peak with which to match. This analysis is given in Figure \ref{fig:AcosmologyTest}, and shows that for both spline and polynomial methods outlined above, statistical uncertainty in fits far exceeds any difference in matching results due to the change in dewiggling process. The polynomial method was selected to be the final method, due to the observed roughness in spline fitting which is the result of the changing dependence on the positive smoothing factor.

\chapter{Power Spectrum to Correlation Function} \label{app:pk2xi}

The monopole moment of the power spectrum obtained in model creation is analytically transformed to a correlation function via the first order three dimensional Fourier transformation 
\begin{align} \label{eq:she}
\xi(s) = \frac{1}{(2 \pi)^3} \int 4 \pi k^2 \, P(k)\,  \frac{\sin(ks)}{ks}.
\end{align}
Unfortunately, non-linear growth of the power spectrum at high $k$ hinders convergence of numerical computation of the correlation function. In this section, two found methodologies to increase convergence, respectively from \citet{BlakeDavis2011} and \citet{AndersonAubourg2012}, will be tested against a high quality (and thus exceedingly slow) numerical method to determine the effect the modified algorithms have on the final model.

\begin{figure}[h!]
  \begin{center}
    \includegraphics[width=\textwidth]{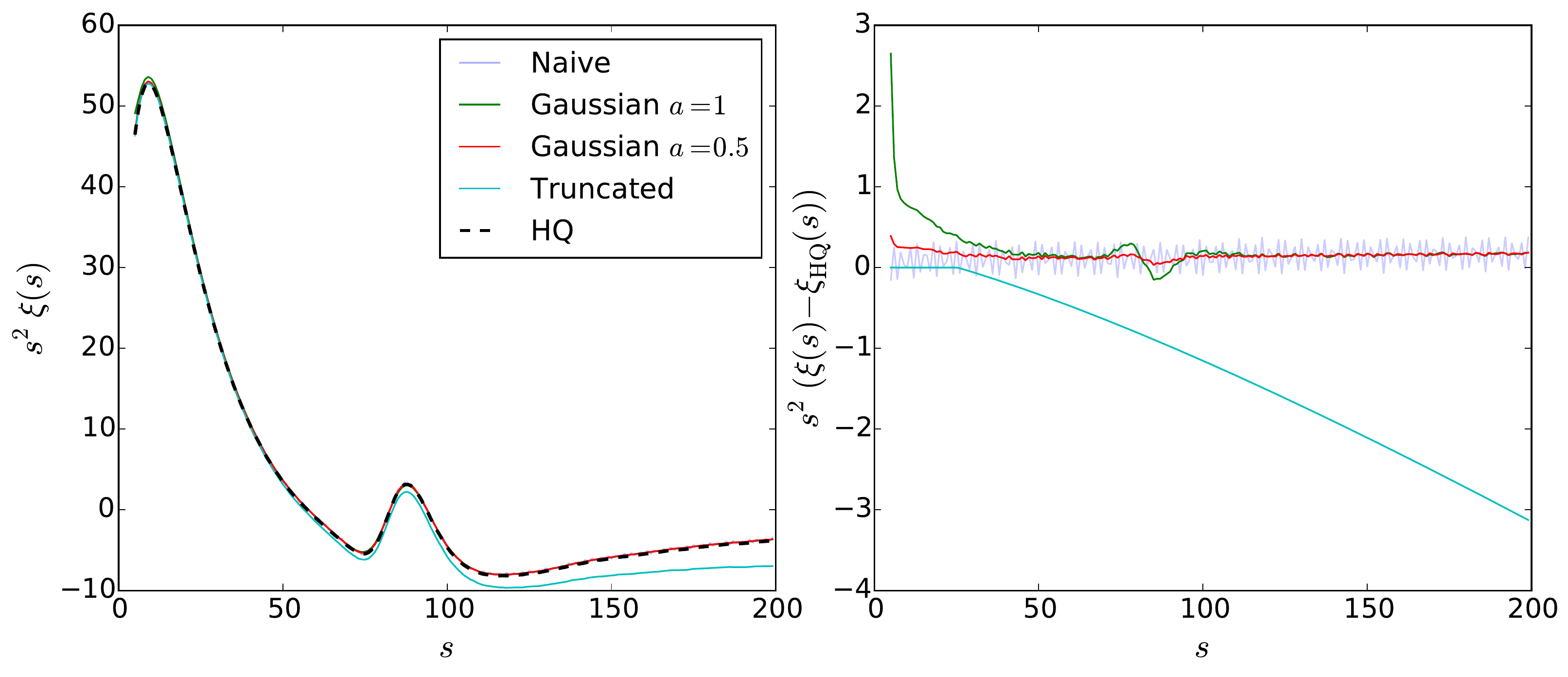}
  \end{center}
  \caption{A comparison of the different algorithms used to perform the numerical Fourier transformation. The power spectrum supplied to the algorithms consisted of 732 data points ranging up to $k = 223.56 \ h/$Mpc. The oscillations present in the naive spectrum are due to the fact the integration bounds are not infinite, and form due to the cumulative effect of the $\sin(ks)$ term when the integration bounds truncate the calculation when $\sin(ks) \neq 0$. The Gaussian method used by \citet{AndersonAubourg2012} with $a=1$ provided good convergence to the high quality algorithm at $s > 20 h^{-1}$ Mpc, with a slight deviation around the BAO peak itself and a general positive offset of approximately of $\Delta = 0.1 s^2 \xi(s)$ (which has negligible impact on cosmological fitting due to the marginalisation over power amplitude from $b^2$). Both the initial deviation and the peak deviation were greatly reduced in magnitude when $a$ was set to $0.5$ instead of $1$. The greatest deviation from the high quality algorithm was found by the truncated algorithm used in \citet{BlakeDavis2011}, where the truncation increased divergence as we go to larger separation.}
  \label{fig:pk2xicomp}
\end{figure}

The method employed by \citet{BlakeDavis2011} increases convergence by truncating the numerical integral after a certain point, corresponding to 900 periods of the $\sin(ks)$ term found in \eqref{eq:she}, whilst the method employed by \citet{AndersonAubourg2012} adds a Gaussian dampening term to equation \eqref{eq:she} such that it becomes
\begin{align}
\xi(s) = \frac{1}{(2 \pi)^3} \int 4 \pi k^2 \, P(k)\,  \frac{\sin(ks)}{ks} e^{-a^2 k^2},
\end{align}
where $a$ was set to $1 h^{-1}$ Mpc to damp signal at high high $k$. In addition to these two methods, a naive approach in which a supersampled power spectrum is integrated via trapezoids, and a high quality version, which supersamples each $\sin(ks)$ oscillation independently and sums the contributions from each period. This method, whilst providing high quality integration, takes too much computational time to be viable when using MCMC analysis (approximately one second per point in the correlation function). The results of the comparison are shown in in Figure \ref{fig:pk2xicomp}, which suggests the optimal algorithm to recover a high quality numerical integration whilst retaining sufficient speed is to the Gaussian dampening algorithm with $a=0.5$. A value of $a=0.1$ was also tested with positive results to ensure that this value of $a$ was robust to differing cosmological models with shifted BAO peaks, however it is recommended that any analysis which involves a highly varying BAO peak location should use a dynamic $a$ value. As this is not the case in the analysis found in this document, $a$ is simply fixed to $0.5$.

\begin{figure}[h!]
  \begin{center}
    \includegraphics[width=0.6\textwidth]{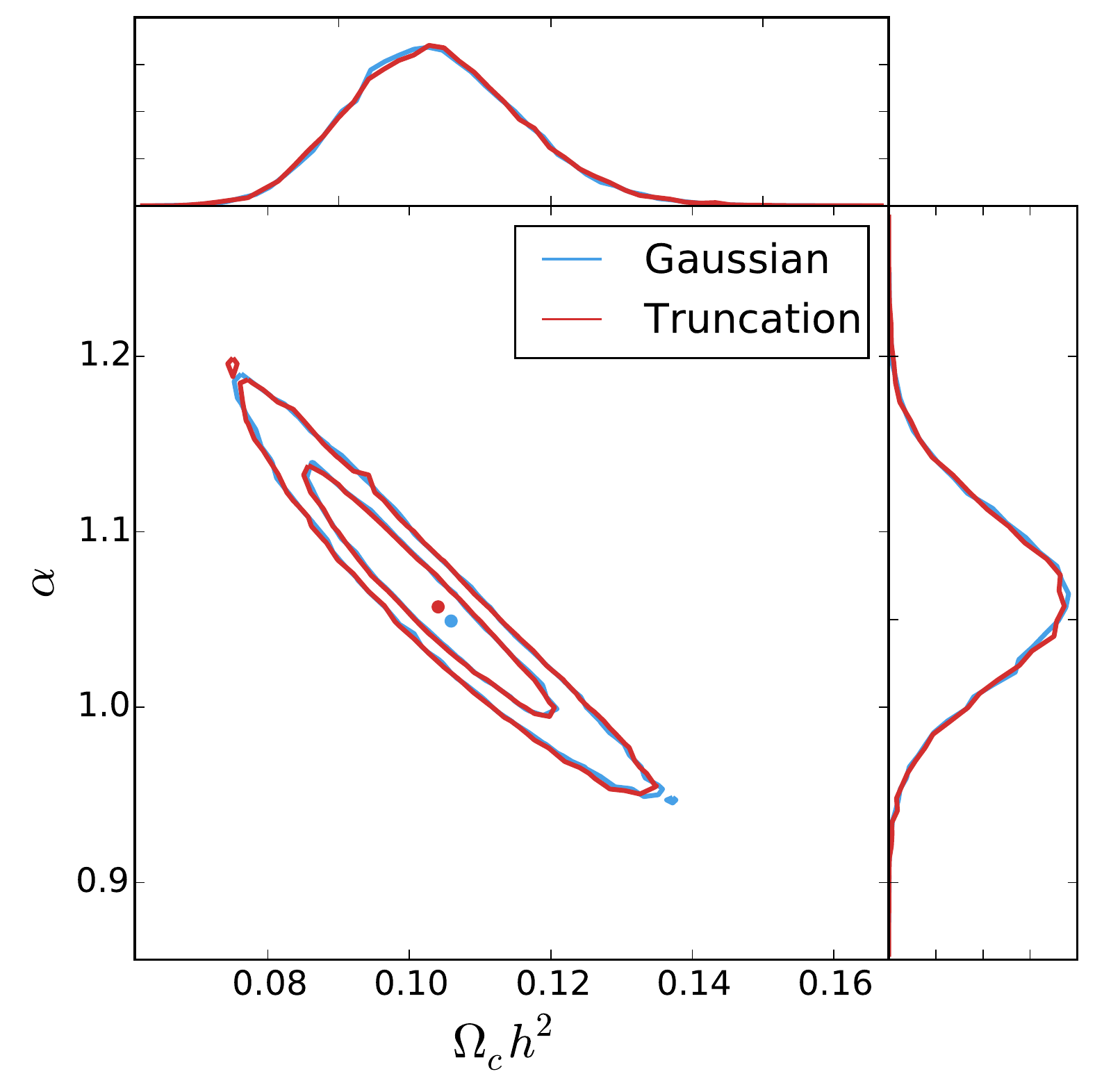}
  \end{center}
  \caption{Likelihood surfaces for $1\sigma$ and $2\sigma$ confidence levels and marginalised distributions were created using both the Gaussian $a=0.5$ and truncated method of moving from a power spectrum to a correlation function. Models were compared to the combined monopole moment of the 600 WizCOLA simulations in the $z=0.4$ redshift bin. Parameters $\beta, b^2, k_*, \sigma_v H(z)$ are marginalised over in these MCMC fits. Data noisiness exists from halting the MCMC algorithm early (2 million steps combined) after it became clear the two methods gave negligible differences.}
  \label{fig:BcosmoTest}
\end{figure}

Whilst Figure \ref{fig:pk2xicomp} shows what appears to be significance difference between the alternate methods (Gaussian with $a=1$ and truncation), we should realise the plots display $s^2 \xi(s)$, and at the scales of divergence ($\sim 100\, h^{-1}$ Mpc), this means any deviations are exaggerated by approximately four orders of magnitude. Considering this, it is unclear if the difference presented in Figure \ref{fig:pk2xicomp} is in any way significant, so a cosmological comparison was run using the combined 600 realisations of the WizCOLA simulation, to test the limits of these differences with data that should give tight constraints. The resulting likelihood surfaces and marginalised distributions detailed in Figure \ref{fig:BcosmoTest} show that the difference between these two algorithms is completely negligible.

\chapter{Effects of dataset truncation} \label{app:truncation}

\begin{figure}[h!]
  \begin{center}
    \includegraphics[width=0.7\textwidth]{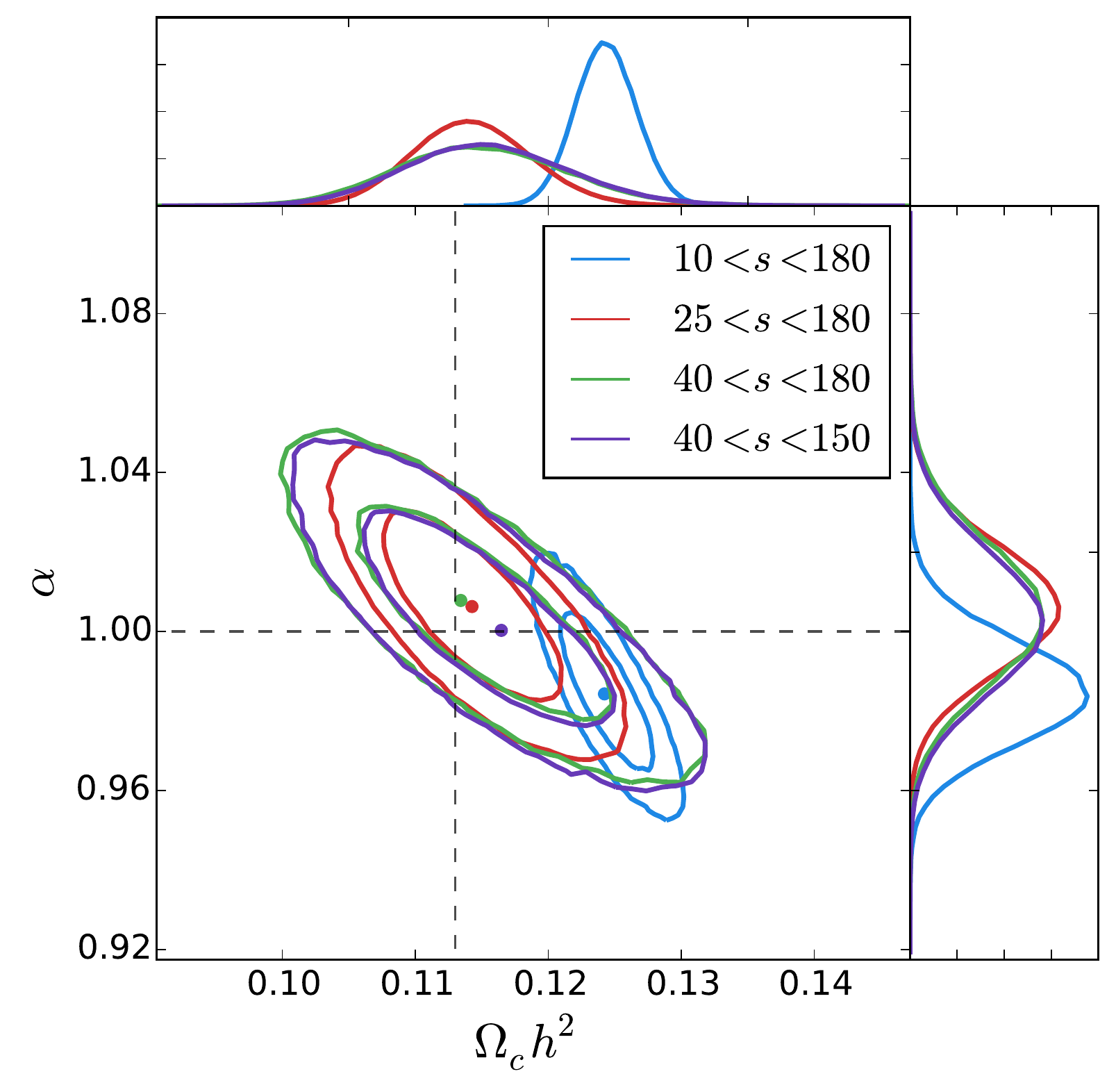}
  \end{center}
  \caption{Four different dataset truncation values are used in fitting to the WizCOLA $z=0.6$ mean dataset. Utilising the $10<s<180 h^{-1}$ Mpc range employed by \citet{BlakeDavis2011} provided strong constraints on the parameters $\Omega_c h^2$ and $\alpha$, but recovered values more than $3-\sigma$ away from the desired outcomes. Increasing the lower bound of the data due to inaccuracies in the model at low separation (following \citet{ChuangWang2012} shifted the recovered parameters to be well below $1-\sigma$ in deviation from the desired outcome, at the cost of larger uncertainty in the likelihood surfaces. A reduced upper bound was tested as well due to its presence in prior literature, however minimal impact was found by reducing the upper limit.}
  \label{fig:CdatasetTrunc}
\end{figure}

The failure of modern cosmological models at small separations and their similarity at large separations often lead to the use of truncated data sets when analysing the BAO signal, as detailed in \S\ref{sec:trunc}. As there is no agreement in prior literature as to what data ranges the standard BAO model is valid in, I investigate the effect data truncation has on the recovered parametrisations when fitting to the WizCOLA simulation data. In order to constrain statistical uncertainty as much as possible, fits were performed to the combined dataset, in which the input values are determined from the mean of all 600 realisations of the WizCOLA simulation. Multiple data ranges are then used in fits, with the desired output parameters for $\Omega_c h^2$ and $\alpha$ compared to the output fit parametrisations. These plots are shown in Figure \ref{fig:CdatasetTrunc}, and the outcome of the comparison is the decision to use a restricted dataset range of $25 < s < 180 \ h^{-1}$ Mpc.

\chapter{WizCOLA Covariance} \label{app:wizDiff}

\begin{figure}[h!]
  \begin{center}
    \includegraphics[width=\textwidth]{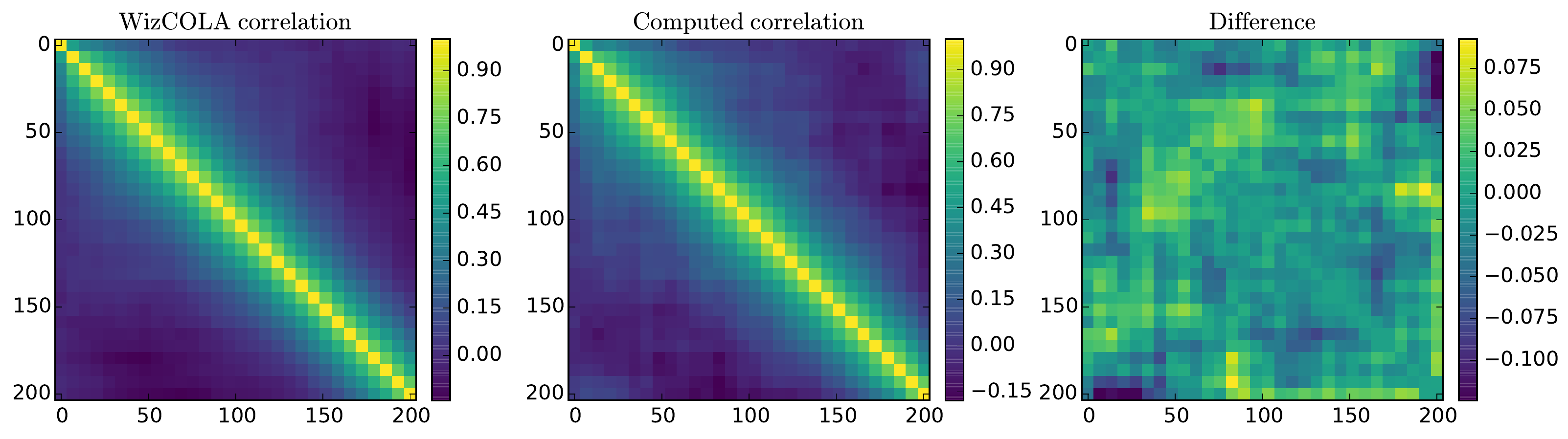}
  \end{center}
  \caption{Plots of the supplied WizCOLA correlation function and the computed correlation function. The difference is shown in the third plot, where the greatest differences can be observed in the off diagonal terms.}
  \label{fig:DfullCorrelationDifferences}
\end{figure}

When comparing the computed covariance matrix using the provided 600 WizCOLA realisations to the supplied covariance matrix from the WizCOLA data release, I found discrepancies in the off diagonal terms, which are illustrated in Figure \ref{fig:DfullCorrelationDifferences}. Contact with academics involved in the simulation creation has not been able to resolve the source of these differences. Visually, the WizCOLA correlation matrix appears smoother than the computed correlation matrix, and so a potential source of discrepancy would be if the WizCOLA covariance was calculated using more than the 600 realisations available in the data release, or if steps were deliberately taken to smooth the covariance matrix. Covariance smoothing is a topic of growing interest, and new efficient methodologies for smoothing \citep{OconnellEisenstein2015} makes this a topic that should be revisited with the WizCOLA mocks.\\

\begin{figure}[h!]
  \begin{center}
    \includegraphics[width=1\textwidth]{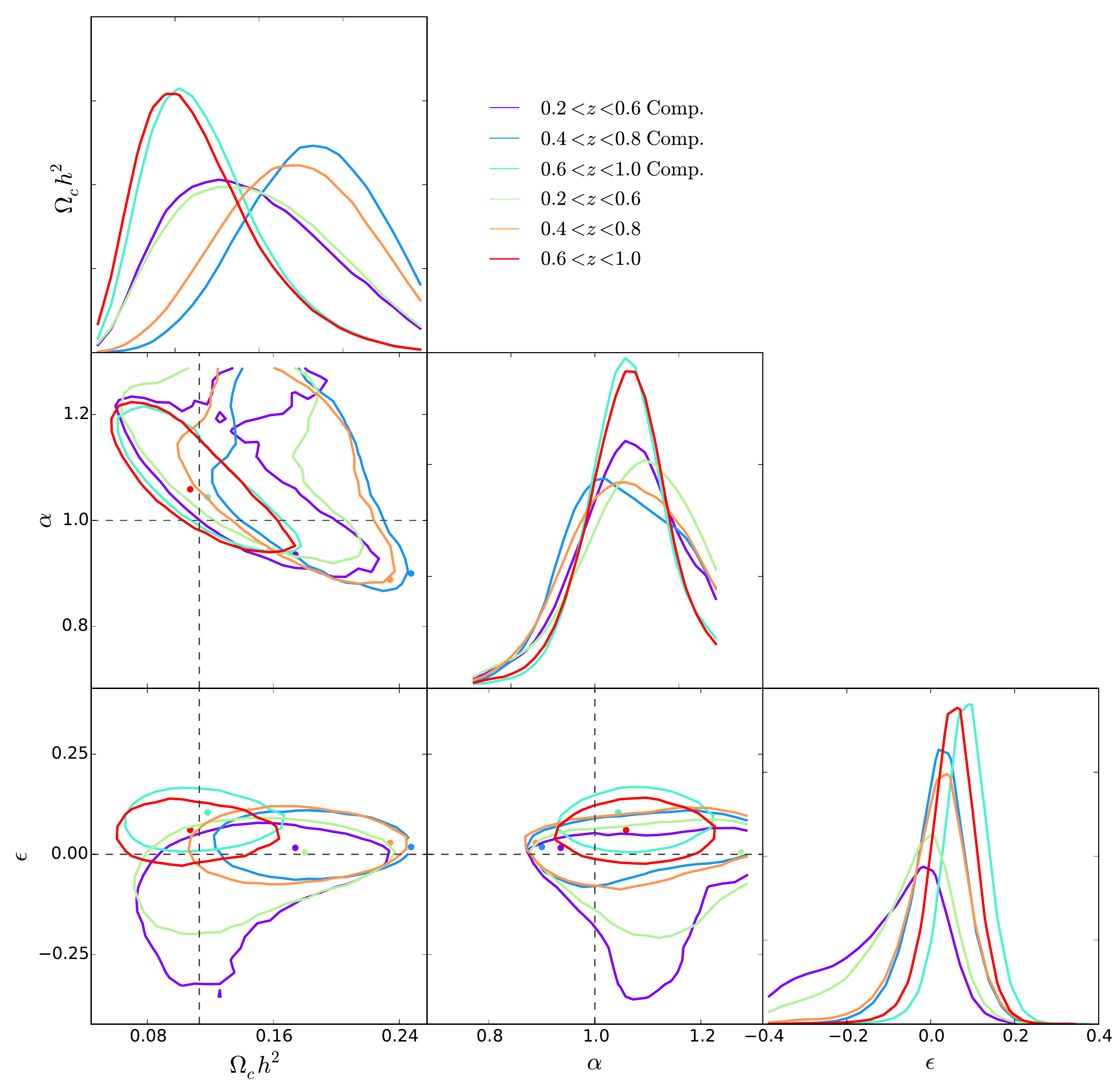}
  \end{center}
  \caption{Parameter fits to the WiggleZ multipole data using both the supplied and computed (labeled `Comp.' in the legend) covariance matrices.}
  \label{fig:DcomputedDiff}
\end{figure}

In order to determine the significance of difference in covariance, fits to the WiggleZ multipole data were formed with both covariance matrices, with the output marginalised distributions and likelihood surfaces shown in Figure \ref{fig:DcomputedDiff}. The output parameters are detailed in Table \ref{tab:Ddiff}, and we can see that changes in final parameters are well below the $1\sigma$ level, with a mean and median deviation of $0.18\sigma$ and $0.14\sigma$ respectively across all parameters and all bins. These results are strong enough that the cause of the covariance difference should be investigated, but as the differences fall well below a $1\sigma$ threshold it should not effect final output significantly.

\begin{table}[h]
\centering
\caption{A comparison between a multipole analysis when using either the provided WizCOLA covariance or the computed variance from WizCOLA realisations.}
\begin{tabular}{cc|ccc|ccc}
\specialrule{.1em}{.05em}{.05em} 
Sample & $z_{\rm{eff}}$ & \multicolumn{3}{c}{Provided}  & \multicolumn{3}{c}{Computed}\\
&  & $\Omega_m h^2$ &$\alpha$ & $\epsilon$ & $\Omega_m h^2$ & $\alpha$ & $\epsilon$ \\
\specialrule{.1em}{.05em}{.05em} 
$0.2 < z < 0.6$ & $0.44$ & $0.128^{+0.058}_{-0.036}$ & $1.12^{+0.13}_{-0.12}$ & $0.00^{+0.07}_{-0.13}$ & $0.127^{+0.053}_{-0.039}$ & $1.08^{+0.13}_{-0.11}$ & $-0.01^{+0.07}_{-0.17}$ \\
$0.4 < z < 0.8$ & $0.60$ & $0.172^{+0.042}_{-0.042}$ & $1.06^{+0.16}_{-0.11}$ & $0.03^{+0.06}_{-0.08}$ & $0.180^{+0.042}_{-0.034}$ & $1.02^{+0.18}_{-0.08}$ & $0.03^{+0.06}_{-0.07}$ \\
$0.6 < z < 1.0$ & $0.73$ & $0.098^{+0.038}_{-0.027}$ & $1.08^{+0.09}_{-0.10}$ & $0.06^{+0.05}_{-0.05}$ & $0.103^{+0.037}_{-0.026}$ & $1.08^{+0.09}_{-0.10}$ & $0.09^{+0.05}_{-0.05}$ \\
\specialrule{.1em}{.05em}{.05em} 
\end{tabular} \label{tab:Ddiff}
\end{table}

\end{appendices}


\end{document}